\documentclass[onecolumn,superscriptaddress,aps]{revtex4-1}
\usepackage{graphicx}
\usepackage{amsmath}
\usepackage{amsthm}
\usepackage{amssymb}
\usepackage{latexsym}
\usepackage{array}
\usepackage{hyperref}
\usepackage{float}
\usepackage{amsfonts}
\usepackage{dsfont}
\usepackage{mathrsfs}
\usepackage{verbatim}
\usepackage{bbold}
\usepackage{lipsum}
\usepackage[normalem]{ulem}

\usepackage{upgreek}

\usepackage{makecell}
\usepackage{adjustbox,lipsum}

\usepackage{algorithm}
\usepackage{algpseudocode}

\usepackage{color}

\newcommand{\bra}[1]{\left<#1\right|}
\newcommand{\ket}[1]{\left|#1\right>}
\newcommand{\abs}[1]{\bigl|#1\bigr|}
\newcommand{\norm}[1]{\left\lVert#1\right\rVert}

\newcommand{\braket}[2]{\left<{#1}|{#2}\right>}
\newcommand{\ketbra}[2]{\ket{#1}\!\!\bra{#2}}

\newtheorem{theorem}{Theorem}
\newtheorem{proposition}{Proposition}
\newtheorem{lemma}{Lemma}
\newtheorem{corollary}{Corollary}
\newtheorem{definition}{Definition}

\newtheorem{remark}{Remark}

\newcommand{\tr}[1]{\mbox{Tr}{#1}}

\begin{document}

\title{Learning at the Edge of Causality: Optimal Learning-Sample Complexity from No-Signaling Constraints}

\author{Jeongho~Bang}\email{jbang@yonsei.ac.kr}
\affiliation{Institute for Convergence Research and Education in Advanced Technology, Yonsei University, Seoul 03722, Republic of Korea}
\affiliation{Department of Quantum Information, Yonsei University, Incheon 21983, Republic of Korea}

\author{Kyoungho~Cho}
\affiliation{Institute for Convergence Research and Education in Advanced Technology, Yonsei University, Seoul 03722, Republic of Korea}
\affiliation{Department of Statistics and Data Science, Yonsei University, Seoul 03722, Republic of Korea}

\author{Jeongwoo~Jae}\email{jwjae@hanyang.ac.kr}
\affiliation{Department of Physics, Hanyang University, Seoul, 04763, Republic of Korea}

\date{\today}

\begin{abstract}
What ultimately fixes the sample cost of quantum learning---algorithmic ingenuity or physical law? We study this question in an arena where computation, learning, and causality collide. A twist on Grover's search that reflects about an \emph{a priori} unknown state can collapse the query complexity from $O(\sqrt{N})$ to $O(\log N)$ over a search space $N$, i.e., an exponential speedup. Yet, standard quantum theory forbids such a unknown-state reflection (no-reflection theorem). We therefore build a state-learning-assisted architecture, called ``amplify--learn,'' which alternates the coherent amplitude amplification with state learning. Embedding this amplify--learn into the Bao--Bouland--Jordan no-signaling framework, we show that the logarithmic-round dream would open a super-luminal communication channel unless each round expends the learning-sample and reflection-circuit budgets scaling at least as $\Omega(\sqrt{N}/\log N)$. In parallel, we derive tight computational learning-theoretic sample bounds for learning circuit-generated pure states, revealing a state-universal ansatz ``lock'' at order $N$ in the worst case. The dramatic closure is that no-signaling does not merely veto the unphysical primitive, but it fixes the only consistent reflection-circuit complexity, and feeding this causality-enforced complexity into the computational learning bound makes it collapse onto the very same $\sqrt{N}/\log N$ scaling demanded by no-signaling alone. No-signaling thus acts as a \emph{regulator of learnability}: a constraint that mediates between physics and computation, welding query, gate, and sample complexities into a single causality-compatible triangle.
\end{abstract}

\maketitle

\setcounter{tocdepth}{-100}

How far can computation and learning be accelerated, and what ultimately fixes their cost? Conventional answers are phrased in computational terms---hypothesis-class complexity, oracle lower bounds, and/or information-theoretic limits~\cite{Bernstein1993,Kearns1994}. Yet any information-processing device is also a physical system, and physics imposes the constraints that are invisible in purely computational models~\cite{Landauer1961,Bekenstein1981,Lloyd2000}. A paradigmatic example is relativistic causality, operationally captured by the no-signaling principle: local actions cannot transmit information to a space-like separated partner~\cite{Simon2001,Brukner2014}. While no-signaling is best known as a constraint on nonlocal correlations, it can also be viewed as a ``meta--complexity'' principle that rules out some primitives which would collapse established complexity barriers~\cite{brassard2006,botteron2024}. This perspective motivates the central question we ask here: can no-signaling regulate learnability?

A paradigmatic example of the interplay between causality and quantum complexity is Grover’s quantum search~\cite{Grover1997}. Grover's algorithm finds a marked item in an $N$-element space using $O(\sqrt{N})$ oracle queries, and the Bennett-Bernstein-Brassard-Vazirani hybrid argument shows this scaling is optimal in standard quantum theory~\cite{Bennett1997,Zalka1999}. More recently, a complementary viewpoint has emerged: in broad classes of modified quantum theories, any primitive that achieves a super-Grover scaling $o(\sqrt{N})$ can be repurposed (up to polynomial overhead) into a super-luminal signaling protocol, and conversely~\cite{Bao2016}. In this sense, the familiar $\sqrt{N}$ barrier can be regarded not only as a query-model theorem, but as a causality constraint: beating it would open a faster-than-light classical channel.

In this work, we extend this causality-complexity connection to another layer, i.e., the cost of quantum state learning. As a controlled thought experiment, we consider a modified amplitude-amplification dynamics in which the usual fixed reflection about the initial state is replaced, round by round, by a reflection about the previous output generated inside the algorithm. This seemingly mild change induces a qualitatively different geometry: the target overlap can grow ``cubically'' per round, compressing the number of oracle-query rounds from $O(\sqrt{N})$ down to $O(\log{N})$. However, realizing this dynamics would require a programmable reflection about an a priori unknown quantum state, which is forbidden by a no-reflection theorem~\cite{Kumar2011,Yee2020}. We therefore introduce an explicit ``amplify--learn'' architecture~\cite{Baek2025}: each amplification round is paired with a state-learning step that consumes many fresh copies of the intermediate state to synthesize (approximately) the reflection needed for the next round.

Our main results show that once no-signaling is imposed, the learning resources become sharply determined. Embedding the amplify--learn architecture into the no-signaling framework yields a causality-enforced lower bound: to maintain constant target accuracy per round, the learning-sample budget and the complexity of the learned reflection circuits must scale at least as $\sqrt{N}/\log N$. This bound is significant precisely because a generic state learning is much harder: in a state-universal regime, where the learning ansatz must approximate arbitrary pure states in a Hilbert space of dimension $N$ (equivalently, explore $\mathrm{SU}(N)$), computational learning theory alone locks the worst-case sample complexity at order $N$, and no algorithm can do better. The amplify-learn setting does not demand this universal power; instead, causality selects a narrower corridor of physically admissible intermediate states and forces a matching reflection complexity, thereby ``unlocking'' the admissible learning-sample scaling down from $N$ to $\sqrt{N}/\log N$. Taken together, our results align three a priori distinct constraints---oracle queries, unknown-state reflection complexity, and learning circuit samples---into a single scaling hierarchy enforced by relativistic causality.

\section*{Result and Theorems}

\subsection*{Previous-output reflections and logarithmic rounds}

We begin with a controlled thought experiment that exposes the key geometric mechanism behind our later bounds. Consider unstructured search over an $N$-dimensional computational basis $\{ \ket{x} \}_{x=0}^{N-1}$ with a unique marked element $x^\star$, encoded as the target state $\ket{\tau}:=\ket{x^\star}$. The oracle is the phase flip, defined as $\hat{R}_\tau = \hat{\mathds{1}} - 2\ketbra{\tau}{\tau}$. In standard amplitude amplification, the second reflection is fixed as a reflection about the initial state $\ket{\psi_0}$ (e.g., the uniform superposition), which leads to the familiar $\Theta(\sqrt{N})$ query complexity.

Now modify only this second reflection: at each round, reflect not about the fixed $\ket{\psi_0}$, but about the previous output state itself. Let $\ket{\psi_r}$ denote the state at round $r$ and define the (state-dependent) reflection: $\hat{R}_{\psi_r}=\hat{\mathds{1}} - 2\ketbra{\psi_r}{\psi_r}$~\cite{householder1958}. The idealized update rule is then
\begin{eqnarray}
\ket{\psi_{r+1}} = e^{-i\pi}\hat{R}_{\psi_r}\hat{R}_\tau\ket{\psi_r},
\label{eq:prev_output_update}
\end{eqnarray}
where the global phase $e^{-i\pi}$ is physically irrelevant.  This process is perfectly well-defined at the level of linear algebra, because for any fixed $|\psi_r\rangle$ the product of two reflections is a rotation in the two-dimensional subspace spanned by $\ket{\tau}$ and $\ket{\psi_r}$.

To make this explicit, let us introduce the orthonormal basis $\{ \ket{\tau}, \ket{\tau_\perp} \}$, where $\ket{\tau_\perp}$ is the normalized superposition of all non-marked basis states. Then, parameterize the round-$r$ state by an angle $\theta_r \in (0, \pi/2)$: $\ket{\psi_r} = \sin\theta_r \ket{\tau} + \cos\theta_r \ket{\tau_\perp}$. A direct calculation shows that applying $\hat{R}_{\psi_r}\hat{R}_\tau$ maps the target overlap as
\begin{eqnarray}
\braket{\tau}{\psi_{r+1}} = \sin(3\theta_r).
\label{eq:cubic_amp}
\end{eqnarray}
Hence, as long as $\theta_r<\pi/6$, the angle itself triples $\theta_{r+1}=3\theta_r$; thereby, $\theta_r=3^r\theta_0$. Starting from the usual uniform initialization, $\theta_0 = \arcsin(1/\sqrt{N}) \approx 1/\sqrt{N}$. Therefore, the target overlap reaches a constant (i.e., success probability of order unity) after
\begin{eqnarray}
r = \Theta(\log N)
\label{eq:log_rounds}
\end{eqnarray}
rounds, because $3^r\theta_0$ becomes $O(1)$ when $r=O(\log N)$. Crucially, each round invokes the oracle $\hat{R}_\tau$ only once, so the query complexity of this idealized previous-output-reflection protocol would be $O(\log N)$, an exponential improvement over Grover scaling. This dramatic acceleration hinges entirely on the availability of the programmable reflection $\hat{R}_{\psi_r}$ about an a priori unknown state produced within the algorithm.

\subsection*{No-reflection theorem and an ``amplify--learn'' realization}

The above logarithmic-round dynamics relies on a primitive, that is, a programmable reflection about an a priori unknown state. However, the obstruction is captured by a no-go statement:
\begin{theorem}[no-reflection theorem~\cite{Kumar2011}]
There exists no unitary $\hat{U}$ that, for all normalized $\ket{\chi}$ and all target states $\ket{\phi}$, implements
\begin{eqnarray}
\hat{U}\Big(\ket{\chi}_c \otimes \ket{\phi}_t\Big) = \ket{\chi}_c \otimes \Big(\hat{\mathds{1}} - 2\ketbra{\chi}{\chi}\Big)\ket{\phi}_t .
\end{eqnarray}
\end{theorem}
The reason is ultimately linearity: demanding the above action on multiple nonorthogonal program states forces contradictory constraints when $\hat{U}$ is applied to their superpositions. Therefore, the previous-output reflection cannot be an exact one-copy-controlled gate in quantum theory, and any physical implementation must approximate $\hat{R}_{\psi_r}$ by expending additional resources.

We implement this approximation via state learning, which turns many copies of an unknown intermediate state into a classical circuit description. Specifically, assume access to a learning routine $\mathcal{F}$ that takes $M_s$ independent copies of an unknown $n$-qubit (and hence, $N=2^n$ dimensional) pure state $\ket{\psi}$ and outputs the classical parameters $\boldsymbol\theta_\psi$ specifying an ansatz circuit $\hat{A}(\boldsymbol\theta_\psi)$, such that $\hat{A}(\boldsymbol\theta_\psi)\ket{\mathbb{0}} \simeq \ket{\psi}$ with $\ket{\mathbb{0}}=\ket{0}^{\otimes n}$. Given $\hat{A}(\boldsymbol\theta_\psi)$, we synthesize an approximate reflection about $\ket{\psi}$ by conjugating a fixed reference reflection $\hat{R}_0 = \hat{\mathds{1}} - 2 \ketbra{\mathbb{0}}{\mathbb{0}}$:
\begin{eqnarray}
\hat{R}(\boldsymbol\theta_\psi) = \hat{A}(\boldsymbol\theta_\psi)\hat{R}_0\hat{A}^\dagger(\boldsymbol\theta_\psi) \simeq \hat{R}_\psi.
\end{eqnarray}

\begin{figure}[t]
  \centering
  \includegraphics[width=1.00\textwidth]{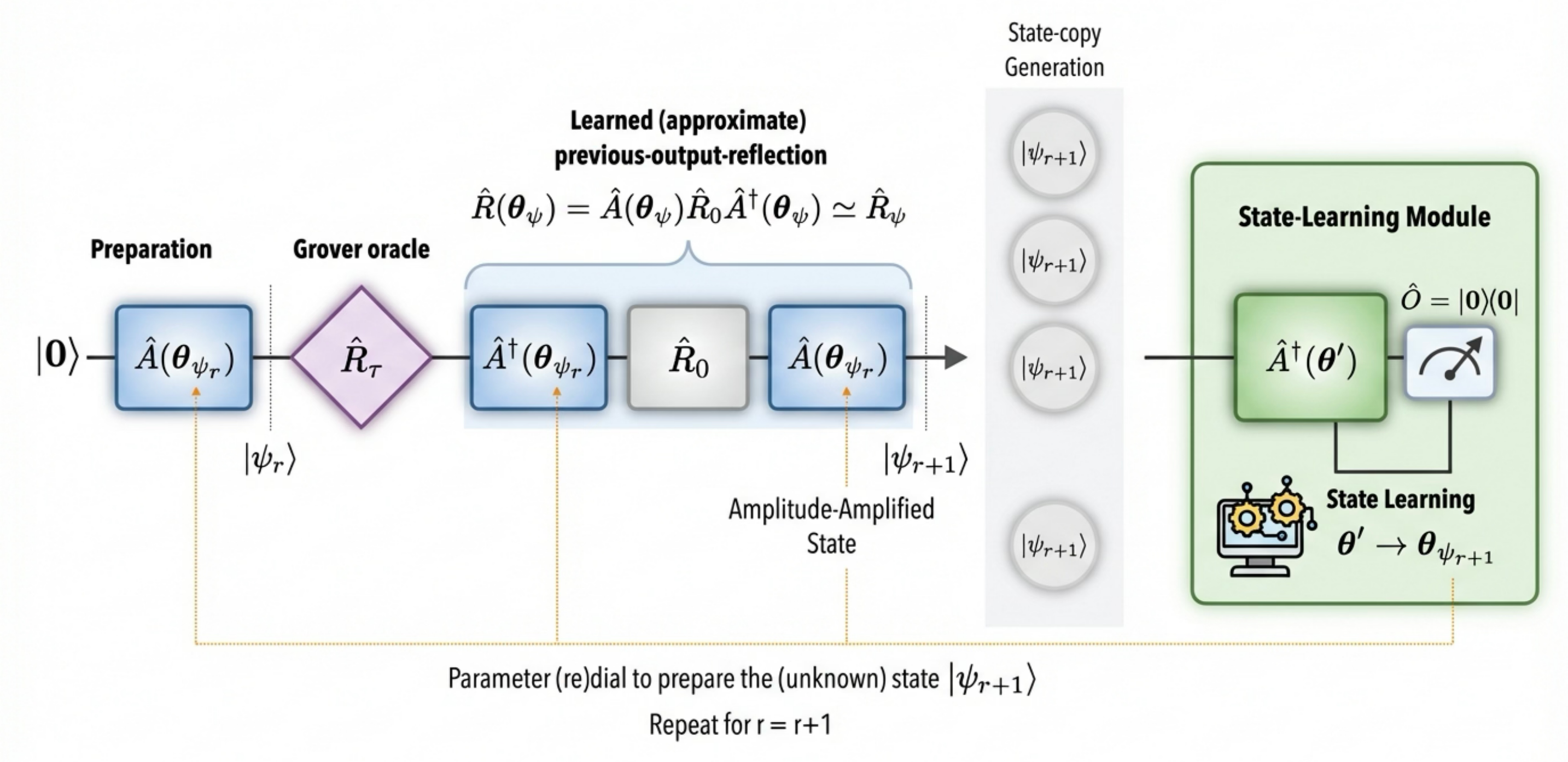}
  \caption{{\bf Schematic of the ``amplify--learn'' protocol.} In a certain round $r$, $\hat{A}(\boldsymbol{\theta}_{\psi_r})\hat{R}_0\hat{A}(\boldsymbol{\theta}_{\psi_r})^\dagger$ consists of $\hat{R}_{\psi_r}$ with the learned parameter $\boldsymbol{\theta}_{\psi_r}$; i.e., $\hat{A}(\boldsymbol{\theta}_{\psi_r})\ket{0} = \ket{\psi}$. With the target reflection $\hat{R}_\tau$ and $\hat{A}(\boldsymbol{\theta}_{\psi_r})\hat{R}_0\hat{A}(\boldsymbol{\theta}_{\psi_r})^\dagger$ can generate the target-amplitude amplified state $\ket{\psi_{r+1}}$. Here, note that the state $\ket{\psi_{r+1}}$ is unknown. We generate the copies of $\ket{\psi_{r+1}}$ and feed into a state-learning module, in which $\boldsymbol{\theta}_{\psi_r}$ is updated to $\boldsymbol{\theta}_{\psi_{r+1}}$. Concretely, the circuit $\hat{A}(\boldsymbol{\theta}')^\dagger$ (incorporated in the state-learning module with observable $\hat{O}=\ketbra{0}{0}$) learns $\hat{A}(\boldsymbol{\theta}_{\psi_r})\hat{R}_0\hat{A}(\boldsymbol{\theta}_{\psi_r})^\dagger\hat{R}_\tau$, so that $\hat{A}(\boldsymbol{\theta}')^\dagger\ket{\psi_{r+1}} = \ket{0}$, or equivalently, $\hat{A}(\boldsymbol{\theta}')\ket{0} = \ket{\psi_{r+1}}$. By identifying $\boldsymbol{\theta}'$ and (re)dialing $\boldsymbol{\theta}_{\psi_r} \to \boldsymbol{\theta}' := \boldsymbol{\theta}_{\psi_{r+1}}$, we can repeat the aforementioned processes.}
  \label{fig:amplify-learn-protocol}
\end{figure}

This yields an explicit two-stage architecture---dubbed as ``amplify--learn''~\cite{Baek2025} (see Fig.~\ref{fig:amplify-learn-protocol}). In the amplify stage of round $r$, one applies $\hat{R}(\boldsymbol\theta_{\psi_r})\hat{R}_\tau$ to the current state to produce $\ket{\psi_{r+1}}$. In the learn stage, one prepares (or collects) the fresh copies of $\ket{\psi_{r+1}}$ and runs $\mathcal{F}$ to update $\boldsymbol\theta_{\psi_{r+1}}$, which is then fed forward to the next round. Under perfect learning (idealized), $\hat{R}(\boldsymbol\theta_{\psi_r})=\hat{R}_{\psi_r}$ and the physical loop reproduces the previous-output-reflection map; with imperfect learning, the deviation is controlled by the learning accuracy and the available sample budget $M_s$.

\subsection*{Causality-enforced lower bounds}

The remaining question is then not \emph{whether} we can approximate $\hat{R}_{\psi_r}$, but \emph{how expensive} this must be so that the overall protocol cannot become the super-Grover. The key physical input is that the super-Grover search and super-luminal signaling are operationally equivalent in a broad class of theories that modify standard quantum dynamics by an extra primitive operation (Bao-Bouland-Jordan~\cite{Bao2016}). In particular, any primitive that reduces the unstructured-search query cost below $\Theta(\sqrt{N})$ can be wrapped into a protocol that transmits a classical bit across a space-like separation; conversely, any such signaling capability implies an oracle speedup.  Thus, enforcing no-signaling is equivalent to enforcing the Grover threshold as a physical constraint, not merely an algorithmic result.

We now embed our amplify--learn architecture into this causality framework.  Let $N=2^n$ and let $r(N)=\Theta(\log N)$ be the number of amplification rounds required by the ideal previous-output-reflection dynamics. In each round $r$, the intermediate state $\ket{\psi_r}$ depends on the hidden marked item $\tau$, hence cannot be prepared without invoking the oracle $\hat{R}_\tau$. However, once the previous round has been learned, additional copies of $\ket{\psi_r}$ can be generated by running the already learned circuit for round $r-1$ and applying a constant number of fixed amplification steps.  Consequently, there exists a constant $c_{\rm prep}>0$ such that preparing one fresh copy of $\ket{\psi_r}$ costs at least $c_{\rm prep}$ oracle uses, independent of $r$. If the learner consumes $M_{\rm s}$ copies per round to synthesize $\hat{R}_{\psi_r}$ to a fixed constant accuracy, then the training-phase query cost obeys
\begin{eqnarray}
Q_{\rm train}(N) \ge c_{\rm prep}M_{\rm s}r(N).
\end{eqnarray}
The final ``production'' run uses only $O(r(N))$ additional oracle calls and is negligible compared to $M_{\rm s}r(N)$ in the regime of interest.  Therefore the total oracle usage satisfies
\begin{eqnarray}
Q_{\rm tot}(N) \ge c_{\rm prep}M_{\rm s}r(N).
\end{eqnarray}
No-signaling, via the Bao-Bouland-Jordan equivalence, enforces the Grover lower bound in operational form,
\begin{eqnarray}
Q_{\rm tot}(N) \ge c_Q\sqrt{N}
\end{eqnarray}
for some constant $c_Q>0$.  Combining these relations with $r(N)=\Theta(\log N)$ yields the first central constraint: the learning sample budget per round must scale as $\sqrt{N}/\log N$, otherwise the total query count collapses below the Grover threshold and the protocol would open a super-luminal channel.

A parallel causality argument constrains the gate cost of the unknown-state reflections.  Let $G_{\rm ref}(n)$ be the (two-qubit) gate complexity required to implement a single learned $\hat{R}_{\psi_r}$ (to constant precision). Since the logarithmic-round search invokes $\hat{R}_{\psi_r}$ a constant number of times per round, the reflection depth of the full circuit is at most
\begin{eqnarray}
D_{\rm ref}(N) \le c_DG_{\rm ref}(n)r(N)
\end{eqnarray}
for a constant $c_D>0$.  In the signaling embedding, if $D_{\rm ref}(N)=o(\sqrt{N})$, then for sufficiently large $N$ Alice would complete the distinguishing experiment in time shorter than the light-travel time from Bob, violating no-signaling.  Hence, $D_{\rm ref}(N) \ge c_{\rm NS}\sqrt{N}$ for some constant $c_{\rm NS}>0$, which again forces $G_{\rm ref}(n) \gtrsim \sqrt{N}/\log N$.
\begin{theorem}[No-signaling enforces a unique $\sqrt{N}/\log N$ scaling]
\label{thm:causal_bounds}
Fix a constant target accuracy for synthesizing each previous-output reflection $\hat{R}_{\psi_r}$ in the amplify--learn architecture, and let $r(N)=\Theta(\log N)$ denote the number of rounds in the ideal previous-output-reflection search. If the overall protocol is required to respect operational no-signaling (in the Bao-Bouland-Jordan sense), then for all sufficiently large $N=2^n$ the following per-round lower bounds hold:
\begin{eqnarray}
M_{\rm s} \ge \Omega\left(\frac{\sqrt{N}}{\log N}\right),
\quad
G_{\rm ref}(n) \ge \Omega\left(\frac{\sqrt{N}}{\log N}\right).
\label{eq:causal_bounds}
\end{eqnarray}
\end{theorem}

Here, note that although no-signaling is not uniquely quantum, the $\sqrt{N}/\log N$ constraint derived here is: it follows from enforcing operational no-signaling in an entanglement-enabled oracle setting (Bao-Bouland-Jordan), where any $o(\sqrt{N})$ unstructured-search speedup would constitute a superluminal channel. In a purely classical black-box setting the baseline limit is $\Omega(N)$, so the present amplify--learn and no-reflection tradeoff should be read as a constraint on simulating a forbidden ``quantum'' primitive (see Methods).

\subsection*{Computational learning theory viewpoint: the universal-ansatz ``lock'' and the no-signaling ``unlock''}

The no-signaling bounds in Eq.~(\ref{eq:causal_bounds}) can be read in a complementary way: it identifies which learning problem is actually being posed by our amplify--learn architecture. From the perspective of computational learning theory, one fixes a hypothesis class of the circuit-generated pure states,
\begin{eqnarray}
\mathcal{S}_{n,G} := \Big\{ \ket{\psi}=\hat{U}\ket{0}^{\otimes n} : \hat{U}\in\mathcal{U}_{n,G}\Big\},
\end{eqnarray}
where $\mathcal{U}_{n,G}$ denotes the family of $n$-qubit circuits with at most $G$ two-qubit gates, and one asks for the worst-case number of copies $M_s$ needed to learn $\hat{\rho}=\ket{\psi}\bra{\psi}$ within the trace-distance error $\epsilon$ and failure probability $\delta$. In this purely computational setting (arbitrary collective POVMs and post-processing), the dominant scaling is governed by the effective size of the class:
\begin{theorem}[Computational learning-sample lower bound for circuit-generated states]
\label{thm:comp_bound}
For $0 < \epsilon < 1$ and constant $\delta$, any learner that outputs $\hat{\rho}_h$ satisfying $D(\hat{\rho}, \hat{\rho}_h) \le \epsilon$ with probability at least $1-\delta$ for all $\hat{\rho}\in\mathcal{S}_{n,G}$ must use
\begin{eqnarray}
\tilde{M}_s(n,G,\epsilon,\delta) \ge \frac{c}{\epsilon^2}\min\{2^n,G\},
\end{eqnarray}
for a universal constant $c>0$.
\end{theorem}

This theorem already exposes a dramatic dichotomy. If one insists on a universal learning ansatz---meaning that the circuit family must approximate an $\epsilon$-net of all $n$-qubit pure states, equivalently explore $\mathrm{SU}(2^n)$ up to finite accuracy---then the circuit complexity itself is exponential, $G_{\rm univ}(n)=\Theta(2^n)$. Feeding this into the computational bound, we yield the familiar tomography ``lock''~\cite{haah2016}:
\begin{corollary}[Universal-ansatz lock at $N=2^n$ samples]
If the learner is state-universal, then
\begin{eqnarray}
\tilde{M}_s(n,G_{\rm univ},\epsilon,\delta) \ge \tilde{\Omega}\Big(\frac{2^n}{\epsilon^2}\Big) = \tilde{\Omega}\Big(\frac{N}{\epsilon^2}\Big).
\end{eqnarray}
\end{corollary}

What completes our story is that the amplify--learn protocol does not demand this universal power. It only ever needs reflections about the highly structured intermediate states $\ket{\psi_r}$ produced by the search dynamics. Here, the causality enters as a selection rule: no-signaling forbids the reflections from becoming ``too cheap,'' because cheap reflections would reinstate the logarithmic-round search and thereby a super-luminal channel. Quantitatively, a gate lower bound $G_{\rm ref}(n)\ge\Omega(\sqrt{N}/\log N)$ in Eq.~(\ref{eq:causal_bounds}) is enforced for synthesizing the unknown-state reflections used in the rounds. Thus, by substituting this \emph{causality-enforced} gate complexity into the \emph{computational bounds}, we can yield the corresponding learning ``unlock'',
\begin{eqnarray}
\tilde{M}_s(n,G_{\rm ref},\epsilon,\delta) \ge \Omega\Big(\frac{1}{\epsilon^2}\frac{\sqrt{N}}{\log N}\Big),
\end{eqnarray}
matching the no-signaling-derived scaling (up to the accuracy factor). In this sense, no-signaling does more than rule out an unphysical primitive: it carves out a causality-compatible corridor of learnability, collapsing a priori admissible scalings into the square-root law that keeps query, gate, and sample costs aligned. 

\subsection*{Physics-computation concordance: $G_{\rm ref}$ as the bridge and ``Three Bounds'' unification}

\begin{figure}[t]
  \centering
  \includegraphics[width=0.75\textwidth]{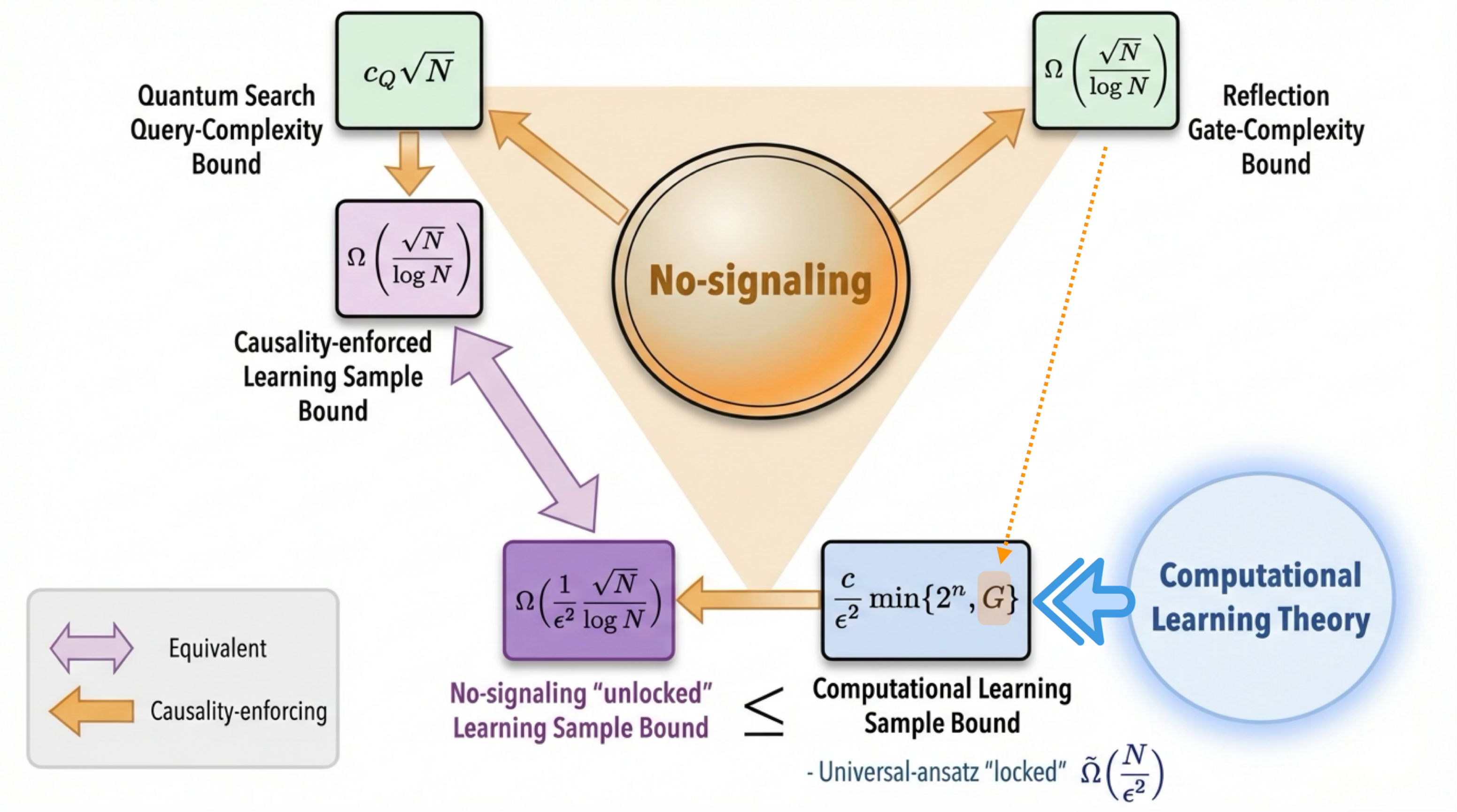}
  \caption{{\bf Causality triangle and physics-computation concordance} Embedding our amplify--learn loop within the Bao--Bouland--Jordan equivalence (super-Grover search $\Leftrightarrow$ superluminal signaling) upgrades no-signaling into an operational complexity principle: $Q_{\rm tot}(N)$ must remain $\Omega(\sqrt{N})$, which forces the per-round costs $M_s$ and $G_{\rm ref}$ to scale at least $\Omega(\sqrt{N}/\log N)$. In parallel, the computational learning theory yields $\tilde M_s(n,G,\epsilon,\delta)\ge (c/\epsilon^2)\min\{2^n,G\}$, giving the universal-ansatz tomography lock $\tilde{M}_s=\tilde\Omega(N/\epsilon^2)$ for $G=\Theta(2^n)$. The crucial closure is that in amplify--learn the learned reflection circuit is the hypothesis: no-signaling fixes $G=G_{\rm ref}$, and the computational bound collapses to $\tilde{M}_s \ge \Omega((1/\epsilon^2)\sqrt{N}/\log N)$, matching the no-signaling-only-derived bound. Equivalently, the logarithmic number of rounds is exactly compensated by the per-round reflection cost, restoring the Grover $\sqrt{N}$ barrier. No-signaling thus acts as a regulator of learnability, carving out the causality-compatible corridor and aligning query, gate and sample complexities into one triangle.}
  \label{fig:unified_view}
\end{figure}

We are now in a position to complete the loop in the most dramatic way: the physics of no-signaling yields the \emph{causality-enforced} bound $M_s$ by forbidding any overall search resource from dropping below Grover's $\sqrt{N}$ threshold, while computational learning theory yields the \emph{computational} learning-sample bound $\tilde{M}_s$ as a function of the hypothesis-class complexity (here, the circuit complexity needed to realize unknown-state reflections). The crucial point is that in amplify--learn, these are not independent: the learned reflection circuit is the hypothesis, and its gate complexity is precisely the parameter $G$ that controls $\tilde{M}_s$. No-signaling therefore fixes the only consistent choice of $G_{\rm ref}$, and this forces $\tilde{M}_s$ to collapse onto $M_s$.
\begin{corollary}[``Three Bounds'' unification into a causality-compatible triangle]
Under {\bf Theorem~\ref{thm:causal_bounds}} and~{\bf \ref{thm:comp_bound}}, the resources of (i) oracle queries, (ii) reflection-circuit complexity, and (iii) learning samples cannot be chosen independently.  Instead, they are compelled to satisfy, simultaneously,
\begin{eqnarray}
Q_{\rm tot}(N)\ge\Omega(\sqrt{N}),
\quad
G_{\rm ref}(n)\ge\Omega\left(\frac{\sqrt{N}}{\log N}\right),
\quad
M_{\rm s}\ge\Omega\left(\frac{\sqrt{N}}{\log N}\right),
\end{eqnarray}
so that the logarithmic number of rounds is exactly compensated by the per-round cost of implementing the unknown-state reflections.
\end{corollary}

Conceptually, this ``triangle'' is stronger than what computational learning theory alone dictates. Without causality, one can a priori imagine many admissible scalings for learning pure states, ranging from linear in an ansatz size up to the full $N$-dimensional tomography rate.  No-signaling removes this ambiguity: in the amplify--learn setting, any attempt to make reflections and learning ``too cheap'' would immediately translate into an o$(\sqrt{N})$ unstructured search and hence into super-luminal signaling.  In this sense, no-signaling acts as a \emph{regulator of learnability}, selecting the causality-compatible scaling law and aligning three ostensibly distinct lower bounds into one structured resource hierarchy. In other words, no-signaling does not merely add \emph{one more} inequality; it is the selector that aligns the computational and physical notions of difficulty. Note that the universal-ansatz world remains exponentially hard---$2^n$ samples are unavoidable when one demands arbitrary-state learnability---but causality pins down $G_{\rm ref}$ and thereby pins down the admissible learning-sample scaling. This view is summarized in Fig.~\ref{fig:unified_view}.

\section*{Implications and Outlook}

Our results sharpen a simple but underexplored moral: in quantum information processing, the ultimate cost of learning is not dictated solely by clever algorithm design, but can be fixed by the same physical principles that enforce causality. We made this point concrete by treating the state learning as a first-class resource inside a Grover-type search architecture. This led to an explicit amplify--learn architecture, where each amplification round is coupled to a state-learning step that synthesizes an approximate reflection from many copies of the intermediate state. The central conclusion is strikingly rigid: if one tries to make this learning assistance ``too cheap,'' the resulting effective query complexity drops below the Grover's $\sqrt{N}$ threshold (at fixed per-round accuracy), which is operationally equivalent to enabling faster-than-light signaling in the Bao-Bouland-Jordan framework. Enforcing no-signaling therefore compels a specific per-round resource investment, scaling as $\sqrt{N}/\log N$ in both the learning-sample budget and the gate complexity required to implement (approximate) unknown-state reflections. In this sense, causality does not merely forbid unphysical primitives; it quantitatively calibrates how closely they may be simulated. 

From the viewpoint of computational learning theory, the causality-enforced bounds are even more revealing. In a universal-ansatz setting where one attempts to learn arbitrary pure states in an $N$-dimensional Hilbert space (equivalently, to explore an $\mathrm{SU}(N)$-scale hypothesis class), the worst-case sample lower bound is locked at order $N$ and cannot be reduced by purely computational arguments. The no-signaling principle changes the game by singling out a compatible scaling law that lies strictly below the universal bound. The outcome is a unified ``triangle'' of constraints---query complexity, reflection-gate complexity, and learning-sample complexity---that align into one structured scaling when relativistic causality is imposed. Importantly, the unified learning-sample bound $\Omega(\sqrt{N}/\log N)$ is strictly smaller than the universal state-learning barrier: in the fully state-universal regime, the sample complexity is locked at $\Omega(N)$, whereas causality admits a substantially reduced scaling for the restricted corridor relevant to amplify--learn.

Looking ahead, the broader implication is architectural. Any future quantum algorithm that hopes to use learning to emulate powerful primitives must account for a causality-enforced budget: learning cannot be treated as a free subroutine without risking an unphysical collapse of established lower bounds. Extending this causality-aware perspective, the channel simulation, adaptive control, and fault-tolerant settings may reveal additional instances where the physical principles act as regulators of the learnability. More generally, our work suggests a route to deriving sharp complexity constraints not only from computational models, but from the operational demands of spacetime itself: the feasible region of quantum learning and computation is carved out as much by light cones as by algorithms.

\section*{METHODS}

\subsection*{Problem setting and notation}

We consider an $n$-qubit Hilbert space of dimension $N=2^n$ with a unique marked computational basis state $\ket{\tau}$, and we use the phase-flip oracle (target reflection)
\begin{eqnarray}
\hat{R}_\tau := \hat{\mathds{1}}-2\ketbra{\tau}{\tau}.
\label{eq:M_oracle_reflection}
\end{eqnarray}
For a reference state $\ket{\mathbb{0}}=\ket{0}^{\otimes n}$ we define the fixed reflection
\begin{eqnarray}
\hat{R}_0 := \hat{1}-2\ketbra{\mathbb{0}}{\mathbb{0}}.
\label{eq:M_ref_reflection}
\end{eqnarray}
For any (generally unknown) pure state $\ket{\psi}$, the corresponding Householder reflection is~\cite{householder1958}
\begin{eqnarray}
\hat{R}_\psi := \hat{1}-2\ketbra{\psi}{\psi}.
\label{eq:M_unknown_reflection}
\end{eqnarray}

We quantify learning accuracy by trace distance between density operators~\cite{helstrom1969},
\begin{eqnarray}
D(\hat{\rho},\hat{\sigma}) := \frac{1}{2}\norm{\hat{\rho}-\hat{\sigma}}_1,
\label{eq:M_trace_distance}
\end{eqnarray}
and reflection-synthesis accuracy by operator norm,
\begin{eqnarray}
\norm{\hat{X}}_\infty := \sup_{\ket{\phi}\neq 0}\frac{\norm{\hat{X}\ket{\phi}}_2}{\norm{\ket{\phi}}_2}.
\label{eq:M_op_norm}
\end{eqnarray}
The leading-order scaling statements are asymptotic in $N\to\infty$ and are insensitive to the base of $\log$.

We count three resources: (i) \emph{oracle queries} $Q$, i.e., the number of invocations of $\hat{R}_\tau$; (ii) \emph{reflection gate complexity} $G(n)$, the two-qubit-gate count required to implement a single learned approximation of $\hat{R}_\psi$ at fixed constant precision; (iii) \emph{learning samples} $M_s$, the number of fresh copies of the relevant intermediate state consumed by a state-learning subroutine per amplification round.

\subsection*{Amplify--learn architecture and reflection synthesis from a learned ansatz}

We model a learning routine $\mathcal{F}$ that, given $M_s$ independent copies of an unknown pure state $\ket{\psi}$, outputs classical parameters $\theta_\psi$ specifying an ansatz circuit $\hat{A}(\theta_\psi)$ such that
\begin{eqnarray}
D\Bigl(\ketbra{\psi}{\psi}, \hat{A}(\boldsymbol\theta_\psi)\ketbra{\mathbb{0}}{\mathbb{0}}\hat{A}(\boldsymbol\theta_\psi)^\dagger\Bigr) \le \epsilon,
\label{eq:M_learning_goal}
\end{eqnarray}
for a fixed target accuracy $0<\epsilon<1$. Once $\hat{A}(\boldsymbol\theta_\psi)$ is available, an approximate reflection about $\ket{\psi}$ is synthesized by conjugating the fixed reflection $\hat{R}_0$:
\begin{eqnarray}
\hat{R}(\theta_\psi) := \hat{A}(\boldsymbol\theta_\psi)\hat{R}_0\hat{A}(\boldsymbol\theta_\psi)^\dagger.
\label{eq:M_reflection_synthesis}
\end{eqnarray}
When Eq.~(\ref{eq:M_learning_goal}) holds, $\hat{R}(\boldsymbol\theta_\psi)$ approximates $\hat{R}_\psi$ at constant precision on the relevant subspace (and in operator norm under standard continuity bounds), which is sufficient for constant-bias search and for the causality-based reductions below.

An \emph{amplify--learn} round consists of~\cite{Baek2025}: (i) a coherent amplification step that applies $\hat{R}(\theta_{\psi_r})\hat{R}_\tau$ (or its phase-equivalent iterate) to produce $\ket{\psi_{r+1}}$; and (ii) a learning step that consumes $M_s$ fresh copies of $\ket{\psi_{r+1}}$ to output $\theta_{\psi_{r+1}}$ for the next round. The idealized logarithmic-round behavior is recovered only in the limit where the learning stage is treated as ``free'' (Fig.~1).

\subsection*{No-signaling reduction and the lower bound $M_s=\Omega(\sqrt{N}/\log N)$}

The key resource-tradeoff is obtained by embedding the amplify--learn-based logarithmic-round search into an operational no-signaling framework: any protocol that would solve unstructured search with $o(\sqrt{N})$ oracle queries can be converted into a super-luminal classical signaling protocol in suitable bipartite settings, and conversely. Enforcing no-signaling therefore enforces the Grover lower bound $Q_{\rm tot}(N)=\Omega(\sqrt{N})$ for any physically realizable algorithm.

To connect this to learning samples, we account for the oracle-query cost of producing training data. In the amplify--learn loop, each fresh copy of $\ket{\psi_{r+1}}$ must be generated by applying at least one amplification step that invokes $\hat{R}_\tau$ a constant number of times, because $\tau$ enters the dynamics only through the oracle. Thus there exists a constant $c_q>0$ such that producing $M_s$ training copies at round $r$ costs at least
\begin{eqnarray}
Q_r^{({\rm train})} \ge c_q M_s.
\end{eqnarray}
If the intended number of amplification rounds is $r(N)=\Theta(\log N)$, the total training-query cost satisfies
\begin{eqnarray}
Q_{\rm train}(N) \ge \sum_{r=1}^{r(N)}Q_r^{({\rm train})} \ge c_q M_s r(N).
\end{eqnarray}
The final ``production'' run contributes only $O(r(N))$ additional oracle calls, which is negligible compared with $M_s r(N)$ in the regime of interest. Therefore, for sufficiently large $N$,
\begin{eqnarray}
Q_{\rm tot}(N) \ge c_q M_s r(N).
\end{eqnarray}
Imposing the no-signaling-implied Grover lower bound $Q_{\rm tot}(N)\ge c_G\sqrt{N}$ (for some constant $c_G>0$) and using $r(N)=\Theta(\log N)$ yields
\begin{eqnarray}
M_s \ge \Omega\Bigl(\frac{\sqrt{N}}{\log N}\Bigr).
\end{eqnarray}
This is the causality-selected learning-sample lower bound: if $M_s$ were asymptotically smaller, the amplify--learn construction would enable a super-Grover search and hence super-luminal signaling.

\subsection*{Gate-depth version: reflection complexity $G(n)=\Omega(\sqrt{N}/\log N)$}

An analogous argument applies to the circuit depth (or gate count) required to implement each learned reflection. Let $G(n)$ denote the two-qubit gate complexity of implementing $\hat{R}(\theta_\psi)$ at fixed constant precision. In a logarithmic-round protocol, reflections are invoked $O(r(N))$ times, so the total depth contributed by unknown-state reflections scales as
\begin{eqnarray}
D_{\rm ref}(N) &\le& c_d G(n) r(N),
\label{eq:M_total_ref_depth}
\end{eqnarray}
for a constant $c_d>0$ that accounts for a constant number of reflection calls per round and constant-depth wrappers. In the same no-signaling embedding, a depth-$o(\sqrt{N})$ implementation with constant success bias would enable super-luminal signaling by executing ``too quickly'' relative to the light-crossing time. Thus $D_{\rm tot}(N)=\Omega(\sqrt{N})$ implies
\begin{eqnarray}
G(n) &\ge& \Omega\Bigl(\frac{\sqrt{N}}{\log N}\Bigr)=\Omega\Bigl(\frac{2^{n/2}}{n}\Bigr).
\label{eq:M_gate_lower_bound}
\end{eqnarray}

\subsection*{Remark (Classical vs quantum scope)}
Although no-signaling is a relativistic principle shared by classical physics, the $\Omega(N/\log{N})$ per-round bound derived here is not a generic “classical” consequence of no-signaling alone. It follows from combining (i) the Grover threshold $\Theta(\sqrt{N})$, enforced operationally by no-signaling via the Bao-Bouland-Jordan equivalence between super-Grover search and superluminal signaling, with (ii) the specifically quantum obstruction to programmable reflections about unknown states (the no-reflection theorem), which forces any implementation to proceed via state learning and reflection synthesis. The classical (stochastic) theories do not feature an analogue of unknown-state reflections or amplitude amplification; conversely, postulating an ensemble-dependent/nonlinear classical primitive capable of mimicking previous-output reflections would generically re-enable signaling and thus be excluded by no-signaling. Hence, our results should be read as a causality constraint on quantum (or quantum-extended) learning primitives.

\subsection*{Computational learning-theoretic bounds for circuit-generated states}

Independently of no-signaling, we also use standard information-theoretic tools from quantum learning theory to bound the sample complexity of learning circuit-generated states. Let $\mathsf{U}_{n,G}$ be the set of $n$-qubit circuits with at most $G$ two-qubit gates from a fixed universal gate set, and define the induced hypothesis class of pure states
\begin{eqnarray}
\mathsf{S}_{n,G} := \Bigl\{\ket{\psi}=\hat{U}\ket{0}^{\otimes n} : \hat{U}\in\mathsf{U}_{n,G}\Bigr\}.
\label{eq:M_state_class}
\end{eqnarray}
A learner receives $M_s$ copies of $\hat{\rho}=\ket{\psi}\bra{\psi}$ with $\ket{\psi}\in\mathsf{S}_{n,G}$, performs an arbitrary joint POVM, and outputs a hypothesis $\hat{\rho}_h$.

We define the worst-case sample complexity $M_s(n,G,\epsilon,\delta)$ as the smallest number of copies for which there exists a learner such that, for every $\hat{\rho}\in\mathsf{S}_{n,G}$,
\begin{eqnarray}
{\rm Pr}\Bigl[D(\hat{\rho},\hat{\rho}_h)\le\epsilon\Bigr] &\ge& 1-\delta.
\label{eq:M_worst_case_def}
\end{eqnarray}
Two standard ingredients yield matching bounds (up to log factors): (i) a covering-number upper bound for $\mathsf{S}_{n,G}$ in trace distance, scaling as $\log{\abs{\mathcal{N}_\epsilon}}=O(G\log(G/\epsilon))$ because depth-$G$ circuits form an $O(G)$-dimensional manifold; and (ii) a packing-net lower bound with $\log{\abs{\mathcal{P}_\epsilon}}=\Omega(\min\{2^n,G\})$ combined with Fano's inequality and the Holevo bound on accessible information. These give, for fixed constant $\delta$ and $0<\epsilon<1$,
\begin{eqnarray}
M_s(n,G,\epsilon,\delta) &=& O\Bigl(\frac{1}{\epsilon^2}\min\{2^n,G\log(G/\epsilon)\}\Bigr),
\label{eq:M_comp_upper}
\\
M_s(n,G,\epsilon,\delta) &=& \Omega\Bigl(\frac{1}{\epsilon^2}\min\{2^n,G\}\Bigr).
\label{eq:M_comp_lower}
\end{eqnarray}
These bounds are in line with those obtained in a recent independent study (Ref.~\cite{Zhao2024}).

\subsection*{Universal ansatz ``lock'' and causality-selected scaling}
A universal state-preparation (or universal learning-ansatz) architecture must cover the full $n$-qubit pure-state manifold, which has real dimension $2(2^n-1)$. Parameter counting therefore enforces $G=\Theta(2^n)$ in the worst case, and Eq.~(\ref{eq:M_comp_lower}) yields the familiar tomography-like ``lock'':
\begin{eqnarray}
M_s(n,G,\epsilon,\delta) = \Omega\Bigl(\frac{2^n}{\epsilon^2}\Bigr).
\label{eq:M_universal_lock}
\end{eqnarray}
In contrast, the amplify--learn setting does not require learning arbitrary states; it requires learning the structured intermediate states that arise along the amplification trajectory. The no-signaling analysis constrains the reflection complexity for these states to $G(n)=\Omega(\sqrt{N}/\log N)$ [Eq.~(\ref{eq:M_gate_lower_bound})]. Substituting this physically dictated $G$ into the computational lower bound (\ref{eq:M_comp_lower}) recovers the reduced sample scaling
\begin{eqnarray}
M_s(n,G,\epsilon,\delta) = \Omega\Bigl(\frac{1}{\epsilon^2}\frac{\sqrt{N}}{\log N}\Bigr),
\label{eq:M_causality_selected}
\end{eqnarray}
matching the direct causality-enforced argument in {\bf Theorem~\ref{thm:causal_bounds}}. In this sense, computational learning theory alone admits many a priori scalings (between polynomial and exponential in $N$), while enforcing no-signaling selects a unique causality-compatible scaling law that aligns query, gate, and sample complexities.

Full proofs and additional technical refinements (including constants, robustness to constant-precision reflections, and the explicit no-signaling embedding) are provided in the Supplementary Information.

\section*{Acknowledgement}
This work was supported by the Ministry of Science, ICT and Future Planning (MSIP) by the National Research Foundation of Korea (RS-2024-00432214, RS-2025-03532992, and RS-2025-18362970) and the Institute of Information and Communications Technology Planning and Evaluation grant funded by the Korean government (RS-2019-II190003, ``Research and Development of Core Technologies for Programming, Running, Implementing and Validating of Fault-Tolerant Quantum Computing System''), the Korean ARPA-H Project through the Korea Health Industry Development Institute (KHIDI), funded by the Ministry of Health \& Welfare, Republic of Korea (RS-2025-25456722). We acknowledge the Yonsei University Quantum Computing Project Group for providing support and access to the Quantum System One (Eagle Processor), which is operated at Yonsei University.

\section*{Declarations}

\medskip\noindent
{\bf Data Availability.}---The data that support the findings of this study are available from the corresponding author upon request.

\medskip\noindent
{\bf Code Availability.}---Not applicable.

\medskip\noindent
{\bf Author contributions.}---JB and JJ conceived the initial idea of the work. This idea was subsequently developed in collaboration with KC. JB carried out the theoretical proofs with KC and JJ. JB wrote the manuscript, and all authors jointly reviewed and refined the structure, logic, and presentation of the paper. Correspondence and requests for materials should be addressed to JB and JJ.

\medskip\noindent
{\bf Competing interests.}---The authors declare no competing interests.

\clearpage

\setcounter{section}{0}
\renewcommand{\thesection}{S\arabic{section}}

\setcounter{equation}{0}
\renewcommand{\theequation}{S\arabic{equation}}

\setcounter{figure}{0}
\renewcommand{\thefigure}{S\arabic{figure}}

\setcounter{table}{0}
\renewcommand{\thetable}{S\arabic{table}}

\begin{center}
{\Large\bfseries Supplementary Information}
\end{center}

\vspace{1em}

This Supplementary Information (SI) is written as a self-contained and complete companion paper, so that readers can fully reproduce and understand the calculations, logical flow, and results of the main manuscript from this document alone. 

\section*{------------------------------------------------------------}

\setcounter{tocdepth}{2}
\tableofcontents 

\section{Introduction}\label{Sec:Intro}

An intrinsic question at the interface of physics and information science is to what extent the capabilities and/or limitations of computing are fixed not by engineering ingenuity, but by the underlying physical laws. Classical examples already hint at some answers: Landauer’s principle ties the minimum energy cost of erasing a bit to thermodynamics~\cite{landauer1961irreversibility}, Bekenstein-type bounds limit how much information can be stored in a finite region of space~\cite{bekenstein1981universal}, and proposals for ultimate computers are constrained by relativistic and quantum speed limits~\cite{bremermann1967quantum,margolus1998maximum,lloyd2000ultimate}. Quantum information science has sharpened this perspective. Quantum superposition and entanglement enable qualitatively novel information processing tasks---secure key distribution~\cite{ekert1991quantum,bennett2014quantum}, exponential-dimensional Hilbert-space exploration~\cite{feynman2018simulating,lloyd1996universal}, and exponential speedups for specific problems~\cite{shor1994algorithms,bernstein1993quantum}---yet they also introduce the impossibilities, such as, no-cloning theorem and stringent trade-offs between information gain and disturbance~\cite{wootters1982single,dieks1982communication, fuchs1998information}. In this light, it is natural to ask a sharper question: when we design quantum learning and computing architectures that operate far beyond today’s hardware, can we predict their optimal performance solely from computational arguments, or must we also enforce fundamental physical principles? In particular, can a principle, as basic as no-signaling, which forbids faster-than-light communication, also determine the ultimate computational complexity for learning?

The relativistic causality can be encoded operationally in the no-signaling principle: any local action on one subsystem cannot be used to transmit information to a space-like separated partner. Within quantum theory, this requirement tightly constrains admissible operations and correlations. Together with the linearity, the no-signaling excludes the tasks, such as, cloning of unknown states~\cite{gisin1998quantum,gisin2009quantum}, perfect discrimination of nonorthogonal states~\cite{ivanovic1987differentiate,peres1988differentiate}, or post-selection onto arbitrary outcomes that would otherwise enable signaling~\cite{gisin1989stochastic,polchinski1991weinberg}. At the same time, these no-go theorems underpin some constructive protocols: the security of quantum-key-distribution can be framed as the impossibility of super-luminal information leakage from a measurement choice, even in the presence of entanglement and adversarial devices~\cite{barrett2005no, acin2007device}. More exotic hypothetical theories, with stronger-than-quantum nonlocal correlations that still obey no-signaling, illustrate just how delicate this balance is: if one relaxes quantum structure too far while maintaining no-signaling, many complexity-theoretic separations can collapse~\cite{popescu1994quantum, van2013implausible}. These examples collectively suggest that ``physics-aware'' information theory---one that builds the no-signaling and linearity into its axioms---can both forbid impossible information-processing tasks and calibrate the power of the possible ones.

Grover’s quantum search algorithm provides a paradigmatic case where the computational complexity and causality already meet. On the algorithmic side, Grover’s procedure solves an unstructured search over an $N$-element database using $O(\sqrt{N})$ oracle queries~\cite{grover1996fast}, and the argument of Bennett, Bernstein, Brassard, and Vazirani shows that this quadratic speedup is optimal within the standard query model~\cite{bennett1997strengths}. On the physics side, recent work by Bao {\it et al.} demonstrated that this optimality is not merely a technical property of the proof, but is in fact equivalent to the no-signaling principle~\cite{bao2016grover}. Roughly, if one extends quantum mechanics by a primitive operation, say $\mathcal{M}$, acting on a bounded number of qubits, then the ability to construct a ``super-Grover'' search with $o(\sqrt{N})$ queries is polynomially equivalent to the ability to send a classical bit super-luminally using $\mathcal{M}$ and entanglement, and vice versa. Thus, requiring that no such super-luminal channel exist rederives the Grover’s lower bound from relativistic causality alone. In this sense, the familiar $\sqrt{N}$ scaling is already a manifestation of a physical light-cone structure: namely, any attempt to squeeze the query complexity below this threshold would open a super-luminal channel.

In this work, we bring a third layer into this picture: the cost of learning unknown quantum states. In particular, we ask how many samples are fundamentally required when the standard quantum mechanics and no-signaling principle are imposed simultaneously. To make this question concrete, we consider an imaginary strategy, state-learning-assisted amplitude amplification, that serves as a controlled thought experiment. On the purely algebraic level, this protocol modifies the Grover iteration by replacing the usual reflection about a fixed initial state with a reflection about the state produced in the previous round. This previous-output reflection yields a cubic enhancement of the target overlap and compresses the number of oracle-query rounds from order $\sqrt{N}$ to $\log N$, suggesting an exponential quantum-search speedup. The no-reflection theorem, however, shows that such a previous-output reflection is not realizable in theory~\cite{kumar2011quantum}, so any admissible scheme must only approximate this map at the price of extra resources. To capture this trade-off, we introduce an explicit ``amplify-learn'' architecture in which each round alternates a coherent amplitude-amplification with a state-learning structure that consumes many fresh copies of the current state to synthesize a circuit implementing an approximate reflection about that state. The central question we then address is how large the associated learning-sample budget must be so that such an amplify-learn scheme cannot be promoted to a super-Grover search, and hence, to a super-luminal-signaling scheme.

Our main results show that the answer to the aforementioned question cannot be derived from the computational learning theory alone, but only emerges when the causality is imposed. From a purely computational or statistical viewpoint, the worst-case sample complexity for learning pure states prepared by finite-depth learning-ansatz circuits is governed by the effective size of the hypothesis class: we prove sharp upper and lower bounds in which the leading term scales linearly with the gate complexity of the learning-ansatz circuit and only quadratically with the inverse accuracy. In contrast, by embedding our amplify-learn scheme into the Bao-Bouland-Jordan signaling framework, we show that any attempt to implement the ideal logarithmic-round quantum search within standard quantum mechanics must pay a state-learning cost of order $\sqrt{N}/\log N$ per round; otherwise, the total query count would drop below Grover threshold and thereby enable a super-luminal signaling. Combining these two perspectives, we obtain a coherent ``triangle'' of lower bounds: the query complexity of unstructured search, the gate complexity of unknown-state reflections, and the sample complexity of learning are all forced to share the highly structured resource scaling when the causality is respected. In this way, the no-signaling principle provides a sharper and more restrictive lower bound on learning than computational arguments and logic alone can offer, and it acts as the physical thread that ties together three distinct complexity notions while selecting a unique scaling law among many computationally admissible ones. In a more broad sense, our results support a view in which the ultimate performance and architecture of future computing and IT technologies---classical or quantum, learning-based or algorithmic---are not merely matters of algorithm design and/or engineering, but are carved out by the physical theories that underwrite them.

\section{State-learning-assisted amplitude amplification and logarithmic oracle-query complexity}\label{sec:SLAA}

In this section, we reformulate the amplitude amplification so that the second reflection is not fixed by the initial state, but by the previous output state of the algorithm. This seemingly innocent modification drastically changes the geometry of the iteration. At the level of formal quantum mechanics, it allows an exponential acceleration of the Grover-type search, reducing the number of oracle-query rounds from the standard order of $\sqrt{N}$ to that of $\log{N}$ for the Hilbert-space dimension $N$. The price to pay, however, is that such a reflection about an unknown quantum state is forbidden in theory. Here, we show how the state-learning can be used to approximate this no-go principle and thereby implement the same abstract process within a physically meaningful architecture.

\subsection{Standard amplitude amplification revisited}

For later comparison it is convenient to recall the standard amplitude amplification in its simplest form. Firstly, let
\begin{eqnarray}
\ket{s} = \frac{1}{\sqrt{N}} \sum_{x=0}^{N-1} \ket{x}
\label{eq:def-s}
\end{eqnarray}
be the uniform superposition over all basis states $\{\ket{x}\}_{x=0}^{N-1}$. The oracle is represented as the phase flip
\begin{eqnarray}
\hat{R}_\tau = \hat{\mathds{1}} - 2 \ketbra{\tau}{\tau},
\label{eq:R-tau}
\end{eqnarray}
and the diffusion operator is a reflection about the initial state,
\begin{eqnarray}
\hat{R}_s  = \hat{\mathds{1}} - 2\ketbra{s}{s},
\label{eq:R-s}
\end{eqnarray}
where a unique marked element $x^\star$ is encoded in a ``target'' state $\ket{\tau} := \ket{x^\star}$.
The Grover iterate is then~\cite{grover1996fast}
\begin{eqnarray}
\hat{G} = e^{-i \xi} \hat{R}_s \hat{R}_\tau, \quad \xi = \pi.
\label{eq:G-def}
\end{eqnarray}

The dynamics of $G$ is confined to the two-dimensional subspace spanned by the target state $\ket{\tau}$ and its orthogonal complement inside the initial state. To make this structure explicit, let us define
\begin{eqnarray}
\ket{\tau_\perp}
  = \frac{1}{\sqrt{N-1}} \sum_{x\neq x^\star} \ket{x} ,
\end{eqnarray}
so that $\{\ket{\tau},\ket{\tau_\perp}\}$ is an orthonormal basis for the relevant two-dimensional subspace. In this basis,
\begin{eqnarray}
\ket{s}
 = \sin\theta_0 \ket{\tau} + \cos\theta_0 \ket{\tau_\perp}
\label{eq:s-theta0}
\end{eqnarray}
with 
\begin{eqnarray}
\sin{\theta_0} = \frac{1}{\sqrt{N}}, \quad \cos{\theta_0} = \sqrt{\frac{N-1}{N}}.
\label{eq:theta_0}
\end{eqnarray}
Here it is convenient to regard $\theta_0 \approx \tfrac{1}{\sqrt{N}}$ as the initial angle between $\ket{s}$ and the non-target direction $\ket{\tau_\perp}$. This original architecture of the amplitude amplification follows:
\begin{theorem}[Standard amplitude amplification]
\label{thm:standard_G}
Let $\ket{\psi_r} = \hat{G}^r \ket{s}$ be the state after $r$ Grover iterations. Then,
\begin{eqnarray}
\ket{\psi_r} = \sin{\bigl((2r+1)\theta_0\bigr)}\ket{\tau} + \cos{\bigl((2r+1)\theta_0\bigr)}\ket{\tau_\perp},
\label{eq:grover-angle}
\end{eqnarray}
and the success probability $\abs{\braket{\tau}{\psi_r}}^2$ reaches order unity after $r=O(\sqrt{N})$ oracle calls.
\end{theorem}

\begin{proof}---The proof is straightforward. In the basis $\{ \ket{\tau}, \ket{\tau_\perp} \}$ the projectors are
\begin{eqnarray}
\ketbra{\tau}{\tau} =
	\begin{pmatrix}
	1 & 0 \\
	0 & 0
	\end{pmatrix},
\quad
\ketbra{s}{s} =
	\begin{pmatrix}
	\sin^2\theta_0 & \sin\theta_0\cos\theta_0 \\
	\sin\theta_0\cos\theta_0 & \cos^2\theta_0
	\end{pmatrix}.
\end{eqnarray}
Thus,
\begin{eqnarray}
\hat{R}_\tau
  &=& \hat{\mathds{1}} - 2\ketbra{\tau}{\tau} =
	\begin{pmatrix}
	-1 & 0 \\
	0 & 1
	\end{pmatrix}, \nonumber \\
\hat{R}_s
  &=& \hat{\mathds{1}} - 2\ketbra{s}{s}
	= - \begin{pmatrix}
	2\sin^2\theta_0 -1 & 2\sin\theta_0\cos\theta_0 \\
	2\sin\theta_0\cos\theta_0 & 2\cos^2\theta_0 -1
	\end{pmatrix}.
\end{eqnarray}
Therefore, we have
\begin{eqnarray}
\hat{G} = {e^{-i \pi}}\hat{R}_s \hat{R}_\tau =
	\begin{pmatrix}
	\cos 2\theta_0 & \sin 2\theta_0 \\
	-\sin 2\theta_0 & \cos 2\theta_0
	\end{pmatrix},
\end{eqnarray}
which is exactly a rotation by angle $2\theta_0$ in the $\{ \ket{\tau}, \ket{\tau_\perp} \}$ plane. Since $\ket{\psi_0}=\ket{s}$ has the components $(\sin\theta_0, \cos\theta_0)^T$, after $r$ iterations, we obtain
\begin{eqnarray}
\ket{\psi_r} = \hat{G}^r \ket{\psi_0} = 
	\begin{pmatrix}
	\sin\bigl((2r+1)\theta_0\bigr) \\
	\cos\bigl((2r+1)\theta_0\bigr)
	\end{pmatrix},
\end{eqnarray}
which is Eq.~(\ref{eq:grover-angle}). The success probability is thus
\begin{eqnarray}
p_r = \abs{\braket{\tau}{\psi_r}}^2 = \sin^2\bigl((2r+1)\theta_0\bigr) .
\end{eqnarray}
Since $\theta_0 \approx N^{-1/2}$, choosing $r \sim \frac{\pi}{4}\theta_0^{-1} = O(\sqrt{N})$ makes $(2r+1)\theta_0$ close to $\pi/2$, and hence $p_r$ is close to $1$. This establishes the $O(\sqrt{N})$ query complexity. See Refs.~\cite{grover1996fast} for a more detailed and mathematically complete proofs.
\end{proof}

\subsection{Imaginary protocol with previous-output-state reflection}

We then consider an imaginary variant of amplitude amplification where, in each round, the reflection about the initial state is replaced by a reflection about the previous output state of the algorithm. At round $r$, we denote the current state by $\ket{\psi_r}$. The target reflection $\hat{R}_\tau$ is kept fixed, but the second reflection is taken as
\begin{eqnarray}
\hat{R}_{\psi_r} := \hat{\mathds{1}} - 2\ketbra{\psi_r}{\psi_r} .
\label{eq:R-psir}
\end{eqnarray}
The update rule is then
\begin{eqnarray}
\ket{\psi_{r+1}} = e^{-i \pi} \hat{R}_{\psi_r} \hat{R}_\tau \ket{\psi_r} .
\label{eq:update-imag}
\end{eqnarray}
At the level of formal linear algebra, this map is perfectly well-defined: it is simply the composition of two reflections in the two-dimensional subspace spanned by $\ket{\tau}$ and $\ket{\psi_r}$. The crucial difference with the standard Grover iteration is that the axis of the second reflection changes as the state moves. This turns out to induce a nonlinear evolution for the overlap with the target, in the sense that the update of the angle depends on the angle itself.

To analyze the process of Eq.~(\ref{eq:update-imag}), we consider the basis $\{\ket{\tau},\ket{\tau_\perp}\}$ again. At round $r$, we can write the previous $(r-1)$-th output, hence the current $r$-th input, state
\begin{eqnarray}
\ket{\psi_r}
 = \sin\theta_r\ket{\tau} + \cos\theta_r\ket{\tau_\perp} ,
\label{eq:psir-theta}
\end{eqnarray}
with $0<\theta_r<\pi/2$ and $\theta_0$ is set as in Eq.~(\ref{eq:theta_0}). The target reflection $\hat{R}_\tau$ is as before, i.e., as in Eq.~(\ref{eq:R-tau}). The product of the two reflections is then 
\begin{eqnarray}
\hat{Q}_r = \hat{R}_{\psi_r} \hat{R}_\tau = 
     \begin{pmatrix}
       -\cos{2\theta_r} & -\sin{2\theta_r} \\
       \sin{2\theta_r} & -\cos{2\theta_r}
     \end{pmatrix}.
\label{eq:Qr-matrix}
\end{eqnarray}
Up to the irrelevant global phase $e^{-i\pi}$, $\hat{Q}_r$ is a rotation by angle $-2\theta_r$ in the $\{\ket{\tau},\ket{\tau_\perp}\}$ plane. Thus, we can state the following theorem:
\begin{theorem}[Cubic amplification with previous-output reflections]
\label{thm:cubic-aa}
Let $\ket{\psi_{r+1}} = e^{-i\pi}\hat{R}_{\psi_r} \hat{R}_\tau \ket{\psi_r}$ with $\ket{\psi_r}$ as in Eq.~(\ref{eq:psir-theta}). Then, 
\begin{eqnarray}
\braket{\tau}{\psi_{r+1}} = \sin(3\theta_r) .
\label{eq:sin3theta}
\end{eqnarray}
As long as $0 < \theta_r < \pi/6$, we have $\theta_{r+1} = 3\theta_r$.
\end{theorem}

\begin{proof}---Write $\ket{\psi_r}$ in vector form as $\ket{\psi_r} \equiv \bigl( \begin{smallmatrix} \sin{\theta_r} \\ \cos{\theta_r} \end{smallmatrix} \bigr)$. Applying $Q_r$ from Eq.~(\ref{eq:Qr-matrix}), the update rule Eq.(\ref{eq:update-imag}) becomes
\begin{eqnarray}
\ket{\psi_{r+1}} = e^{-i \pi} \hat{Q}_r \ket{\psi_r} =
        \begin{pmatrix}
          \cos{2\theta_r} & \sin{2\theta_r} \\
          -\sin{2\theta_r} & \cos{2\theta_r}
	\end{pmatrix}
        \begin{pmatrix}
          \sin{\theta_r} \\
          \cos{\theta_r}
        \end{pmatrix}.
\end{eqnarray}
The first component of the product is
\begin{eqnarray}
\sin{\theta_{r+1}} = \cos{2\theta_r} \sin{\theta_r} + \sin{2\theta_r} \cos{\theta_r} = \sin{3\theta_r}.
\end{eqnarray}
Thus the overlap with the target state after $r$-th round is given by Eq.~(\ref{eq:sin3theta}). For $0 < \theta_r < \pi/6$, $\sin{\theta}$ is strictly increasing, and we can take the principal value of the arcsine to conclude that $\theta_{r+1} = 3\theta_r$.
\end{proof}

This cubic amplification of the angle has dramatic algorithmic consequences:
\medskip\noindent
\begin{corollary}[Imaginary logarithmic-query search]
\label{cor:log-search}
Assume that the initial state is the usual uniform superposition, so that $\theta_0 \approx N^{-1/2}$. Suppose that at each round $r$, we can implement a perfect reflection $\hat{R}_{\psi_r}$ about the current state. Then, after $r = O(\log N)$ rounds of the map $\ket{\psi_{r+1}} = e^{-i\pi}\hat{R}_{\psi_r} \hat{R}_\tau\ket{\psi_r}$, the success probability $\abs{\braket{\tau}{\psi_r}}^2$ becomes of order unity. Each round uses a single application of the target reflection $R_\tau$; hence, the query complexity is $O(\log N)$.
\end{corollary}

\begin{proof}---From {\bf Theorem~\ref{thm:cubic-aa}}, as long as $\theta_r < \tfrac{\pi}{6}$, we have the exact recursion
\begin{eqnarray}
\theta_r = 3^r \theta_0.
\end{eqnarray}
For $N \gg 1$, $\theta_0 = \arcsin{\tfrac{1}{\sqrt{N}}} \approx \tfrac{1}{\sqrt{N}}$. We choose $r$ minimal such that $\theta_r$ exceeds some fixed constant $c$ with $0 < c < \tfrac{\pi}{6}$. Then, the following holds:
\begin{eqnarray}
3^r \theta_0 \ge c
\Longleftrightarrow
 r \ge \frac{\log{\frac{c}{\theta_0}}}{\log 3} = O(\log N) .
\end{eqnarray}
At that round, the overlap satisfies $\left|\braket{\tau}{\psi_r}\right|^2 = \sin^2\theta_r \ge \sin^2 c$, where $c$ a constant bounded away from $0$. Since each round of Eq.~(\ref{eq:update-imag}) uses a single application of $R_\tau$, the number of oracle-call is $O(\log N)$.
\end{proof}

Taken at face value, {\bf Corollary~\ref{cor:log-search}} describes an algorithm with an exponential speedup over the quadratic bound. However, this comes at the cost of assuming that at each round, we can implement a reflection $\hat{R}_{\psi_r}$ about the current state, which is \emph{unknown} to the algorithm. The state $\ket{\psi_r}$ is defined by the dynamics of Eq.~(\ref{eq:update-imag}); it is not specified in advance as a classical description. The following no-go theorem shows that such a universal previous-output-state reflection is fundamentally impossible in quantum theory~\cite{kumar2011quantum,PhysRevLett.79.321}.
\begin{theorem}[No-reflection theorem]
\label{thm:no-reflection}
There does not exist a unitary process $\hat{U}$ acting on two copies of the system such that, for all normalized states $\ket{\chi}$ and all target states $\ket{\phi}$,
\begin{eqnarray}
\hat{U} \ket{\chi}_c \otimes \ket{\phi}_{\tau} = \ket{\chi}_c \otimes \bigl(\hat{\mathds{1}} - 2\ketbra{\chi}{\chi}\bigr)\ket{\phi}_{\tau},
\label{eq:no-reflection-spec}
\end{eqnarray}
where the subscript $c$ denotes the ``control'' register that supplies the unknown state about which the reflection is taken, and the subscript $\tau$ the ``target'' register on which the reflection acts.
\end{theorem}

\begin{proof}---Assume, to the contrary, that such the unitary $\hat{U}$ satisfying Eq.~(\ref{eq:no-reflection-spec}) exists. Let $\ket{\psi}$ and $\ket{\varphi}$ be two distinct normalized states with nonzero overlap ${\cal L}=\braket{\psi}{\varphi}$ and $0 < \abs{{\cal L}} < 1$. Firstly, consider first the action of $\hat{U}$ on $\ket{\psi}_c \ket{\psi}_t$. By Eq.~(\ref{eq:no-reflection-spec}),
\begin{eqnarray}
\hat{U} \ket{\psi}_c \ket{\psi}_{\tau} = - \ket{\psi}_c \ket{\psi}_{\tau}.
\label{eq:U-psipsi}
\end{eqnarray}
and similarly,
\begin{eqnarray}
\hat{U} \ket{\varphi}_c \ket{\varphi}_{\tau} = -\ket{\varphi}_c \ket{\varphi}_{\tau}.
\label{eq:U-phiphi}
\end{eqnarray}
Next, consider the cross terms. For $\ket{\psi}_c \ket{\varphi}_t$, we obtain
\begin{eqnarray}
\hat{U} \ket{\psi}_c \ket{\varphi}_{\tau} = \ket{\psi}_c \ket{\varphi}_{\tau} - 2{\cal L} \ket{\psi}_c \ket{\psi}_{\tau}.
\label{eq:U-psiphi}
\end{eqnarray}
Analogously, for $\ket{\varphi}_c \ket{\psi}_{\tau}$, we have
\begin{eqnarray}
\hat{U} \ket{\varphi}_c \ket{\psi}_{\tau} = \ket{\varphi}_c \ket{\psi}_{\tau} - 2{\cal L}^\ast \ket{\varphi}_c \ket{\varphi}_{\tau}.
\label{eq:U-phipsi}
\end{eqnarray}
Now, let $\ket{\chi} = \tfrac{1}{\sqrt{2}}\bigl(\ket{\psi} + \ket{\varphi}\bigr)$. Applying $\hat{U}$ to $\ket{\chi}_c \ket{\chi}_{\tau}$ in two different ways must yield the same state. On the one hand, by linearity, we can write
\begin{eqnarray}
\ket{\chi}_c\ket{\chi}_{\tau} = \frac{1}{2}\Bigl( \ket{\psi}_c\ket{\psi}_{\tau} + \ket{\psi}_c\ket{\varphi}_{\tau} + \ket{\varphi}_c\ket{\psi}_{\tau} + \ket{\varphi}_c\ket{\varphi}_{\tau} \Bigr),
\end{eqnarray}
and using Eqs.~(\ref{eq:U-psipsi})--(\ref{eq:U-phipsi}),
\begin{eqnarray}
\hat{U} \ket{\chi}_c\ket{\chi}_{\tau} = \frac{1}{2}\Bigl( - (1+2{\cal L})\ket{\psi}_c\ket{\psi}_{\tau} + \ket{\psi}_c\ket{\varphi}_{\tau} + \ket{\varphi}_c\ket{\psi}_{\tau} - (1+2{\cal L}^\ast)\ket{\varphi}_c\ket{\varphi}_{\tau}
     \Bigr) .
\label{eq:U-chi-linear}
\end{eqnarray}
On the other hand, applying Eq.~(\ref{eq:no-reflection-spec}) directly with $\ket{\chi}_c \ket{\chi}_{\tau}$, we attain
\begin{eqnarray}
\hat{U} \bigl(\ket{\chi}_c\ket{\chi}_{\tau}\bigr)  = -\ket{\chi}_c\ket{\chi}_{\tau}  = -\frac{1}{2}\Bigl(\ket{\psi}_c\ket{\psi}_{\tau} + \ket{\psi}_c\ket{\varphi}_{\tau} + \ket{\varphi}_c\ket{\psi}_{\tau} + \ket{\varphi}_c\ket{\varphi}_{\tau} \Bigr).
\label{eq:U-chi-direct}
\end{eqnarray}
Since both Eq.~(\ref{eq:U-chi-linear}) and Eq.~(\ref{eq:U-chi-direct}) are expansions in the orthogonal product basis
\begin{eqnarray} \bigl\{\ket{\psi}_c\ket{\psi}_{\tau},\; \ket{\psi}_c\ket{\varphi}_{\tau},\; \ket{\varphi}_c\ket{\psi}_{\tau},\; \ket{\varphi}_c\ket{\varphi}_{\tau} \bigr\},
\end{eqnarray}
their coefficients must agree. Comparing the coefficient of $\ket{\psi}_c\ket{\varphi}_{\tau}$ in Eq.~(\ref{eq:U-chi-linear}) with that in Eq.~(\ref{eq:U-chi-direct}), we can find immediately that there is contradiction and no generality. Therefore, no such unitary $\hat{U}$ can exist, and the assumption underlying Eq.~(\ref{eq:no-reflection-spec}) is false.
\end{proof}

Physically, {\bf Theorem~\ref{thm:no-reflection}} expresses a deep constraint stemming from the linearity of the quantum theory. To reflect a state about itself, one must know which direction in Hilbert-space to treat as the mirror axis. When this information is encoded only in a single unknown copy of the state, no universal unitary can extract it without disturbing the state in a way that spoils the reflection property. The situation is analogous to the no-cloning theorem: the linearity that forbids universal cloning also forbids universal state-dependent reflections.

From the quantum algorithm viewpoint, {\bf Theorem~\ref{thm:no-reflection}} tells us that the imaginary protocol of Eq.~(\ref{eq:update-imag}) cannot be implemented \emph{using only a single copy of $\ket{\psi_r}$} per round. The $O(\log N)$ query complexity of {\bf Corollary~\ref{cor:log-search}}, therefore, implies not merely that it breaks the known lower bounds for black-box search, but is incompatible with the fundamental structure of quantum theory.

\subsection{State-learning-assisted realization of previous-output-state reflection}

The imaginary protocol discussed above assumes that we can implement an exact reflection $\hat{R}_{\psi_r}$ about the current state $\ket{\psi_r}$ at each round. The no-reflection theorem shows that such an operation is impossible when $\ket{\psi_r}$ is available only as an unknown quantum state. However, if we are allowed to perform a \emph{state-learning} procedure on many copies of $\ket{\psi_r}$, we can hope to reconstruct a classical description of a unitary circuit that (approximately) prepares $\ket{\psi_r}$ from an easy reference state. Once such a circuit is available, we can implement an approximate reflection about $\ket{\psi_r}$ by conjugating a fixed reflection about the reference state.

Formally, we assume the following kind of state-learning primitive: Given access to many copies of an unknown pure state $\ket{\psi}$ in an $N$-dimensional Hilbert space, a learning algorithm outputs a classical description of a unitary circuit $\hat{A}(\boldsymbol{\theta}_\psi)$ such that
\begin{eqnarray}
\hat{A}(\boldsymbol{\theta}_\psi)\ket{0} \simeq \ket{\psi} ,
\label{eq:Apsi-preparation}
\end{eqnarray}
where $\ket{0}$ is a simple reference state (for example, the all-zero computational basis state), and $\boldsymbol{\theta}_\psi$ is a set of control parameter when the state-learning is completed. The learning algorithm (i.e., the rule of $\boldsymbol{\theta}$-update) may be fully quantum, fully classical, or hybrid~\cite{brassard2000quantum, aaronson2007learnability, Baek:2025asd}. Then, we simply assume that such a primitive exists: once  $\hat{A}(\boldsymbol{\theta}_\psi)$ is identified, we can implement approximately a reflection about $\ket{\psi}$ such that
\begin{eqnarray}
\hat{R}_\psi \simeq \hat{R}(\boldsymbol{\theta}_\psi) := \hat{A}(\boldsymbol{\theta}_\psi) \bigl( \hat{\mathds{1}} - 2\ketbra{0}{0} \bigr) \hat{A}(\boldsymbol{\theta}_\psi)^\dagger.
\label{eq:R-psi-learned}
\end{eqnarray}
If Eq.~(\ref{eq:Apsi-preparation}) holds exactly, then $\hat{R}_\psi$ coincides with $\hat{R}(\boldsymbol{\theta}_\psi)$ on the entire Hilbert-space; if it holds approximately, then $\hat{R}(\boldsymbol{\theta}_\psi)$ is an approximate reflection whose action whenever the state $\hat{A}(\boldsymbol{\theta}_\psi)\ket{0}$ has a large overlap with $\ket{\psi}$. 

Motivated by this structure, we now formalize a state-learning-assisted amplitude amplification protocol that mirrors the imaginary map of Eq.~(\ref{eq:update-imag}). For conceptual clarity, we separate each round into a coherent ``amplification stage'' and an incoherent ``state-learning stage.''

{\em Amplification stage.}---At round $r$, we assume access to a fixed (learned) set of control parameter $\boldsymbol{\theta}_{\psi_r}$ satisfying $\hat{A}(\boldsymbol{\theta}_{\psi_r})\ket{0} = \ket{\psi_r}$ (or approximately so). Using $\hat{A}(\boldsymbol{\theta}_{\psi_r})$, we implement the reflection $\hat{R}(\boldsymbol{\theta}_{\psi_r})$ as in Eq.~(\ref{eq:R-psi-learned}). We then apply $\hat{R}(\boldsymbol{\theta}_{\psi_r}) \hat{R}_\tau$ to $\ket{\psi_r}$, and obtain the output state $\ket{\psi_{r+1}}$ up to global phase $e^{-i\pi}$.

{\em State-learning stage.}---Assume that many fresh copies of $\ket{\psi_{r+1}}$ are given. Then, we run the state-learning primitive, which yields a new parameter set $\boldsymbol{\theta}_{\psi_{r+1}}$, so that $\hat{A}((\boldsymbol{\theta}_{\psi_{r+1}}))\ket{0} = \ket{\psi_{r+1}}$. This circuit is then fed back into the next round as the basis for the reflection $\hat{R}(\boldsymbol{\theta}_{\psi_{r+1}})$.

These two stages can be incorporated as a single loop and it consists of the algorithm for the state-learning-assisted amplitude amplification (The schematic is drawn in our main manuscript). At the abstract level, this algorithm implements exactly the same update rule as the imaginary previous-output-state reflection, provided that the learning stages are ideal. Then, we can consider a lemma.
\begin{lemma}[Equivalence with an imaginary process under perfect learning.]
\label{lemma:equiv-perfect-learn}
Assume that the state-learning primitive is perfect in the sense that, at every round $r$, the learned circuit satisfies $\hat{A}(\boldsymbol{\theta}_{\psi_r})\ket{0}=\ket{\psi_r}$ exactly. Then, the sequence of states generated by the state-learning-assisted algorithm coincides with that of the imaginary process: $\ket{\psi_{r+1}}  =e^{-i\pi} \hat{R}_{\psi_r} \hat{R}_\tau \ket{\psi_r}$.
\end{lemma}

\begin{proof}---The proof is straightforward: if $\hat{A}(\boldsymbol{\theta}_{\psi_r})\ket{0}=\ket{\psi_r}$ exactly, then for any state $\ket{\phi}$, we have $\hat{A}(\boldsymbol{\theta}_{\psi_r})\bigl( \hat{\mathds{1}} - 2\ketbra{0}{0} \bigr) \hat{A}(\boldsymbol{\theta}_{\psi_r})^\dagger \ket{\phi}  = \hat{R}_{\psi_r}\ket{\phi}$. Thus, the update rule reduces to that of Eq.~(\ref{eq:update-imag}).
\end{proof}

As a direct consequence, if we further idealize the situation by counting only the number of uses of the target reflection $R_\tau$ and treating the cost of the state-learning as ``free,'' i.e., as $O(1)$, we obtain the same logarithmic-query behavior as in {\bf Corollary~\ref{cor:log-search}}. Thus, we state that
\begin{corollary}[Logarithmic query rounds under free state-learning]
\label{cor:logN-free-sl}
Under the assumptions of {\bf Lemma~\ref{lemma:equiv-perfect-learn}} and assuming $O(1)$ resources associated with the state-learning stages, the state-learning-assisted amplitude amplification finds the marked target with a constant success probability after $O(\log N)$ oracle calls.
\end{corollary}

\begin{proof}---By {\bf Lemma~\ref{lemma:equiv-perfect-learn}} the state sequence coincides with that of the imaginary previous-output-state protocol. The number of uses of $\hat{R}_\tau$ (i.e., the oracle calls) is identical to the number of rounds, since each round applies exactly a single application of $\hat{R}_\tau$. The claim then follows directly from {\bf Corollary~\ref{cor:log-search}}.
\end{proof}

Actually, {\bf Corollary~\ref{cor:logN-free-sl}} is deliberately idealized: in any realistic setting, the learning primitive consumes a large number of state copies and produces only an approximate description of the state. However, this idealized scenario is extremely useful from a conceptual point of view. It shows that, if state learning is fundamentally ``free'' in the sense of not costing any physically relevant oracle-query, or equivalently, of exhibiting $O(1)$, then the composition of the state-learning and amplitude amplification would lead to an exponential speedup over the original amplitude amplification schemes. Such a speedup would contradict the known optimality of Grover's algorithm within the black-box model~\cite{bennett1997strengths}. This observation allows a critical question about the state-learning; how expensive state-learning must be performed in order to prevent the kind of a theoretically forbidden computational speedup.

\section{No-Signaling, Super-Grover Speedups, and State-Learning Lower Bounds}\label{sec:NoSignaling}

Here we bring into a single framework three a priori distinct layers in the description of quantum information processing tasks. On the algorithmic side, Grover’s original quantum search algorithm establishes that the unstructured search over an $N$-element database admits a quadratic speedup, and no better, within the standard oracle model. On the physics side, relativistic causality is encoded in the no-signaling principle, which forbids any super-luminal transmission of classical information, even in the presence of entanglement. On the state-learning side, we derive sharp sample-complexity bounds for learning unknown quantum states prepared by finite-depth quantum circuits~\cite{haah2016sample,huang2020predicting}. We show that these three perspectives---algorithmic, physical, and statistical learning---are tightly interwoven and, when taken together, a common set of fundamental limits on the state-learning can be derived.

\subsection{Grover speedup versus super-luminal signaling}\label{sec:NoSignaling_Bao}

As described in Sec.~\ref{sec:SLAA}, the Grover's algorithm can solve the unstructured search problem on $N=2^n$ items with $O(\sqrt{N})$ queries. The hybrid method of Bennett, Bernstein, Brassard, and Vazirani shows that, within standard quantum theory, this scaling is optimal: no algorithm can solve the worst-case search problem with $o(\sqrt{N})$ queries~\cite{bennett1997strengths}. Recently, Bao {\it et al.} showed that this lower bound is not merely a technical artifact of the hybrid argument, but is in fact operationally equivalent to the no-signaling principle within broad classes of modified quantum theories~\cite{bao2016grover}.

To state the previous results in the form useful for us, it is helpful to separate the algorithmic resources (e.g., the number of qubits and operations) from the relativistic resources (e.g., the communication channel capacity and spacelike separation). We consider a general modification of the quantum theory obtained by adding a single extra operation $\mathcal{M}$ to the usual unitary and measurement dynamics.  This $\mathcal{M}$ may be a nonunitary map or a special post-selection, depending on the model. Here, we introduce a critical theorem:

\begin{theorem}[Bao-Bouland-Jordan result (reformulated)]
\label{thm:Bao_equivalence}
Consider a theory obtained by supplementing the standard finite-dimensional quantum theory with an additional operation $\mathcal{M}$ acting on at most $m$ qubits at a time.  Then, within each of the models casted in Ref.~\cite{bao2016grover}, the following are equivalent up to the polynomial overhead in the resources $m$ and the number of uses of $\mathcal{M}$:
\begin{itemize}
\item[(i)] The ability to send a classical bit ``super-luminally'' with channel capacity at least a fixed constant using an entangled state of $m$ qubits and $O(m)$ applications of $\mathcal{M}$.
\item[(ii)] The ability to solve the unstructured search on a database of size $N$ using $O\bigl( \mathrm{poly}(m)\log{N} \bigr)$ applications of $\mathcal{M}$ and standard quantum gates.
\end{itemize}
In particular, if a modification $\mathcal{M}$ allows a ``super-Grover speedup'' (i.e., an algorithm with $o(\sqrt{N})$ oracle queries), then it also allows the super-luminal signaling with polynomially related physical resources, and vice versa.
\end{theorem}

\begin{figure}[t]
  \centering
  \includegraphics[width=0.75\textwidth]{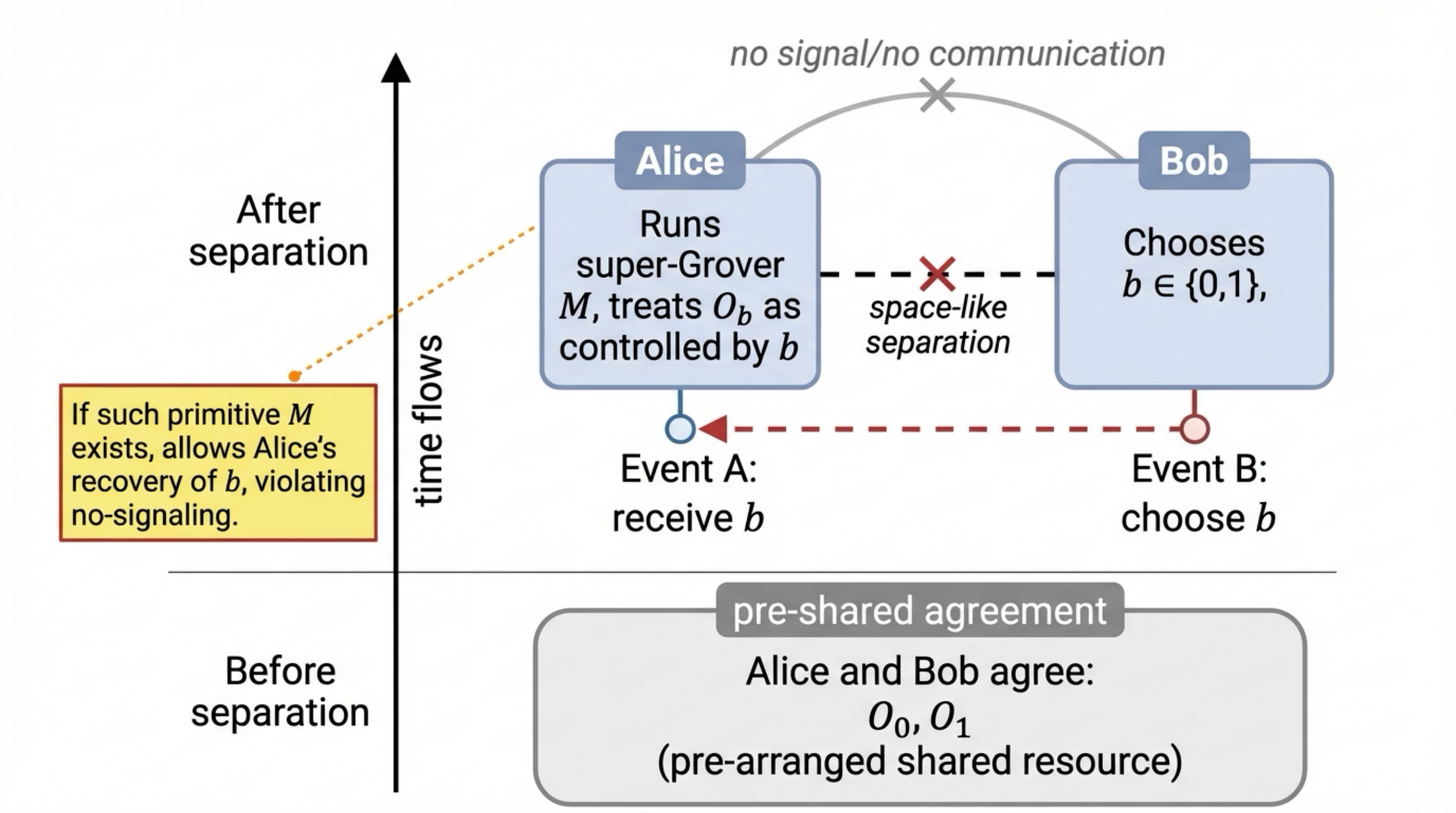}
\caption{Schematic construction of a super-luminal classical channel from a hypothetical super-Grover search algorithm. Bob and Alice are placed at space-like separation. Before they part, they agree on two candidate oracles $O_0$ and $O_1$ acting on Alice's query register, and on a primitive operation $\mathcal{M}$ that, when available to Alice in addition to standard quantum gates, yields an $o(\sqrt{N})$-query super-Grover search. The shared resource is arranged so that Bob's local choice of $b \in \{0,1\}$—implemented by acting only on his subsystem—selects which oracle $O_b$ is effectively realized on Alice's side, without transmitting any physical system between them. After the parties are separated, Bob encodes his bit by choosing $b$ and thereby fixing $O_b$, while Alice, who has no direct access to Bob's choice, runs the super-Grover search that uses $\mathcal{M}$ and treats $O_b$ as the underlying search oracle. If such a primitive $\mathcal{M}$ existed within the standard quantum theory, Alice would recover $b$ in time $o(\sqrt{N})$, i.e., before any light signal from Bob could arrive, thus realizing a super-luminal signaling protocol. For more detailed and self-contained scenario, see Appendix~\ref{append:bao_equiv} or Ref.~\cite{bao2016grover}.}
  \label{fig:super-luminal-channel-AA}
\end{figure}

\begin{proof}---The proof of Theorem~\ref{thm:Bao_equivalence} proceeds in two steps.  First, given a super-Grover search algorithm that uses $\mathcal{M}$, one embeds the searching oracle into a bipartite entangled state shared by two distant parties, say, Alice and Bob. Here, Bob's local choice of whether to encode a solution induces, after Alice's application of the super-Grover routine, measurably different outcome distributions on her side alone, even when Alice and Bob are space-like separated (see Fig.~\ref{fig:super-luminal-channel-AA}). This protocol implements a super-luminal channel whose capacity can be estimated in terms of the search success bias. Conversely, given any protocol that produces a super-luminal channel of nonzero capacity using $\mathcal{M}$, we show that the same primitive can be used to implement a nonunitary transformation on a register encoding the search space.  A refined hybrid argument then implies that this transformation can be used to separate two oracle-marked states whose overlap would be indistinguishable by any
standard-quantum search algorithm in $O(\sqrt{N})$ queries. The detailed and complete proof of this theorem is given in Appendix~\ref{append:bao_equiv}.
\end{proof}

Then, for the purposes of our convenience, we take {\bf Theorem~\ref{thm:Bao_equivalence}} and its constructions as a black box and isolate the operational statement about unstructured search that follows when we forbid the super-luminal signaling. We thus formulate as follow:

\begin{corollary}[Grover lower bound from no-signaling]
\label{cor:Grover_from_NoSignaling}
Assume an operational version of the no-signaling principle: no experiment using at most $\mathrm{poly}(n)$-qubit states and $\mathrm{poly}(n)$ applications of any physically allowed operation can transmit a classical bit between two space-like separated parties with nonzero channel capacity.  Then, within any modified physical theories (Bao-type~\cite{bao2016grover}), every algorithm that solves the unstructured search on $N=2^n$ items using at most $\mathrm{poly}(n)$ qubits must make at least $\Omega(\sqrt{N})$ oracle queries in the worst case.
\end{corollary}

\begin{proof}---Suppose, to the contrary, that there exists a family of search algorithms $\{\mathcal{A}_N\}$ that succeeds with bounded error using $q(N)=o(\sqrt{N})$ oracle queries and $\mathrm{poly}(n)$-size states, for some modification $\mathcal{M}$ of quantum theory.  For any polynomial bound $q(N)\le N^{1/2-\varepsilon}$ with $\varepsilon>0$, {\bf Theorem~\ref{thm:Bao_equivalence}} implies that the same resources $(m, q(N))$ suffice to construct a protocol that transmits a classical bit ``super-luminally'' with constant channel capacity. This contradicts the operational no-signaling assumption.  Therefore no such family $\{\mathcal{A}_N\}$ can exist, and the standard Grover lower bound $q(N) \ge c\sqrt{N}$ for some constant $c > 0$ follows within the classes of physical theories analyzed.
\end{proof}

Actually, {\bf Corollary~\ref{cor:Grover_from_NoSignaling}} is a more applicable form in our work, which invokes {\bf Theorem~\ref{thm:Bao_equivalence}}.  Rather than viewing Grover's lower bound as a purely computational statement about the quantum query, we regard it as an operational consequence of relativistic causality: that is, any attempt to beat the $\Omega(\sqrt{N})$ scaling necessarily opens a super-luminal signaling channel if the deviation from standard quantum mechanics is exploitable with polynomial resources.

\subsection{No-signaling-based lower bounds on state-learning complexity}\label{sec:NoSignaling_LB}

{\bf Corollary~\ref{cor:log-search}} shows that logarithmic-query search is achievable if the previous-output reflection $\hat{R}_{\psi_r}$ [see Eq.~(\ref{eq:R-psir})] is treated as an abstract primitive. In physically realistic settings, however, this a priori unknown-state reflection must itself be implemented, for example, by incorporating a state-learning routine into the amplitude-amplification process, as described in Fig.~1 of our main manuscript. We now show that under the no-signaling principle, any such state-learning necessarily incurs a large resource overhead, thereby the logarithmic-round advantage should be negated. This naturally leads to a fundamental lower bound on the resources required for the state learning.

\subsubsection*{A state-learning model for unknown-state reflection} 

We model a generic state-learning-assisted implementation of the previous-output reflection $\hat{R}_{\psi_r}$ as follows. Let $\ket{\psi}$ be an unknown $n$-qubit pure state, and $\mathcal{F}$ be a learning algorithm that has access to $M_\mathrm{s}$ independent copies of $\ket{\psi}$. $\mathcal{F}$ yields a classical description $\boldsymbol{\theta}_\psi$ of an ansatz circuit $\hat{A}(\boldsymbol{\theta}_\psi)$ with the gate complexity $G(n)$, such that
\begin{eqnarray}
\left\| \hat{A}(\boldsymbol{\theta}_\psi)\ket{0}^{\otimes n} - \ket{\psi} \right\| \le \varepsilon
\end{eqnarray}
for some fixed accuracy $\varepsilon<1$. Using $\hat{A}(\boldsymbol{\theta}_\psi)$ and its adjoint $\hat{A}(\boldsymbol{\theta}_\psi)^\dagger$, one can synthesize the reflection $\hat{R}_\psi$ approximately, i.e., with the operator-norm error $O(\varepsilon)$.

We are interested in the dependence of the required sample budget $M_\mathrm{s}$ and the gate complexity $G$ on the problem size $N=2^n$ in the worst case over all $\ket{\psi}$. We assume that $\mathcal{F}$ itself is implemented by a polynomial-time classical and/or quantum computation, but, for the lower bounds below, it suffices to count only the number of copies $M_\mathrm{s}$ consumed and the gate complexity $G$ of the synthesized circuit.

\subsubsection*{A sample lower bound from query complexity} 

To see how the no-signaling principle constrains $M_\mathrm{s}$, we consider the implementation of the ideal logarithmic-round search algorithm of {\bf Corollary~\ref{cor:log-search}} by substituting the state-learning-assisted reflections. For notational simplicity, we write $N=2^n$ and denote by $r(N) = \Theta(\log N)$.

At the $r$-th round of the protocol, the ideal previous-output state $\ket{\phi_r}$ is a function of the unknown marked index $\tau$, and the corresponding reflection must be implemented by calling $\mathcal{F}$ on
$\ket{\phi_r}$. Here, the crucial observation is that, once the state at round $r-1$ has been learned and encoded into an ansatz circuit $\hat{A}_{r-1} = \hat{A}(\boldsymbol{\theta}_{\psi_{r-1}})$, further preparations of $\ket{\phi_{r-1}}$ can be performed by applying $\hat{A}_{r-1}$, without any additional oracle queries (see Fig.~1 of our main manuscript). To obtain a single copy of $\ket{\phi_r}$, one may therefore prepare $\ket{\phi_{r-1}}$ via $\hat{A}_{r-1}$ and then apply one fixed amplitude-amplification step that invokes the oracle $\hat{R}_\tau$ a constant number of times. Hence, there exists a constant $c_1 > 0$ such that, for every round $r$, preparing $M_\mathrm{s}$ copies of $\ket{\phi_r}$ for the state-learning requires at least
\begin{eqnarray}
Q_r^\mathrm{(train)} \ge c_1 M_\mathrm{s}
\end{eqnarray}
oracle queries, independently of $r$. Summing over all rounds up to $r(N)$, the total number of oracle queries devoted to training the reflections satisfies
\begin{eqnarray}
Q_\mathrm{train}(N) \ge \sum_{r=1}^{r(N)} Q_r^\mathrm{(train)} \ge c_1 M_\mathrm{s} r(N).
\label{eq:AAAA}
\end{eqnarray}
The final ``production'' run, which uses the learned reflections without further state-learning, adds only $O(r(N))$ additional oracle queries, which is negligible compared to $O(M_\mathrm{s} r(N))$ learning budget for large $M_\mathrm{s}$. Combining these estimates gives, for all sufficiently large $N$,
\begin{eqnarray}
Q_\mathrm{tot}(N) \ge c_1' M_\mathrm{s} r(N)
\label{eq:Qtot_vs_Ns}
\end{eqnarray}
for some constant $c_1'>0$ that absorbs the lower-order contributions.

On the other hand, {\bf Corollary~\ref{cor:Grover_from_NoSignaling}} asserts that any physically realizable quantum search must satisfy
\begin{eqnarray}
Q_\mathrm{tot}(N) \ge c_Q \sqrt{N}
\end{eqnarray}
for some constant $c_Q > 0$, otherwise one can use the algorithm to construct a super-luminal signaling protocol with polynomial resources. Thus, substituting Eq.~(\ref{eq:Qtot_vs_Ns}) and recalling
$r(N)=\Theta(\log N)$, we can yield the following theorem:
\begin{theorem}[No-signaling sample lower bound in state learning]
\label{thm:SampleLB}
Let $N=2^n$ and $r(N)=\Theta(\log N)$ be the number of amplification rounds in the ideal previous-output-reflection search algorithm of {\bf Theorem~\ref{thm:cubic-aa}}. Suppose that for each round $r$ this
algorithm is implemented in standard quantum theory by a state-learning subroutine $\mathcal{F}$ that, given $M_\mathrm{s}$ copies of the $n$-qubit previous-output state $\ket{\phi_r}$, produces an approximate reflection $\hat{R}_{\psi_r}$ with fixed accuracy $O(\varepsilon)$.

If the overall search algorithm respects the no-signaling principle in the sense of {\bf Corollary~\ref{cor:Grover_from_NoSignaling}}, then for all sufficiently large $N$, the sample budget must satisfy
\begin{eqnarray}
M_\mathrm{s} \ge \Omega\left( \frac{\sqrt{N}}{\log N} \right).
\end{eqnarray}
Equivalently, in terms of the number $n=\log_2 N$ of qubits,
\begin{eqnarray}
M_\mathrm{s}(n) \ge \Omega\left( \frac{2^{n/2}}{n} \right).
\end{eqnarray}
\end{theorem}

\begin{proof}---Combining Eq.~(\ref{eq:Qtot_vs_Ns}) with the Grover lower bound $Q_\mathrm{tot}(N)\ge c_Q\sqrt{N}$ implied by no-signaling gives the following: $c_1' M_\mathrm{s} r(N) \ge c_Q \sqrt{N}$. Since $r(N)=\Theta(\log N)$, there exists a constant $c_3>0$ such that
$r(N)\le c_3 \log N$ for all sufficiently large $N$. Therefore, we have
\begin{eqnarray}
M_\mathrm{s} \ge \frac{c_Q}{c_1' c_3} \frac{\sqrt{N}}{\log N},
\end{eqnarray}
which establishes the claimed $\Omega(\sqrt{N}/\log N)$ scaling. The statement in terms of $n$ follows from $N=2^n$.
\end{proof}

{\bf Theorem~\ref{thm:SampleLB}} indicates that any state-learning-assisted implementation of the logarithmic-round previous-output-reflection protocol must consume at least $\Omega\bigl(\tfrac{\sqrt{N}}{\log N}\bigr)$ copies of each intermediate state (up to constant factors), otherwise the total number of oracle queries would drop below the Grover threshold and the resulting search algorithm could be used to send super-luminal signaling.

\subsubsection*{Gate complexity lower bound} 

A similar reasoning gives a lower bound on the gate complexity $G$ of the ansatz circuit $\hat{A}$ used to implement the unknown-state reflections. For this purpose, it is also useful to count the elementary gates together with the oracle queries.

Let $D_\mathrm{ref}$ denote the depth (or, equivalently, the number of sequential layers of one- and two-qubit gates) of the synthesized reflection $\hat{R}_{\phi_r}$ treated by $\mathcal{F}$; by assumption $D_\mathrm{ref}$ is $O(G)$. The logarithmic-round search algorithm calls a previous-output reflection a constant number of times per round, so its total logical circuit depth due to these reflections is at most
\begin{eqnarray}
D_\mathrm{tot}(N) \le c_4 G(n) r(N)
\end{eqnarray}
for some constant $c_4>0$, where we made explicit the dependence $G=G(n)$ on the number of data qubits $n$. Additional gates implementing $\hat{R}_\tau$ and simple single-qubit rotations contribute at most $O(r(N))$ depth and are negligible in the regime where $G(n)$ grows super-logarithmically with $N$.

Now imagine embedding this search algorithm into the bipartite signaling scenario of Bao {\it et al.}~\cite{bao2016grover}, in which Alice and Bob share an entangled state and only Alice executes the search circuit on her local subsystem, while Bob's local choice (encoding the message bit) determines whether or not an effective solution exists. Assuming each elementary gate (layer) acts within a fixed time $t$ and that Alice and Bob are separated by a distance $L$, the total time between Bob's choice and Alice's measurement is approximately $t D_\mathrm{tot}(N)$. If $D_\mathrm{tot}(N)$ were $o(\sqrt{N})$, then for sufficiently large $N$ one could choose $K$ such that
\begin{eqnarray}
t D_\mathrm{tot}(N) < \frac{L}{c^\star} \quad (c^\star: \text{speed of light}),
\end{eqnarray}
so that Alice would learn Bob's bit before a classical light signal could cross the distance $L$, in violation of the no-signaling. This simple and conceptual reasoning yields:
\begin{theorem}[No-signaling gate-complexity lower bound]
\label{thm:GateLB}
Under the assumptions of {\bf Theorem~\ref{thm:cubic-aa}}, suppose that each previous-output reflection is implemented by an ansatz circuit $\hat{A}$ with gate complexity $G(n)$ and depth $D_\mathrm{ref}=O(G(n))$. If the resulting logarithmic-round search algorithm respects the operational no-signaling principle for all $N=2^n$, then for all sufficiently large $N$, one must have
\begin{eqnarray}
G(n) \ge \Omega\left( \frac{\sqrt{N}}{\log N} \right) = \Omega\left( \frac{2^{n/2}}{n} \right).
\end{eqnarray}
\end{theorem}

\begin{proof}---We defer a more detailed and self-contained proof to Appendix~\ref{append:nosig_lower}.
\end{proof}


\subsection{Physical remarks}\label{sec:NoSignaling_Physics}

The bounds derived in {\bf Theorems~\ref{thm:SampleLB}} and {\bf Theorem~\ref{thm:GateLB}} admit a natural physical interpretation. The extended theory in which an ideal programmable previous-output reflection exists is structurally similar to the discrete modifications of quantum mechanics studied by Bao {\it et al.}, such as universal cloning or postselection onto an arbitrary state~\cite{bao2016grover}. In each case, adding a single
nonstandard operation to the usual unitary dynamics suffices to collapse the Grover lower bound and to unlock super-luminal signaling with only polynomial overhead in the required resources. Our state-learning-assisted logarithmic-search protocol (of {\bf Theorem~\ref{thm:cubic-aa}}) is precisely of this form: in the limit where the state-learning is treated as ``free,'' it acts as an effective previous-output reflection and therefore falls into the same class of acausal modifications.

From this perspective, {\bf Theorems~\ref{thm:SampleLB}} and {\bf Theorem~\ref{thm:GateLB}} quantify how standard quantum mechanics protects relativistic causality. {\bf Theorem~\ref{thm:no-reflection}} forbids an exact previous-output reflection, but one might hope to approximate it extremely well using a small number of copies of the program state and a shallow ansatz circuit. Our bounds show that this hope is incompatible with no-signaling whenever the approximation is strong enough to support the logarithmic-round search. Either the sample budget $M_\mathrm{s}$ or the gate complexity $G(n)$ (or both) must grow at least on the order of $\tfrac{\sqrt{N}}{\log N} = \tfrac{2^{n/2}}{n}$, up to constant factors, so that when multiplied by the $r(N)=\Theta(\log N)$ rounds of amplitude amplification the total physical resources required to implement previous-output reflections restore the effective $\Omega(\sqrt{N})$ scaling of the unstructured search.

\section{Physics-and-computation concordance in state-learning bounds}\label{sec:consistency}

We now forget about the relativistic causality. Here, we consider an unknown $n$-qubit state, promised only to lie in the manifold of states preparable by a circuit with at most $G$ two-qubit gates, and we seek to characterize how many samples are information-theoretically necessary and sufficient to reconstruct this state up to inaccuracy $\varepsilon$. Our aim is twofold: first, to derive sharp computational upper and lower bounds on this sample complexity; and second, to show that, when these bounds are combined with the no-signaling-based gate lower bounds of Sec.~\ref{sec:NoSignaling}, the resulting picture is fully consistent with the physically derived constraints.

\subsection{Computational sample complexity for circuit-generated pure states} \label{subsec:comp_bounds}

We begin by defining the learning problem. Fix $n \in \mathbb{N}$ and let $N=2^{n}$ denote the Hilbert-space dimension. Let $\mathcal{U}_{n,G}$ be the set of $n$-qubit unitaries that can be realized by a circuit consisting of at most $G$ two-qubit gates drawn from a fixed universal gate set. We define the corresponding family of learning-target pure states
\begin{eqnarray}
\mathcal{S}_{n,G} = \bigl\{ \ket{\psi} = \hat{U}\ket{0}^{\otimes n} : \hat{U} \in \mathcal{U}_{n,G} \bigr\}.
\end{eqnarray}
In this setting, a state-learning algorithm (learner) $\mathcal{F}$ receives $M_{\mathrm{s}}$ copies of the unknown state $\hat{\rho}=\ketbra{\psi}{\psi}$ with $\ket{\psi} \in \mathcal{S}_{n,G}$, performs an arbitrary joint POVM on these copies, and outputs a classical description of a hypothesis state $\hat{\rho}_h$. We measure accuracy in the trace distance $D(\hat{\rho},\hat{\rho}_h)=\tfrac{1}{2} \bigl\| \hat{\rho} - \hat{\rho}_h \bigr\|_{1}$; in particular, we write $D(\hat{\rho},\hat{\rho}_h)$ for the distance between the true state and the hypothesis produced by $\mathcal{F}$.

We now formalize the worst-case learning sample complexity.

\begin{definition}[Worst-case sample complexity]
\label{def:worst-case-SC}
For fixed $n, G, \varepsilon, \delta$ with $0< \varepsilon < 1$ and $0< \delta < \tfrac{1}{2}$, we define $M_{\mathrm{s}}(n, G, \varepsilon, \delta)$ as the smallest integer such that there exists a learner (equivalently, a learning algorithm) $\mathcal{F}$ that outputs $\hat{\rho}_h$ together with a measurement strategy with the following property: for every $\hat{\rho} \in \mathcal{S}_{n,G}$,
\begin{eqnarray}
\Pr\bigl[ D(\hat{\rho}, \hat{\rho}_h) \le \varepsilon \bigr] \ge 1-\delta.
\end{eqnarray}
The probability is taken over the measurement outcomes and any internal randomness of $\mathcal{F}$.
\end{definition}

Throughout this subsection, we assume that the learner $\mathcal{F}$ is allowed to perform arbitrary collective POVMs on all $M_{\mathrm{s}}$ copies and arbitrary classical post-processing, subject only to the standard quantum-information and computation model (i.e., with no additional depth constraint coming from physical principles, such as no-signaling). Under these assumptions, we obtain the following theorem.

\begin{theorem}[Computational state-learning sample bounds]
\label{thm:CompSL}
Let $0< \varepsilon \le \tfrac{1}{4}$ and $0< \delta \le \tfrac{1}{10}$, and consider the family $\mathcal{S}_{n,G}$ of $n$-qubit learning-target states generated by circuits $\hat{A}$ with at most $G$ two-qubit gates. Then the following holds: 
\begin{itemize}
\item[$\mathrm{(i)}$] \emph{Upper bound.}---There exists a (generally inefficient) learning algorithm $\mathcal{F}$ such that 
\begin{eqnarray}
M_{\mathrm{s}}(n,G,\varepsilon,\delta) \le \frac{c_{1}}{\varepsilon^{2}} \min\Bigl\{ 2^{n}\log\frac{1}{\delta}, ~G \log{\frac{G}{\varepsilon}} + \log{\frac{1}{\delta}} \Bigr\}.
\label{eq:CompUB}
\end{eqnarray}

\item[$\mathrm{(ii)}$] \emph{Lower bound.}---For any learning algorithm $\mathcal{F}$ (with arbitrary POVMs and post-processing)\footnote{This result is consistent with the (information-theoretic) Thm.~1 of Ref.~\cite{zhao2024learning}, up to constant factors.},
\begin{eqnarray}
M_{\mathrm{s}}(n,G,\varepsilon,\delta) \ge \frac{c_{2}}{\varepsilon^{2}} \Bigl( \min\bigl\{ 2^{n}, ~G \bigr\} + \log{\frac{1}{\delta}} \Bigr).
\label{eq:CompLB-info}
\end{eqnarray}
\end{itemize}
Here, $c_{1}, c_{2} > 0$ are universal constants.
\end{theorem}

\begin{proof}[Proof sketch.]---We outline the main ingredients; a detailed and self-contained derivation is given in Appendix~\ref{append:CompBounds}.

(i) Upper bound: The upper bound follows from a standard ``net + hypothesis selection'' strategy~\cite{vershynin2009high,buadescu2021improved}. First, we bound the covering number of $\mathcal{S}_{n,G}$ in trace distance and fix a universal gate set with a finite number $p = O(1)$ of real parameters per gate. A depth-$G$ layout (i.e., a choice of qubit locations and gate sequence) can be specified in at most $(\kappa_{\mathrm{arch}} n^{2})^{G}$ ways for some architecture-dependent constant $\kappa_{\mathrm{arch}}>0$. For such a layout, the map
\begin{eqnarray}
\boldsymbol\theta \in \mathbb{R}^{pG} \longmapsto \ket{\psi(\boldsymbol\theta)} = \hat{U}(\boldsymbol\theta)\ket{0}^{\otimes n}
\label{eq:theta_map}
\end{eqnarray}
is Lipschitz in the Euclidean metric on the parameter cube: changing a single parameter perturbs at most one two-qubit gate, and a telescoping product over the $G$ gates yields
\begin{eqnarray}
D\bigl( \ketbra{\psi(\boldsymbol\theta)}{\psi(\boldsymbol\theta)}, ~\ketbra{\psi(\boldsymbol\theta')}{\psi(\boldsymbol\theta')} \bigr) \le C_{\mathrm{L}} G \bigl\| \boldsymbol\theta -\boldsymbol\theta' \bigr\|_{2},
\end{eqnarray}
where $C_{\mathrm{L}} > 0$ is a constant. Thus, by discretizing each parameter on a grid of mesh size $\sim {\varepsilon}/{(C_{\mathrm{L}} G)}$, an $\tfrac{\varepsilon}{4}$-net on the image of that layout is produced. The number of grid points per layout is at most $\bigl[{(C_{p} G)}/{\varepsilon}\bigr]^{pG}$ for some constant $C_{p}>0$ depending only on the choice of gate set, so the total number of net points to cover $\mathcal{S}_{n,G}$ is bounded by
\begin{eqnarray}
\abs{\mathcal{N}_{\varepsilon/4}} \le \bigl(\kappa_{\mathrm{arch}} n^{2}\bigr)^{G} \left( \frac{C_{p} G}{\varepsilon} \right)^{pG} \le \exp\Bigl(\xi G\log{\frac{G}{\varepsilon}}\Bigr),
\end{eqnarray}
where we have absorbed the polynomial factors of $n$ and $G$ into the logarithm. Here, $\xi > 0$ is a constant. Thus the metric entropy of $\mathcal{S}_{n,G}$ with respect to the trace distance satisfies
\begin{eqnarray}
\log\abs{\mathcal{N}_{\varepsilon/4}} = O\Bigl( G \log{\frac{G}{\varepsilon}} \Bigr).
\label{eq:metric-entropy-SnG}
\end{eqnarray}

Given this finite hypothesis class, we can perform hypothesis selection assisted by classical-shadow tomography and other sample-optimal primitives~\cite{buadescu2021improved}. Specifically, with
\begin{eqnarray}
M_{\mathrm{s}} = O \left( \frac{1}{\varepsilon^{2}} \Bigl(\log{\abs{\mathcal{N}_{\varepsilon/4}}} + \log{\frac{1}{\delta}} \Bigr) \right)
\end{eqnarray}
copies, one can estimate a suitable family of observables on all candidates in $\mathcal{N}_{\varepsilon/4}$ and then select the candidate closest to the empirical data. Then, the standard concentration inequalities (e.g., Markov's and Chebyshev's) imply that, with probability at least $1-\delta$ (say, confidence), the chosen hypothesis is within the inaccuracy, i.e., the trace distance, $\varepsilon$ of the true state, which yields Eq.~(\ref{eq:CompUB}) upon substituting Eq.~(\ref{eq:metric-entropy-SnG}). The term proportional to $2^{n}$ in Eq.~(\ref{eq:CompUB}) is obtained by specializing to the regime where $G$ is so large that $\mathcal{S}_{n,G}$ essentially fills the full $N$-dimensional pure-state manifold, recovering the usual optimal tomography rate $M_{\mathrm{s}} = \Theta(2^{n}\log(1/\delta)/\varepsilon^{2})$~\cite{haah2016sample,o2016efficient}.

(ii) Lower bound: We reduce state learning to a multi-hypothesis discrimination problem over a large packing net inside $\mathcal{S}_{n,G}$. In particular, we invoke Holevo-type information bounds in the proof.

When $G\ge c\,2^{n}$ for some constant $c>0$, the family $\mathcal{S}_{n,G}$ contains (up to arbitrarily small error) all $n$-qubit pure states, and standard tomography lower bounds imply~\cite{haah2016sample,o2016efficient}
\begin{eqnarray}
M_{\mathrm{s}}(n,G,\varepsilon,\delta) \ge \Omega\Bigl(\frac{2^{n} + \log(1/\delta)}{\varepsilon^{2}}\Bigr),
\end{eqnarray}
which matches the first term in Eq.~(\ref{eq:CompLB-info}).

In the circuit-limited regime $G\ll 2^{n}$, we proceed as follows. Fix $k$ such that $2^{k} \asymp G$ (for instance, $k=\lfloor\log_{2}G\rfloor$), and restrict attention to circuits $\hat{U} \in \mathcal{U}_{n,G}$ that act nontrivially only on the first $k$ qubits while leaving the remaining $n-k$ qubits in $\ket{0}$. For suitable layouts, the depth-$G$ circuits on $k$ qubits generate a set of states whose image contains an $\varepsilon$-net of the full $k$-qubit (unit) sphere. Then, a volumetric packing argument on this sphere produces a subset $\bigl\{ \hat{\rho}_{1}, \ldots, \hat{\rho}_{K} \bigr\} \subset \mathcal{S}_{n,G}$ with
\begin{eqnarray}
K = e^{\Omega(G)},
\end{eqnarray}
exhibiting $D(\hat{\rho}_{i}, \hat{\rho}_{j}) > 2\varepsilon$ for all $i \neq j$. In other words, $\mathcal{S}_{n,G}$ contains a $2\varepsilon$-packing net of size exponential in $G$. 

Now, place a uniform prior on the index $J \in \{1, \ldots, K\}$ and consider any learning algorithm $\mathcal{F}$ that, using $M_{\mathrm{s}}$ copies of $\hat{\rho}_{J}$, yields $\hat{\rho}_h$ such that $D(\hat{\rho}_{J}, \hat{\rho}_h) \le \varepsilon$ with success probability at least $1-\delta$ for every $J$. By the triangle inequality, whenever $D(\hat{\rho}_{J},\hat{\rho}_h)\le\varepsilon$, the nearest net point to $\hat{\rho}_h$ in the trace-distance metric must be $\hat{\rho}_{J}$ itself, because all other net points are more than $2\varepsilon$ away. Thus, from the hypothesis $\hat{\rho}_h$, one can construct a decoder $\tilde{J}$ for the index $J$ whose error probability is at most $\delta$.

Let $X$ denote the random variable representing the true index $J$ (uniform on $\{1, \ldots, K\}$), and let $Y$ denote the classical outcome of the measurement. Fano's inequality then implies
\begin{eqnarray}
I(X:Y) \ge (1-\delta)\log{K} - h_2(\delta),
\end{eqnarray}
where $h_2(\delta)$ is the binary entropy. On the other hand, Holevo's theorem bounds the accessible information per copy of the quantum state: for ensembles of pure states that are pairwise within trace distance $O(1)$, one obtains
\begin{eqnarray}
I(X:Y) \le C_\chi \varepsilon^{2} M_{\mathrm{s}} \quad (\text{for some constant $C_\chi > 0$}),
\end{eqnarray}
reflecting that the distinguishability (and hence the mutual information) scales at most quadratically with the trace-distance separation when many hypotheses must be resolved. By combining these two inequalities and solving for $M_{\mathrm{s}}$, we have
\begin{eqnarray}
M_{\mathrm{s}} \ge \frac{(1-\delta)\log{K} - h_2(\delta)}{C_\chi \varepsilon^{2}} = \Omega\Bigl(\frac{G}{\varepsilon^{2}}\Bigr),
\end{eqnarray}
up to additive $\log(1/\delta)$ terms and constant factors, because $\log{K}=\Omega(G)$ by construction. Merging this gate-limited regime with the $G\ge c\,2^{n}$ regime above gives Eq.~(\ref{eq:CompLB-info}).
\end{proof}

Two qualitative conclusions are worth emphasizing. First, in the quantum-information-theoretic setting (arbitrary POVMs and post-processing), the effective ``size'' of the learning problem is $\min\{2^{n},G\}$: whenever $G\ll 2^{n}$, the relevant dimension is the circuit complexity $G$ rather than the full Hilbert-space dimension $2^{n}$ (see also Ref.~\cite{aaronson2018shadow,huang2020predicting}). Second, both the upper bound~(\ref{eq:CompUB}) and the lower bound~(\ref{eq:CompLB-info}) scale linearly in $G$ (up to the harmless $\log(G/\varepsilon)$ factor in the upper bound) and quadratically in $1/\varepsilon$. In particular, there is no way---even with completely unconstrained collective measurements---to reduce the worst-case sample complexity below order $G/\varepsilon^{2}$.

\subsection{From no-signaling to learnability: matching physical and computational lower bounds}\label{subsec:matching_bounds}

In Sec.~\ref{sec:NoSignaling}, {\bf Theorem~\ref{thm:SampleLB}} and {\bf Theorem~\ref{thm:GateLB}} show that any state-learning-assisted implementation of the quantum search setting must obey
\begin{eqnarray}
M_{\mathrm{s}}(N) \ge \alpha \frac{\sqrt{N}}{\log N},
\quad
G_{\mathrm{ref}}(n) \ge \beta \frac{\sqrt{N}}{\log N},
\label{eq:PhysLB-recap}
\end{eqnarray}
where $\alpha, \beta > 0$ are universal constants. Here, $G_{\mathrm{ref}}(n)$ denotes the gate complexity of the reflection ansatz $\hat{A}$. Note that the bounds in Eq.~(\ref{eq:PhysLB-recap}) are derived using only a physical principle, namely no-signaling. On the other hand, from the computational side, {\bf Theorem~\ref{thm:CompSL}} gives, under the purely quantum-information-theoretic assumptions [i.e., Eq.~(\ref{eq:CompLB-info})],
\begin{eqnarray}
M_{\mathrm{s}}(n, G, \varepsilon, \delta) \ge \frac{c_{2}}{\varepsilon^{2}} \biggl( \min\bigl\{ 2^{n}, G\bigr\} + \log{\frac{1}{\delta}} \biggr).
\label{eq:CompLB-info-again}
\end{eqnarray}
At first sight, Eq.~(\ref{eq:PhysLB-recap}) and Eq.~(\ref{eq:CompLB-info-again}) come from completely different worlds: one is a statement about light cones and super-luminal signaling, the other about metric entropy and accessible information. We now show explicitly how these two levels become tightly connected and fully consistent.

\subsubsection*{State-universal design and the cost of exploring $SU(2^{n})$} 

We begin with a remark on the design principle for the ``worst-case'' ansatz.

\begin{remark}[Worst-case-design principle]
The ansatz circuit $\hat{A}(\boldsymbol{\theta})$ that implements the previous-output reflections is, in principle, required to be able to prepare every $n$-qubit pure state from a fiducial state, e.g., $\ket{0}^{\otimes n}$, up to arbitrarily small error.
\end{remark}

In particular, this remark indicates that the reachable set of $\hat{A}(\boldsymbol{\theta})$ contains an $\varepsilon$-net (for every $\varepsilon>0$) of the full projective space of pure states in $\mathbb{C}^{2^{n}}$. A simple parameter-counting argument shows that the corresponding gate complexity must then be exponential in $n$~\cite{barenco1995elementary,knill1995approximation}. We state this more formally as follows.

\begin{proposition}[Universal state-preparation cost]
\label{prop:G-univ}
Let $\mathcal{G}$ be a fixed universal gate set consisting of one- and two-qubit gates, and let $G_{\mathrm{univ}}(n)$ denote the smallest integer $G$ such that there exists a circuit family with at most $G$ two-qubit gates from $\mathcal{G}$ whose image contains an $\varepsilon$-net of all $n$-qubit pure states for every $\varepsilon > 0$. Then
\begin{eqnarray}
G_{\mathrm{univ}}(n) = \Theta\bigl(2^{n}\bigr) = \Theta(N),
\end{eqnarray}
where $N=2^{n}$ is the Hilbert-space dimension.
\end{proposition}

\begin{proof}---The manifold of $n$-qubit pure states modulo global phase is $\mathbb{CP}^{2^{n}-1}$, of real dimension $2(2^{n}-1) = \Theta(2^{n})$. Fix a circuit layout on $n$ qubits with at most $G$ two-qubit gates from $\mathcal{G}$ and an arbitrary number of one-qubit gates. Each gate carries at most $p=O(1)$ real parameters. Thus, for a fixed layout with $G$ two-qubit gates and at most $\gamma G$ one-qubit gates, the total number of continuous parameters is at most $d_{\mathrm{param}} \le c_{2} G$ for some constant $c_{2}>0$. The map from the parameter space to the set of output states $\Phi: \mathbb{R}^{d_{\mathrm{param}}} \to \mathbb{CP}^{2^{n}-1}$ (specifically, $\theta\mapsto \ket{\psi(\theta)} = \hat{A}(\boldsymbol\theta)\ket{0}^{\otimes n}$ as in Eq.~(\ref{eq:theta_map})) is smooth. The image of a smooth map from a $d_{\mathrm{param}}$-dimensional manifold cannot contain an open set of a manifold of larger dimension. Hence, if $c_{2}G < 2(2^{n}-1)$, the image of $\Phi$ has empty interior in $\mathbb{CP}^{2^{n}-1}$ and cannot contain an $\varepsilon$-net for all sufficiently small $\varepsilon$. Therefore, any state-universal ansatz must satisfy
\begin{eqnarray}
c_{2}G_{\mathrm{univ}}(n) \ge 2(2^{n}-1),
\end{eqnarray}
or equivalently $G_{\mathrm{univ}}(n)\ge c_{2}' 2^{n}$ for some $c_{2}'>0$.
\end{proof}

The known state-synthesis constructions using $O(2^{n})$ two-qubit gates (e.g., based on recursive Householder reflections~\cite{Baek:2025asd}) show that this lower bound is tight, so $G_{\mathrm{univ}}(n)=\Theta(2^{n})$. Substituting $G=G_{\mathrm{univ}}(n)$ into the computational lower bound~(\ref{eq:CompLB-info-again}) and fixing $\delta$ to a constant, we obtain
\begin{eqnarray}
M_{\mathrm{s}}(n,G_{\mathrm{univ}},\varepsilon,\delta) = \Theta\bigl( \tfrac{1}{\varepsilon^{2}} 2^{n} \bigr) = \Theta\bigl( \tfrac{1}{\varepsilon^{2}} N \bigr).
\label{eq:Ms-univ-info}
\end{eqnarray}
In words, if the learner is designed to be able to handle all $n$-qubit target pure states (i.e., state-universal), then purely computational considerations already force an exponential worst-case sample complexity.

Consequently, in the only-computational regime, if one is obliged to assume a state-universal learner and to impose no structural restriction on the learning ansatz $\hat{A}$---that is, no restriction on the family of the learning-target states---then the sample-complexity lower bound remains locked at order $2^{n}$ as in Eq.~(\ref{eq:Ms-univ-info}) and cannot be reduced further.

\subsubsection*{Search-adapted reflections and the $\sqrt{N}$ scale} 

We now turn to the additional structure dictated by the no-signaling principle in the state-learning-assisted amplitude amplification. Let $\Xi_{\mathrm{SL}}(N)$ denote the set of all intermediate pure states that appear in the logarithmic-round amplify-learn protocol of Sec.~\ref{sec:SLAA} when we range over all amplification rounds $1 \le r \le r_{\max}(N)=\Theta(\log N)$. By construction, $\Xi_{\mathrm{SL}}(N)$ is a highly structured subset of the full $2^{n}$-dimensional pure-state manifold.

\begin{definition}[Reflection gate complexity for the amplify-learn protocol]
\label{def:G-ref}
For fixed $n$ and $N=2^{n}$, let $G_{\mathrm{ref}}(n)$ be the smallest integer $G$ such that there exists a family of $n$-qubit learning-ansatz circuits $\{\hat{A}_{{\psi_r}, \tau}\}$ with at most $G$ two-qubit gates, satisfying the following: for every intermediate state $\ket{\psi_{r}} \in \Xi_{\mathrm{SL}}(N)$ and every accuracy parameter $\varepsilon>0$, the circuit $\hat{A}_{{\psi_r},\tau}$ prepares, from $\ket{\mathbb{0}}=\ket{0}^{\otimes n}$, a state within trace distance at most $\varepsilon$ of $\ket{\psi_{r}}$. Here, the reflection about $\ket{\psi_r}$ can be synthesized as [see Eq.~(\ref{eq:R-psi-learned})]
\begin{eqnarray}
\hat{R}_{\psi_{r}} = \hat{\mathds{1}} - 2\ketbra{\psi_{r}}{\psi_{r}} = \hat{A}_{{\psi_r},\tau} \hat{R}_{0} \hat{A}_{{\psi_r},\tau}^{\dagger},
\end{eqnarray}
where $\hat{R}_{0} = \hat{\mathds{1}} - 2\ketbra{\mathbb{0}}{\mathbb{0}}$ is the fixed reference reflection whose cost is independent of $(r, \tau)$.
\end{definition}

By definition, every such intermediate state belongs to the circuit-generated family $\mathcal{S}_{n,G_{\mathrm{ref}}}$, and thus the computational lower bound in Eq.~(\ref{eq:CompLB-info-again}) applies with $G=G_{\mathrm{ref}}(n)$. Combining this with the gate lower bound from no-signaling in Eq.~(\ref{eq:PhysLB-recap}), we can derive a sample lower bound that exactly matches the physics-based one.

\begin{proposition}[No-signaling gate lower bound implies the optimal sample lower bound]
\label{prop:Ms-from-G}
Let $N=2^{n}$ and suppose the state-learning-assisted amplitude amplification protocol of Sec.~\ref{sec:SLAA} is implemented with the reflection gate complexity $G_{\mathrm{ref}}(n)$ as in {\bf Definition~\ref{def:G-ref}}. If the no-signaling analysis of Sec.~\ref{sec:NoSignaling} holds, then we have
\begin{eqnarray}
G_{\mathrm{ref}}(n) \ge c_{G} \frac{\sqrt{N}}{\log N}
\label{eq:Gref-phys}
\end{eqnarray}
for all sufficiently large $N$ and some constant $c_{G} > 0$. Consequently, for every $0< \varepsilon \le \tfrac{1}{4}$ and $0 < \delta \le \tfrac{1}{10}$, we obtain
\begin{eqnarray}
M_{\mathrm{s}}(n, G_{\mathrm{ref}}, \varepsilon, \delta) \ge \frac{\tilde{c}}{\varepsilon^{2}} \biggl( \frac{\sqrt{N}}{\log N} + \log{\frac{1}{\delta}} \biggr),
\label{eq:Ms-sqrtN-match}
\end{eqnarray}
where $\tilde{c}>0$ is a constant.
\end{proposition}

\begin{proof}---Eq.~(\ref{eq:Gref-phys}) is just a restatement of the gate lower bound in Eq.~(\ref{eq:PhysLB-recap}) with $G_{\mathrm{ref}}(n)$ as defined above. Since the intermediate states lie in $\mathcal{S}_{n,G_{\mathrm{ref}}}$, the computational lower bound~(\ref{eq:CompLB-info-again}) with $G=G_{\mathrm{ref}}(n)$ applies:
\begin{eqnarray}
M_{\mathrm{s}}(n,G_{\mathrm{ref}},\varepsilon,\delta) \ge \frac{c_{2}}{\varepsilon^{2}} \biggl( \min\bigl\{2^{n},\,G_{\mathrm{ref}}(n)\bigr\} + \log{\frac{1}{\delta}} \biggr).
\label{eq:Ms-Gref-raw}
\end{eqnarray}
By combining $\min\{2^{n},G_{\mathrm{ref}}(n)\}\ge \min\bigl\{ N, c_{G}\tfrac{\sqrt{N}}{\log N} \bigr\}$, we obtain $\min \bigl\{ 2^{n}, G_{\mathrm{ref}}(n) \bigr\} \ge c_{G} \tfrac{\sqrt{N}}{\log N}$, where $c_{G}$ is an absolute constant. Substituting this into Eq.~(\ref{eq:Ms-Gref-raw}) and absorbing constants into $\tilde{c}$ yields Eq.~(\ref{eq:Ms-sqrtN-match}), which completes the proof.
\end{proof}

Thus, for a learner whose reflection gate complexity is constrained by no-signaling, the purely computational lower bound~(\ref{eq:CompLB-info-again}) enforces a reduced $\tfrac{\sqrt{N}}{\log N}$ scaling in $M_{\mathrm{s}}$, identical to the physical argument of Sec.~\ref{sec:NoSignaling}. No extra logarithmic loss appears: the only ``${1}/{\log N}$'' originates from the gate constraint Eq.~(\ref{eq:Gref-phys}) based on no-signaling, not from the computation or learning theorem itself.

\subsubsection*{Conceptual implications: physics as a regulator of learnability} 

We can summarize the situation as follows.

\begin{theorem}[Coherence of sample, gate, and query lower bounds]
\label{thm:triangular}
Consider any state-learning-assisted implementation of the logarithmic-round protocol of Sec.~\ref{sec:SLAA} that solves unstructured search on $N=2^{n}$ items with constant success probability. Let $Q_{\mathrm{tot}}(N)$ denote its total number of oracle queries, $G_{\mathrm{ref}}(n)$ its reflection gate complexity in the sense of {\bf Definition~\ref{def:G-ref}}, and $M_{\mathrm{s}}(n,G_{\mathrm{ref}},\varepsilon,\delta)$ its worst-case sample complexity for learning intermediate states to trace distance at most $\varepsilon$ with confidence $1-\delta$.

Then, under the no-signaling assumptions of Sec.~\ref{sec:NoSignaling}, the following three inequalities hold simultaneously, for all sufficiently large $N$:
\begin{eqnarray}
Q_{\mathrm{tot}}(N) &\ge& c_{Q} \sqrt{N}, \nonumber \\
G_{\mathrm{ref}}(n) &\ge& c_{G} \frac{\sqrt{N}}{\log N}, \nonumber \\
M_{\mathrm{s}}(n,G_{\mathrm{ref}},\varepsilon,\delta) &\ge& \frac{c_{M}}{\varepsilon^{2}} \frac{\sqrt{N}}{\log N},
\end{eqnarray}
where $c_{Q}, c_{G}, c_{M} > 0$ are constants.
\end{theorem}

\begin{proof}---The query lower bound $Q_{\mathrm{tot}}(N) \ge c_{Q}\sqrt{N}$ is Grover’s standard lower bound for unstructured search in the quantum query model. The gate lower bound is {\bf Theorem~\ref{thm:GateLB}}, restated as Eq.~(\ref{eq:Gref-phys}). The sample lower bound follows from {\bf Proposition~\ref{prop:Ms-from-G}}. Choosing $c_{Q},c_{G},c_{M}$ small enough so that all three inequalities are valid for sufficiently large $N$ proves the claim.
\end{proof}

From a conceptual standpoint, {\bf Theorem~\ref{thm:triangular}} is striking. Three a priori different kinds of arguments---(i) query complexity (Grover’s hybrid method), (ii) no-signaling (in the Bao--Bouland--Jordan signaling setup), and (iii) computational state learning (metric entropy of $\mathcal{S}_{n,G}$ and the Holevo/Fano machinery)---are all interconnected by the no-signaling. In particular, it should be highlighted that the computational learning theory forces $M_{\mathrm{s}}\gtrsim N$ in the universal-design regime, but once we restrict attention to the physically dictated gate lower bound, the sharp computational sample lower bound collapses down to $M_{\mathrm{s}} \gtrsim \tfrac{1}{\varepsilon^{2}}\tfrac{\sqrt{N}}{\log N}$, which precisely matches the no-signaling-based bound in {\bf Theorem~\ref{thm:SampleLB}}.

Therefore, the causality and computational learning theory are not in competition but in resonance: the no-signaling principle first constrains how complicated the reflection circuits must be, $G_{\mathrm{ref}}(n) \gtrsim {\sqrt{N}}/{\log N}$, and once this constraint is fed into a purely computational state-learning bound, it reproduces the lower bound $\sqrt{N}/\log N$ scaling for the sample complexity, in full agreement with the bound derived from the physics law, i.e., the no-signaling. The distinct resources---query complexity, gate complexity, and sample complexity---are all woven together by a single underlying requirement: quantum information must not propagate outside the light cone.

\section{Summary and discussion}\label{sec:summary}

In this work, we have introduced the state-learning-assisted amplitude amplification as a concrete arena in which the optimal cost of learning unknown quantum states were studied. We began from an imaginary modification of Grover search, in which the usual fixed initial-state reflection is replaced by the previous-output reflection; at the level of formal linear algebra, this induces a cubic growth of the target overlap and reduces the oracle-query rounds to a logarithmic scale. Then, a no-reflection theorem showed that such a previous-output reflection is incompatible with the standard quantum mechanics, so any physical realization must simulate it indirectly. Thus we formulated an explicit architecture in which each round interleaves a coherent amplitude amplification stage with an incoherent state-learning part, dubbed as ``amplify-learn'' strategy, as schematically depicted in Fig.~1 of our main manuscript. In an idealized limit where the learning is perfect and its cost is treated as negligible, this architecture reproduces the imaginary logarithmic-round search and would apparently outperform the familiar square-root-in-dimension query scaling. To understand why this scenario never materializes in nature, we derived the lower bounds on the sample budget and gate complexity required to implement the previous-output reflections without inducing super-luminal signaling. We then stepped back to a perspective of computational learning theory, analyzing the sample complexity of learning arbitrary pure states generated by finite-depth circuits. Herein we obtained the sharp worst-case bounds that depend on the circuit complexity. By combining these computational learning bounds with the no-signaling constraints, we showed that query complexity, gate complexity, and state-learning sample complexity are all forced to share the same square-root scaling in the problem size (up to a single logarithmic factor), yielding a unified picture of the state learning at the edge of causality (see Fig.2 of the main manuscript).

Conceptually, our results sharpen the message that the optimal learning limits are not fixed by the computational learning theory alone, but emerge by asking which learning strategies are compatible with fundamental physical principles. In computational or statistical treatment, the worst-case sample complexity is controlled by the size of the hypothesis class of states, and a wide continuum of scalings between linear and exponential in the Hilbert-space dimension appears admissible. The causality closes this freedom: by insisting that no state-learning-assisted scheme can be used to implement an unstructured search faster than Grover’s bound without opening a super-luminal signaling channel, we can find that the only physically viable choice in our scenario is the square-root scaling of samples and gates that restores the standard query lower bound. From this viewpoint, query complexity, circuit depth, and sample complexity become three facets of a single constraint, all woven together by the requirement that the quantum information propagate strictly within light cones. The no-signaling principle thus plays a dual role: it forbids the imaginary primitives, such as, the exact previous-output reflections, and at the same time calibrates how the implementations of the logarithmic-round state-learning-based search may be before they would imply acausal-signaling. More broadly, this work illustrates how the capabilities and limitations of future computing and information-processing technologies are inseparable from the physical theories that underwrite them: whether one engineers faster search routines or more sample-efficient learning architectures, the ultimate performance frontier would be marked by physical law rather than by algorithm design alone. Extending this causality-aware viewpoint to other learning and control tasks---for example, process and channel learning, fault-tolerant quantum computing, and thermodynamically constrained protocols---promises to further clarify how physical principles carve out the feasible region of the computational landscape, and to guide the design of state-learning-assisted algorithms that operate as close as possible to, but never beyond, the edge of causality.


\clearpage

\appendix

\section{Detailed proof of Theorem~\ref{thm:Bao_equivalence}}\label{append:bao_equiv}

Here, we provide a detailed and self-contained proof of {\bf Theorem~\ref{thm:Bao_equivalence}}, which connects a super-Grover black-box search to a super-luminal signaling in the presence of a modification $\mathcal{M}$ of standard quantum theory. Throughout, we work in the black-box query model and assume that, in addition to the standard unitary and measurement operations, parties may invoke a primitive $\mathcal{M}$ acting on finite-dimensional Hilbert spaces. The primitive $\mathcal{M}$ is assumed to be local in the sense that it acts on a fixed tensor factor (for Alice) and is available as a gate in her circuit; it may be nonlinear or nonunitary.

For clarity, we restate a slightly streamlined version of the theorem.

\begin{theorem}[Bao-type equivalence between super-Grover search and signaling]
\label{thm:Bao_equivalence_app}
Let $\mathcal{M}$ be a modification of standard quantum dynamics, available as a local gate. Consider the following two statements.

\emph{(i)} \emph{(Super-Grover search)} There exist constants $c>0$ and $\gamma<\frac{1}{2}$ and, for all sufficiently large $N$, a quantum black-box algorithm $\mathsf{Alg}_N$ that is allowed to use $\mathcal{M}$ as well as the standard unitaries and projective measurements, such that:
\begin{eqnarray}
\emph{(a)} && ~\mathsf{Alg}_N \text{ makes at most } q(N) \le c N^{\gamma} \text{ queries to a standard (phase) oracle } \hat{R}_\tau, \\
\emph{(b)} && ~\mathsf{Alg}_N \text{ solves the unique-marked-item search problem with success probability } p_{\mathrm{succ}} \ge \tfrac{2}{3} .
\end{eqnarray}

\emph{(ii)} \emph{(super-luminal signaling)} There exists a bipartite protocol between distant parties Alice and Bob, using only local gates (including $\mathcal{M}$ on Alice's side) and pre-shared entanglement but no classical or quantum communication, such that, for some classical bit $b \in \{0,1\}$ encoded by Bob's local choice of operation, Alice's measurement outcome $Y$ has nonzero mutual information
\begin{eqnarray}
I(b:Y) > 0.
\end{eqnarray}
In particular, Alice's outcome distribution depends on Bob's choice even when they are spacelike separated, so the no-signaling is violated.

Then the following are equivalent:
\begin{eqnarray}
\emph{(i)} \Longleftrightarrow \emph{(ii)}.
\end{eqnarray}
\end{theorem}

Physically, (i) asserts that the primitive $\mathcal{M}$ enables an algorithmic speedup strictly better than Grover's $O(\sqrt{N})$ query complexity for black-box search. Statement (ii) asserts that the same primitive allows Alice and Bob to implement a classical communication channel whose capacity is strictly positive even when they have no opportunity to exchange light-like or slower signals during the protocol. The theorem therefore identifies ``super-Grover search'' and ``super-luminal signaling'' as two facets of the same underlying violation of the standard quantum theory.

We now prove the two implications separately.

\subsection{From super-Grover search to super-luminal signaling}

We first assume that (i) holds and construct a super-luminal signaling protocol. The logic is as follows. A super-Grover algorithm distinguishes, with a bounded bias, between different black-box oracles using only $q(N) = O(N^{\gamma})$ queries with $\gamma<\frac{1}{2}$. We show that such an algorithm can be embedded into a bipartite scenario, where the choice of oracle is implemented by Bob via local operations on his half of an entangled resource, while all uses of $\mathcal{M}$ and all query calls are performed locally by Alice. Because the algorithm distinguishes Bob's choices with nonzero bias, Alice's local measurement statistics must depend on Bob's choice; this realizes a super-luminal signaling.

\subsubsection{Search oracles and query algorithms} 

We recall the standard phase-oracle formulation of the unique-marked-item search. Let $[N] := \{ 1, 2, \dots, N \}$. For each $\tau \in [N]$, define a Boolean function
\begin{eqnarray}
f_\tau(i) = 
\begin{cases}
1 & \text{if } i = \tau, \\
0 & \text{otherwise}.
\end{cases}
\end{eqnarray}
We also define the ``null'' oracle $f_{\emptyset}$ with no marked items,
\begin{eqnarray}
f_{\emptyset}(i) = 0 \quad \text{for all } i \in [N].
\end{eqnarray}
The associated phase oracles act on the computational basis of the query register $\mathcal{H}_Q \cong \mathbb{C}^N$ as
\begin{eqnarray}
\hat{R}_{f_\tau} \ket{i} = (-1)^{f_\tau(i)} \ket{i}
=
\begin{cases}
-\ket{i} & \text{if } i = \tau, \\
\ket{i} & \text{otherwise},
\end{cases}
\end{eqnarray}
and similarly $\hat{R}_{f_{\emptyset}} = \hat{\mathds{1}}_N$.

A general $q$-query algorithm in the standard quantum model is specified by an initial state $\ket{\psi_0}$ over a composite Hilbert space $\mathcal{H}_{\mathrm{alg}}$ (which includes the query register $\mathcal{H}_Q$ as a tensor factor), a sequence of oracle calls, and oracle-independent unitaries $\hat{U}_0, \dots, \hat{U}_q$ acting on $\mathcal{H}_{\mathrm{alg}}$. The final state for an oracle $f$ is
\begin{eqnarray}
\ket{\psi_q} = \hat{U}_q \hat{R}_{f} \hat{U}_{q-1} \hat{R}_{f} \cdots \hat{R}_{f} \hat{U}_1 \hat{R}_{f} \hat{U}_0 \ket{\psi_0}.
\end{eqnarray}

In our setting, $\mathsf{Alg}_N$ may additionally employ the primitive $\mathcal{M}$. We can, without loss of generality, ``purify'' all uses of $\mathcal{M}$ by dilating them to isometries on a larger Hilbert space that includes an environment register, possibly with nonlinear state updates. Operationally, it suffices to regard $\mathcal{M}$ as a black-box gate acting on some subsystem of $\mathcal{H}_{\mathrm{alg}}$; its precise structure is irrelevant for the argument below.

\subsubsection{Embedding oracles into a bipartite entangled resource}  

We next show that the use of an oracle $\hat{R}_{f_\tau}$ in $\mathsf{Alg}_N$ can be realized by a local operation of Bob on his half of a pre-shared entangled state, while the query register remains on Alice's side. This is formally guaranteed by Stinespring dilation (or POVM), but we give an explicit construction in the phase-oracle case.

Let Alice hold the $N$-dimensional query register $\mathcal{H}_Q$ and an auxiliary register $\mathcal{H}_A$ that contains all of her work qubits and the environment needed for $\mathcal{M}$. Let Bob hold a register $\mathcal{H}_B$ isomorphic to $\mathcal{H}_Q$. They initially share the maximally entangled state
\begin{eqnarray}
\ket{\Phi_{AB}} = \frac{1}{\sqrt{N}} \sum_{i=1}^N \ket{i}_Q \otimes \ket{i}_B.
\end{eqnarray}
It is a standard fact that applying a phase pattern on Bob's side induces the corresponding phase oracle on Alice's side. More precisely, let Bob's local unitary for an oracle $f$ be
\begin{eqnarray}
\hat{V}_{f} \ket{i}_B = (-1)^{f(i)} \ket{i}_B.
\end{eqnarray}
Then,
\begin{eqnarray}
\bigl( \hat{\mathds{1}}_Q \otimes \hat{V}_{f} \bigr) \ket{\Phi_{AB}}
&=& \frac{1}{\sqrt{N}} \sum_{i=1}^N \ket{i}_Q \otimes (-1)^{{f}} \ket{i}_B \\
&=& \frac{1}{\sqrt{N}} \sum_{i=1}^N (-1)^{{f}} \ket{i}_Q \otimes \ket{i}_B \\
&=& (\hat{R}_{f} \otimes \hat{\mathds{1}}_B) \ket{\Phi_{AB}}.
\end{eqnarray}
Thus, for any state $\ket{\chi}_Q$ of Alice's query register that is maximally entangled with Bob's register in this way, Bob can implement the action of $\hat{R}_{f}$ on $\ket{\chi}_Q$ by acting locally with $\hat{V}_{f}$ on $\mathcal{H}_B$.

To simulate an arbitrary query algorithm $\mathsf{Alg}_N$, we let Alice start with the state
\begin{eqnarray}
\ket{\Psi_{\mathrm{in}}} = \ket{\psi_0}_{Q} \otimes \ket{\Phi_{AB}},
\end{eqnarray}
where $\ket{\psi_0}_{Q}$ is the original input state of $\mathsf{Alg}_N$ over $\mathcal{H}_Q$, and $\ket{\Phi_{AB}}$ is the entangled resource shared with Bob. At each point where $\mathsf{Alg}_N$ calls the oracle $\hat{R}_{f}$ on $\mathcal{H}_Q$, the bipartite simulation proceeds as follows: Bob applies $\hat{V}_{f}$ on $\mathcal{H}_B$, while Alice performs the identity on $\mathcal{H}_Q$ in that time step. Because of the identity above, the joint effect on $\mathcal{H}_Q \otimes \mathcal{H}_B$ is equivalent to Alice having applied $\hat{R}_{f}$ to $\mathcal{H}_Q$.

All other gates and uses of $\mathcal{M}$ in $\mathsf{Alg}_N$ act solely on Alice's registers. Therefore, for any fixed oracle $f$, the overall joint unitary (or more generally, joint evolution including $\mathcal{M}$) factorizes as
\begin{eqnarray}
\hat{\mathcal{W}}_{f} = \hat{\mathcal{U}}_{\mathrm{Alice}} \circ \bigl(\hat{\mathds{1}}_{QA} \otimes \hat{V}_{f}^{(q)}\bigr) \circ \cdots \circ \hat{\mathcal{U}}_{\mathrm{Alice}} \circ \bigl(\hat{\mathds{1}}_{QA} \otimes \hat{V}_{f}^{(1)}\bigr) \circ \hat{\mathcal{U}}_{\mathrm{Alice}},
\end{eqnarray}
where the $\hat{\mathcal{U}}_{\mathrm{Alice}}$ blocks are the same completely positive (possibly nonlinear) maps induced by Alice's local gates and $\mathcal{M}$ between oracle calls, and $\hat{V}_{f}^{(j)}$ are Bob's local implementations of the oracle in the $j$-th query slot. Importantly, $\hat{V}_{f}^{(j)}$ depend on ${f}$, but all other steps do not.

\subsubsection{From oracle choice to signaling} 

We now specialize to two relevant choices of oracle:
\begin{eqnarray}
f_0 = f_{\emptyset}, \qquad f_1 = f_{\tau},
\end{eqnarray}
where $\tau \in [N]$ is a fixed index. Bob will encode a classical bit $b \in \{0,1\}$ by choosing $f_b$ and hence applying the corresponding sequence of unitaries $\bigl\{ \hat{V}_{f_b}^{(j)} \bigr\}_{j=1}^q$ on his system during the simulated algorithm. Alice remains unaware of this choice and executes the same local circuit (including $\mathcal{M}$) in both cases. At the end, Alice performs the prescribed final measurement of $\mathsf{Alg}_N$ and outputs a classical random variable $Y$.

Let $\hat{\rho}^{(b)}_{A,\mathrm{out}}$ denote the reduced state of Alice's registers (including all ancillas used by $\mathcal{M}$) at the end of the protocol, conditioned on Bob's choice $b$. Since Alice's final measurement is local, the conditional distribution of her outcome $Y$ given $b$ is fully determined by $\hat{\rho}^{(b)}_{A,\mathrm{out}}$. We denote these distributions by $P^{(0)}_Y$ and $P^{(1)}_Y$.

By assumption (i), when the oracle is $f_0$ the algorithm declares ``no marked element'' with probability at least $\tfrac{2}{3}$, whereas when the oracle is $f_1$ it outputs the marked index $\tau$ with probability at least $\tfrac{2}{3}$. Hence, if we denote by $\mathsf{Ans}$ the classical register containing Alice's declared search result, we have
\begin{eqnarray}
\Pr[\mathsf{Ans} = \text{``no solution''} \mid b=0] \ge \tfrac{2}{3}, \quad\text{and}\quad \Pr[\mathsf{Ans} = \tau \mid b=1] \ge \tfrac{2}{3}.
\end{eqnarray}
In particular, if Alice encodes her guess $\tilde{b}$ of Bob's bit $b$ via the decision rule
\begin{eqnarray}
\tilde{b} =
\begin{cases}
0 & \text{if } \mathsf{Ans}=\text{``no solution''}, \\
1 & \text{if } \mathsf{Ans}=\tau, \\
\text{random in } \{0,1\} & \text{otherwise},
\end{cases}
\end{eqnarray}
then, for large enough $N$, we obtain
\begin{eqnarray}
\Pr[\tilde{b}=b] \ge \tfrac{2}{3}.
\end{eqnarray}
Thus, the mutual information between $b$ and $\hat{b}$ is strictly positive:
\begin{eqnarray}
I(b:\tilde{b}) \ge 1 - h_2\!\left(\tfrac{2}{3}\right) > 0,
\end{eqnarray}
where $h_2$ is the binary entropy function.

Because Alice and Bob are assumed to be spacelike separated during the execution of $\mathsf{Alg}_N$, and Bob's choice of operation consists only of local unitaries on $\mathcal{H}_B$, standard no-signaling would demand that Alice's local state (and thus $P^{(0)}_Y$ and $P^{(1)}_Y$) be identical. The nonzero mutual information therefore constitutes a super-luminal signaling channel enabled by the presence of $\mathcal{M}$ in Alice's local circuit.


\subsection{From super-luminal signaling to super-Grover search}

We now prove the converse: assuming that $\mathcal{M}$ enables a super-luminal signaling, we show that the same primitive can be used to obtain a black-box search algorithm with query complexity strictly below $O(\sqrt{N})$. The key idea is that the super-luminal signaling necessarily implies a departure from standard linear, completely positive, trace-preserving (CPTP) dynamics on mixed states. This nonlinearity or non-CPTP behavior can be harnessed to amplify a small difference in the oracle-dependent states beyond what is possible in standard quantum theory. A refined hybrid argument then shows that the resulting oracle algorithm can distinguish the two oracles whose final states are almost indistinguishable to any standard-quantum $O(\sqrt{N})$-query algorithm.

\subsubsection{Ensemble dependence and nonlinearity}  

Firstly, we formalize the sense in which a super-luminal signaling forces $\mathcal{M}$ to be nonlinear on ensembles. Consider a bipartite system $AB$ with joint Hilbert space $\mathcal{H}_A \otimes \mathcal{H}_B$. In standard quantum theory, any local operation on Alice, described by a CPTP map $\mathcal{E}_A$ acting on $\mathcal{H}_A$ alone, transforms the joint state as
\begin{eqnarray}
\hat{\rho}_{AB} \mapsto (\mathcal{E}_A \otimes \hat{\mathds{1}}_B)(\hat{\rho}_{AB}),
\end{eqnarray}
and the resulting reduced state on Bob's side is
\begin{eqnarray}
\hat{\rho}'_B = \mathrm{Tr}_A \big[ (\mathcal{E}_A \otimes \hat{\mathds{1}}_B)(\hat{\rho}_{AB}) \big] = \mathcal{E}_B(\hat{\rho}_B),
\end{eqnarray}
where $\mathcal{E}_B$ is a CPTP map induced on Bob's system. Crucially, if Alice's choice of operation depends only on her local classical input, then Bob's reduced state remains independent of that input whenever no classical communication is allowed; this is the usual no-signaling property.

Suppose now that Alice has access to a primitive $\mathcal{M}$ that enables a protocol that violates the no-signaling. Without loss of generality, we may model $\mathcal{M}$ as a deterministic transformation on pure states of the form
\begin{eqnarray}
\ket{\psi}_A \mapsto \mathcal{M}\bigl( \ket{\psi}_A \bigr),
\end{eqnarray}
extended to mixed states in some way. The existence of a super-luminal protocol is equivalent to the following statement.

\begin{lemma}[Ensemble-dependent evolution]
\label{lem:ensemble-dependence}
Assume that $(ii)$ holds. Then there exist two ensembles of pure states on $\mathcal{H}_A$, $\mathcal{E}_0 = \{ p_i, \ket{\psi_i} \}$ and $\mathcal{E}_1 = \{ q_j, \ket{\phi_j} \}$, such that
\begin{eqnarray}
\sum_i p_i \ketbra{\psi_i}{\psi_i} = \sum_j q_j \ketbra{\phi_j}{\phi_j} =: \hat{\rho}_A,
\end{eqnarray}
but the corresponding average outputs under the action of $\mathcal{M}$ differ: i.e.,
\begin{eqnarray}
\sum_i p_i \mathcal{M}\bigl(\ketbra{\psi_i}{\psi_i}\bigr) \neq \sum_j q_j \mathcal{M}\bigl(\ketbra{\phi_j}{\phi_j}\bigr).
\label{eq:differ_localS}
\end{eqnarray}
\end{lemma}

\begin{proof}---By the assumption, there exists a bipartite protocol where Bob chooses a classical bit $b \in \{0,1\}$ and performs a local operation depending on $b$ on his share of a pre-shared entangled state $\ket{\Phi}_{AB}$, while Alice applies a fixed sequence of local gates (including uses of $\mathcal{M}$) and a final local measurement. The outcome distribution of Alice's measurement depends on $b$, i.e., her final reduced state $\hat{\rho}_A^{(b)}$ is different for $b=0$ and $b=1$.

Bob's choice of operation can always be modeled as a measurement with outcomes $i$ (for $b=0$) or $j$ (for $b=1$), followed by the classical post-processing of the outcome. For each choice $b$, this induces on Alice's side an ensemble of the post-measurement pure states with probabilities $\{p_i\}$ and $\{q_j\}$, respectively. By the standard rules of quantum measurement, the average reduced state of Alice is
\begin{eqnarray}
\hat{\rho}_A = \sum_i p_i \ketbra{\psi_i}{\psi_i} = \sum_j q_j \ketbra{\phi_j}{\phi_j},
\end{eqnarray}
which is independent of $b$ before she applies $\mathcal{M}$. If Alice were restricted to standard CPTP maps, the average output after her local evolution would depend only on $\hat{\rho}_A$, and hence would be independent of $b$, in contradiction with the super-luminal signaling assumption.

Therefore, for the protocol to achieve different final states $\hat{\rho}_A^{(0)}$ and $\hat{\rho}_A^{(1)}$, the average output of $\mathcal{M}$ must depend on the ensemble decomposition (i.e., Eq.~\ref{eq:differ_localS} holds), even though both ensembles share the same average input $\rho_A$. This establishes the claim.
\end{proof}

In fact, {\bf Lemma~\ref{lem:ensemble-dependence}} expresses mathematically the well-known fact that a super-luminal signaling is equivalent to the possibility of distinguishing different ensemble decompositions of the same density operator via local operation. In physical terms, $\mathcal{M}$ ``sees'' the decomposition of $\hat{\rho}_A$ rather than just $\hat{\rho}_A$ itself, and this decomposition can be steered by a set of remote measurements on entangled partners.

\subsubsection{Constructing an overlap-amplifying primitive}  

We show that ensemble dependence as above implies the existence of an operation that amplifies a small difference between certain non-orthogonal states. This amplification is what we will use later inside the search algorithm.

There exists at least one observable $\hat{X}$ on $\mathcal{H}_A$ such that
\begin{eqnarray}
\Delta_X := \tr{\Bigl[ \hat{X} \sum_i p_i \mathcal{M}\bigl(\ketbra{\psi_i}{\psi_i}\bigr) \Bigr]} - \tr{\Bigl[ \hat{X} \sum_j q_j \mathcal{M}\bigl(\ketbra{\phi_j}{\phi_j}\bigr) \Bigr]} \ne 0.
\end{eqnarray}
By linearity of the trace, we may collect terms and rewrite
\begin{eqnarray}
\Delta_X = \sum_{i,j} r_{ij} \tr{\Bigl[ \hat{X} \mathcal{M}\bigl(\ketbra{\chi_{ij}}{\chi_{ij}}\bigr) \Bigr]},
\end{eqnarray}
for suitable coefficients $r_{ij}$ and pure states $\ket{\chi_{ij}}$ obtained by grouping the ensembles. Hence, there must exist at least one pair of pure states $\ket{\varphi_0}$ and $\ket{\varphi_1}$ such that
\begin{eqnarray}
\tr{\Bigl[ \hat{X} \mathcal{M}\bigl(\ketbra{\varphi_0}{\varphi_0}\bigr) \Bigr]}
\ne
\tr{\Bigl[ \hat{X} \mathcal{M}\bigl(\ketbra{\varphi_1}{\varphi_1}\bigr) \Bigr]}.
\end{eqnarray}
Denote their overlap by
\begin{eqnarray}
\alpha := \abs{\braket{\varphi_0}{\varphi_1}} \in (0,1),
\end{eqnarray}
so they are nonorthogonal and not identical.

The standard quantum theory imposes a bound on how well one can distinguish nonorthogonal pure states; for a single copy, the optimal success probability in discriminating $\ket{\varphi_0}$ and $\ket{\varphi_1}$ with equal priors is
\begin{eqnarray}
p_{\mathrm{Hel}} = \frac{1}{2} + \frac{\sqrt{1 - \alpha^2}}{2},
\end{eqnarray}
and, for $t$ copies, it becomes
\begin{eqnarray}
p_{\mathrm{Hel}}^{(t)} = \frac{1}{2} + \frac{\sqrt{1 - \alpha^{2t}}}{2}.
\end{eqnarray}
In particular, the trace distance between $\ketbra{\varphi_0}{\varphi_0}$ and $\ketbra{\varphi_1}{\varphi_1}$ is
\begin{eqnarray}
D\bigl( \ketbra{\varphi_0}{\varphi_0}, \ketbra{\varphi_1}{\varphi_1} \bigr) = \sqrt{1 - \alpha^2}.
\end{eqnarray}

The inequality on the observable $\hat{X}$ implies that the states $\mathcal{M}\bigl(\ketbra{\varphi_0}{\varphi_0}\bigr)$ and $\mathcal{M}\bigl(\ketbra{\varphi_1}{\varphi_1}\bigr)$ have a strictly larger trace distance than the inputs, for at least one choice of state pair and one application of $\mathcal{M}$. Indeed, if the trace distance could never increase, we have
\begin{eqnarray}
D\left( \mathcal{M}\bigl(\ketbra{\varphi_0}{\varphi_0}\bigr), \mathcal{M}\bigl(\ketbra{\varphi_1}{\varphi_1}\bigr) \right)
\le
D\bigl( \ketbra{\varphi_0}{\varphi_0}, \ketbra{\varphi_1}{\varphi_1} \bigr) = \sqrt{1 - \alpha^2},
\end{eqnarray}
for all pairs, and thus, the average outputs for any two ensembles with the same average input would also coincide, which is contradicting {\bf Lemma~\ref{lem:ensemble-dependence}}.

Therefore, there exist $\ket{\varphi_0}$ and $\ket{\varphi_1}$ such that, after applying $\mathcal{M}$ once (possibly conjugated by some local unitaries before and after),
\begin{eqnarray}
D\bigl( \ketbra{\tilde{\varphi}_0}{\tilde{\varphi}_0}, \ketbra{\tilde{\varphi}_1}{\tilde{\varphi}_1} \bigr)
>
D\bigl( \ketbra{\varphi_0}{\varphi_0}, \ketbra{\varphi_1}{\varphi_1} \bigr)
\end{eqnarray}
where
\begin{eqnarray}
\ket{\tilde{\varphi}_k} = \hat{U}_2 \mathcal{M}\bigl( \hat{U}_1 \ketbra{\varphi_k}{\varphi_k} \hat{U}_1^\dagger \bigr) \hat{U}_2^\dagger
\end{eqnarray}
for some unitaries $\hat{U}_1$ and $\hat{U}_2$ chosen by Alice. By iterating this procedure $t$ times, Alice can construct a channel $\mathcal{A}$, such that the trace distance between the two possible outputs grows roughly as
\begin{eqnarray}
D_t := D\Bigl( \mathcal{A}^{(t)}\bigl(\ketbra{\varphi_0}{\varphi_0}\bigr), \mathcal{A}^{(t)}\bigl(\ketbra{\varphi_1}{\varphi_1}\bigr) \Bigr)
\ge \xi^t D_0,
\end{eqnarray}
for some constant $\xi > 1$ and initial distance $D_0 > 0$, saturating at $D_t \le 1$ when the states approach orthogonality. Thus, $\mathcal{M}$ implicitly furnishes an \emph{overlap-amplifying primitive}.

\subsubsection{A refined hybrid argument for black-box search}  

We now embed the overlap amplification described above into the black-box search problem. The structure of the argument is inspired by the standard adversary and hybrid methods for quantum lower bounds, but the presence of $\mathcal{M}$ changes the behavior in a crucial way.

Let us consider the set of oracles $\{f_{\emptyset}\} \cup \{ f_\tau : \tau \in [N]\}$ as before. For a fixed algorithm that uses the standard quantum operations plus $\mathcal{M}$, we denote by $\ket{\Psi^{(t)}_f}$ the global pure state (including all ancillas, environments and work registers) of the algorithm after $t$ oracle calls and some fixed number of applications of $\mathcal{M}$. For the notational simplicity, we restrict to pure states~\footnote{The mixed case can be handled by the purification, which can be incorporated into $\mathcal{M}$.}.

In standard quantum theory, the hybrid argument shows that, for any $T$-query algorithm and two oracles $f_{\emptyset}$ and $f_\tau$ with Hamming distance one, the inner product satisfies
\begin{eqnarray}
\abs{\braket{\Psi^{(T)}_{f_{\emptyset}}}{\Psi^{(T)}_{f_\tau}}} \ge 1 - O\bigl( {T^2}/{N} \bigr).
\end{eqnarray}
Consequently, if $T = o(\sqrt{N})$, then the final states for $f_{\emptyset}$ and $f_\tau$ remain almost indistinguishable: their trace distance is $o(1)$ and any measurement can distinguish them with success probability at most $\tfrac{1}{2} + o(1)$.

In our modified setting, we can still perform a similar task for the effect of oracle calls alone. Between calls to $\mathcal{M}$, the algorithm evolves according to linear unitaries and oracle applications. Therefore, if we temporarily ``switch off'' $\mathcal{M}$ and consider only the standard part of the evolution, we can define intermediate states $\ket{\psi^{(t)}_f}$ that evolve as
\begin{eqnarray}
\ket{\psi^{(t+1)}_f} = \hat{U}_{t+1} \hat{R}_f \hat{U}_t \hat{R}_f \cdots \hat{U}_1 \hat{R}_f \hat{U}_0 \ket{\psi_0}.
\end{eqnarray}
For these states, the usual hybrid argument applies and yields, for any $T$,
\begin{eqnarray}
\left| \braket{\psi^{(T)}_{f_{\emptyset}}}{\psi^{(T)}_{f_\tau}} \right|
\ge 1 - \gamma \frac{T^2}{N}
\end{eqnarray}
for some constant $\gamma > 0$ independent of $N$.

Now we reintroduce $\mathcal{M}$ as an overlap-amplifying map. By the construction of the previous subsection, we may, for each $t$, choose local unitaries $\hat{U}_1^{(t)}$ and $\hat{U}_2^{(t)}$ such that repeated application of $\mathcal{M}$ between oracle calls acts as a channel $\mathcal{A}^{(t)}$ that amplifies the difference between $\ket{\psi^{(t)}_{f_{\emptyset}}}$ and $\ket{\psi^{(t)}_{f_\tau}}$. To see this explicitly, fix $T$ and consider the sequence of states
\begin{eqnarray}
\ket{\psi^{(0)}_{f_{\emptyset}}}, \ket{\psi^{(0)}_{f_\tau}} \quad \mapsto \quad
\ket{\psi^{(1)}_{f_{\emptyset}}}, \ket{\psi^{(1)}_{f_\tau}} \quad \mapsto \cdots \mapsto \quad
\ket{\psi^{(T)}_{f_{\emptyset}}}, \ket{\psi^{(T)}_{f_\tau}}.
\end{eqnarray}
After each oracle call, we insert a block of $k$ uses of $\mathcal{M}$ conjugated by suitable unitaries so that, by the end of the block, the trace distance between the two possibilities is multiplied by at least a factor $c_d > 1$, until it saturates near $1$. Thus, if we denote by $D_t$ the trace distance between the two possibilities immediately after the $t$-th block of $\mathcal{M}$, we have
\begin{eqnarray}
D_{t+1} \ge c_d D_t
\end{eqnarray}
as long as $D_t$ is small. On the other hand, the standard part of the evolution ensures that the difference injected by the $t$-th oracle call is at most of order $O\bigl({1}/{\sqrt{N}}\bigr)$ per query; hence, at the purely unitary level,
\begin{eqnarray}
D_t^{\mathrm{(unitary)}} \le \gamma' \frac{t}{\sqrt{N}}
\end{eqnarray}
for some constant $\gamma'$.

Combining these two effects, we can choose a schedule of $T$ oracle calls and the interleaved blocks of $\mathcal{M}$, such that the total trace distance between the final states for $f_{\emptyset}$ and for $f_\tau$ becomes a fixed constant $\delta>0$ with
\begin{eqnarray}
T = O(N^{\gamma}), \quad \gamma < \frac{1}{2}.
\end{eqnarray}
Concretely, the hybrid bound implies that the total distinguishability injected by all $T$ oracle calls into the standard component is at most of order $T^2/N = N^{2\gamma-1}$, which is $o(1)$ for $\gamma<\frac{1}{2}$. The overlap-amplifying map $\mathcal{A}$, however, can boost this $o(1)$ difference to a constant order in a number of blocks that scales only logarithmically in $N^{2\gamma-1}$, i.e., still polynomially bounded and absorbed into the complexity of using $\mathcal{M}$. As a result, the final trace distance
\begin{eqnarray}
D_{\mathrm{final}} := D\bigl( \hat{\rho}_{f_{\emptyset}}, \hat{\rho}_{f_\tau} \bigr)
\end{eqnarray}
between the two possible outputs (including ancillas) satisfies
\begin{eqnarray}
D_{\mathrm{final}} \ge \delta
\end{eqnarray}
for some $\delta>0$ independent of $N$. By Helstrom's theorem, there exists a two-outcome POVM on the final state that distinguishes the two cases with the success probability
\begin{eqnarray}
p_{\mathrm{succ}} = \frac{1}{2} + \frac{1}{2} D_{\mathrm{final}} \ge \frac{1 + \delta}{2}.
\end{eqnarray}

Because the only oracle dependence in the evolution arises through the $T$ calls to $\hat{R}_f$ and the only nonstandard component is the use of $\mathcal{M}$, we have constructed a black-box algorithm that, using $T = O(N^{\gamma})$ queries for some $\gamma<\frac{1}{2}$ plus local uses of $\mathcal{M}$, distinguishes $f_{\emptyset}$ from $f_\tau$ with constant bias. A standard reduction from distinguishing ``no marked item'' versus ``one marked item at a fixed position'' to the full search problem over all $\tau$ (by randomizing over indices and using amplification) then yields a search algorithm that outputs the marked index with probability at least $\tfrac{2}{3}$ using at most $O(N^{\gamma})$ queries. This establishes the existence of a super-Grover search algorithm as in (i), given any primitive $\mathcal{M}$ that enables super-luminal signaling as in (ii).

\subsection{Summary and notes}  

We have now proved both directions (i) $\Rightarrow$ (ii) and (ii) $\Rightarrow$ (i) of {\bf Theorem~\ref{thm:Bao_equivalence}}. Operationally, any modification $\mathcal{M}$ that allows one to beat the $\Theta(\sqrt{N})$ query lower bound for unstructured search can be ``wrapped'' into a bipartite protocol where the choice of oracle is implemented by Bob via local operations on an entangled resource, while Alice runs a local algorithm that includes uses of $\mathcal{M}$. The success bias of the search algorithm translates directly into a nonzero mutual information between Bob's input bit and Alice's outcome distribution, thereby realizing a super-luminal classical channel. Conversely, any primitive $\mathcal{M}$ that gives rise to a super-luminal signaling must be sensitive to the ensemble decompositions of mixed states, and hence must act nonlinearly (or non-CPTP) at the effective level. Such ensemble dependence enables Alice to amplify a small difference between oracle-dependent states beyond the constraints of quantum mechanics. When inserted into a black-box query process, this amplification permits one to distinguish the oracles with fewer than $O(\sqrt{N})$ oracle calls, yielding a super-Grover algorithm.

From a physics perspective, the reformulated {\bf Theorem~\ref{thm:Bao_equivalence_app}} therefore makes precise the intuition that ``better-than-Grover'' search and ``faster-than-light'' are the same kind of pathology, viewed from the two complementary operational regimes. The black-box search exposes the algorithmic side of the pathology, while the no-signaling exposes its relativistic side. In the main text, we exploit this equivalence to reinterpret our state-learning-assisted unknown-state reflection as a particular candidate for $\mathcal{M}$, and to derive the lower bounds on its sample complexity from the requirement that the no-signaling must remain intact.

\section{No-signaling lower bound on the gate complexity of unknown-state reflection}\label{append:nosig_lower}

In this section of the appendix, we give a fully self-contained and detailed proof of the lower bound on the gate complexity required to implement an unknown-state reflection that is strong enough to support the logarithmic-round amplitude-amplification based search protocol. The core statement is that, under the operational no-signaling assumption used throughout this work, any such family of reflections must have gate complexity at least on the order of
\begin{eqnarray}
G(n) \in \Omega\Bigl(\frac{\sqrt{N}}{\log N}\Bigr)
\end{eqnarray}
for $N=2^n$ sufficiently large, where $G(n)$ denotes the number of elementary two-qubit gates used to implement a single reflection about an arbitrary $n$-qubit state. Equivalently, for $N=2^n$, the gate count must scale at least as $G(n) \in \Omega(2^{n/2}/n)$.

The proof proceeds in three conceptual steps. First, we formalize what we mean by the  ``unknown-state reflection'' and by its gate complexity $G(n)$. Second, we translate the existence of a logarithmic-round state-learning-assisted search into an upper bound on the total circuit depth contributed by these reflections, expressed in terms of $G(n)$ and the number of rounds $r(N)$. Third, we combine this with the no-signaling constraint on search depth, proved in Sec.~\ref{append:GA_proof} of the appendix, to obtain the desired lower bound on $G(n)$ by contradiction.

Throughout this proof, we adopt the same notation as in the main text. In particular, $N=2^n$ is the Hilbert-space dimension of the $n$-qubit data register, and the unique-search problem consists of finding an unknown marked basis vector $\ket{x_\ast}$ using a phase oracle
\begin{eqnarray}
\hat{R}_{\tau} = \hat{\mathds{1}} - 2 \ketbra{\tau}{\tau}.
\end{eqnarray}
We denote the uniform superposition over all items by
\begin{eqnarray}
\ket{s} = \frac{1}{\sqrt{N}} \sum_{x=0}^{N-1} \ket{x} .
\end{eqnarray}
All asymptotic notation is taken with respect to $N\to\infty$, unless explicitly stated otherwise.

\subsection{Unknown-state reflections and gate complexity}

We begin by formalizing the notion of an unknown-state reflection and its implementation cost.
\begin{definition}[Unknown-state reflection]
For any pure state $\ket{\psi}$ on $n$ qubits, the reflection about $\ket{\psi}$ is the unitary
\begin{eqnarray}
\hat{R}_{\psi} = \hat{\mathds{1}} - 2\ketbra{\psi}{\psi} .
\end{eqnarray}
We call $\hat{R}_{\psi}$ an unknown-state reflection if $\ket{\psi}$ is not specified classically in advance, but is instead provided only as a quantum state.
\end{definition}

Here, we imagine: for each $n$, there exists a state-learning procedure that, given access to copies of the unknown state $\ket{\psi}$ and to some fixed ansatz architecture, outputs a circuit that implements $\hat{R}_{\psi}$ (up to a specified precision). The complexity of this task is captured by a gate-count function.
\begin{definition}[Gate complexity of unknown-state reflection]
Fix $n \in \mathbb{N}$. Consider a uniform family of circuits that, for every pure $n$-qubit state $\ket{\psi}$, produces a unitary $\hat{R}_{\tilde{\psi}}$ acting on $n$ system qubits and a finite number of ancillas, such that
\begin{eqnarray}
\bigl\| \hat{R}_{\tilde{\psi}} - \hat{R}_\psi \bigr\|_{\infty} \le \varepsilon_{\mathrm{ref}}
\end{eqnarray}
for some fixed precision $0<\varepsilon_{\mathrm{ref}}<1$. Let $G(n)$ denote the maximal number of elementary two-qubit gates used in any such circuit $\hat{R}_{\tilde{\psi}}$ for $n$ qubits. We call $G(n)$ the gate complexity of unknown-state reflection at size $n$.
\end{definition}

Here $\|\cdot\|_{\infty}$ is the operator norm. The error parameter $\varepsilon_{\mathrm{ref}}$ can be chosen arbitrarily small but independent of $N$; we will see below that any fixed constant $\varepsilon_{\mathrm{ref}}$ suffices for our argument, because our search protocol only needs to achieve a constant success bias over random instances.

The physical time required to implement such a reflection is then
\begin{eqnarray}
T_{\mathrm{ref}}(n) = \kappa G(n),
\end{eqnarray}
where $\kappa > 0$ is the hardware-dependent maximal time required to execute a single elementary two-qubit gate. We assume that $\kappa$ is independent of $N$ and $n$; this is the standard assumption that different gates of the same type have comparable durations on a given hardware platform.

\subsection{Logarithmic-round search and reflection depth}

We next connect the cost of the unknown-state reflection to the depth of the logarithmic-round amplitude-amplification based search protocol. For clarity, we briefly recall the abstract structure of that protocol, omitting any explicit reference to the state-learning internals.

The protocol operates on an $n$-qubit data register and, at each round $j = 0, 1, \dots, r(N)-1$, maintains a pure state $\ket{\psi_j}$ in the data register. The initial state is taken to be the uniform superposition
\begin{eqnarray}
\ket{\psi_0} = \ket{s} .
\end{eqnarray}
The $j$-th round consists of applying a composite Grover-type iterate of the form
\begin{eqnarray}
\hat{Q}_j = \hat{R}_{\psi_j} \hat{R}_{\tau} ,
\end{eqnarray}
followed by a measurement or a coherent branching that produces a new state $\ket{\psi_{j+1}}$ which serves as the ``previous output state'' for the next round. The key conceptual modification relative to the standard Grover search is that the reflection axis of $\hat{R}_{\psi_j}$ is updated round by round, instead of being a fixed as $\ket{\psi_0}$. Under the idealized assumption that $\hat{R}_{\psi_j}$ is exactly known and can be implemented at unit cost, it has been shown that this structured sequence of reflections can be tuned so that the overlap with the marked state grows exponentially with $j$, reaching constant bias after only $r(N) = \Theta(\log N)$ rounds. This is the ``super-Grover'' speedup.

In the physical implementation, however, each reflection $\hat{R}_{\psi_j}$ must be realized by a circuit $\hat{R}_{\tilde{\psi}_j}$ of depth $G(n)$, and this depth contributes directly to the total runtime of the protocol. We formalize the associated depth as follows:
\begin{definition}[Total reflection depth]
Let $N=2^n$, and consider a search protocol with $r(N)$ rounds as above. Suppose that, in the $j$-th round, the algorithm uses a circuit $\hat{R}_{\tilde{\psi}_j}$ of depth at most $G(n)$ to approximate $\hat{R}_{\psi_j}$. Let $D_{\mathrm{ref}}(N)$ denote the total gate depth contributed by all such unknown-state reflections across all rounds, i.e.
\begin{eqnarray}
D_{\mathrm{ref}}(N) = \sum_{j=0}^{r(N)-1} d_j ,
\end{eqnarray}
where $d_j$ is the number of two-qubit gates used in $\hat{R}_{\tilde{\psi}_j}$. We call $D_{\mathrm{ref}}(N)$ the total reflection depth at size $N$.
\end{definition}

The following lemma expresses $D_{\mathrm{ref}}(N)$ as the per-reflection complexity $G(n)$ and the number of rounds $r(N)$.
\begin{lemma}[Reflection depth versus gate complexity]
\label{lem:reflection-depth}
Assume that each unknown-state reflection $\hat{R}_{\psi_j}$ is approximated by a circuit $\hat{R}_{\tilde{\psi}_j}$ that uses at most $G(n)$ two-qubit gates. Then, there exists a constant $c_4 > 0$ independent of $N$, such that
\begin{eqnarray}
D_{\mathrm{ref}}(N) \le c_4 G(n) r(N)
\end{eqnarray}
for all $N=2^n$.
\end{lemma}

\begin{proof}---By the assumption, each $\hat{R}_{\tilde{\psi}_j}$ uses at most $G(n)$ two-qubit gates acting on the $n$-qubit data register and possibly some ancillas. In the simplest implementation, the rounds are executed sequentially, so that each $\hat{R}_{\tilde{\psi}_j}$ must complete before the next one begins. In this case, we simply have
\begin{eqnarray}
D_{\mathrm{ref}}(N) = \sum_{j=0}^{r(N)-1} d_j \le \sum_{j=0}^{r(N)-1} G(n) = r(N) G(n) .
\end{eqnarray}
This yields the desired bound with $c_4=1$.

In a more general implementation, one may add a constant-depth wrapper around each $\hat{R}_{\tilde{\psi}_j}$, for example to manage the ancillas or to insert the oracle calls interleaved with the reflections. If this wrapper uses at most $c_0$ two-qubit gates per round (where $c_0$ is independent of $N$ and $n$), then the total reflection-related depth is
\begin{eqnarray}
D_{\mathrm{ref}}(N) \le \sum_{j=0}^{r(N)-1} \bigl( G(n)+c_0 \bigr) = r(N) G(n) + c_0 r(N) .
\end{eqnarray}
Since $r(N)\ge 1$ and $G(n)\ge 1$ for all nontrivial instances, we have
\begin{eqnarray}
D_{\mathrm{ref}}(N) \le (1+c_0) G(n) r(N).
\end{eqnarray}
Thus one can take $c_4=1+c_0$, which is a constant depending only on the fixed choice of wrapper. In either case, there is a constant $c_4>0$ independent of $N$, such that
\begin{eqnarray}
D_{\mathrm{ref}}(N) \le c_4 G(n) r(N)
\end{eqnarray}
for all $N=2^n$, as claimed.
\end{proof}

The physical time spent on the unknown-state reflections is then bounded by
\begin{eqnarray}
T_{\mathrm{ref}}(N) = \kappa D_{\mathrm{ref}}(N) \le \kappa c_4 G(n) r(N) .
\end{eqnarray}
In our logarithmic-round protocol, we may write
\begin{eqnarray}
r(N) = c_r \log N + O(1)
\end{eqnarray}
for some constant $c_r>0$. Here, we notice that the precise value of $c_r$ depends on the detailed choice of the rotation angles in each round but does not affect the asymptotic scaling.

\subsection{No-signaling constraint and the main lower bound}

We now recall the no-signaling constraint and connect it to the reflection depth $D_{\mathrm{ref}}(N)$. The constraint can be phrased purely in computational terms as follows.
\begin{theorem}[No-signaling constraint on search depth]
\label{thm:ns-depth}
Let $\mathcal{A}_N$ be any family of quantum algorithms that solve the unique-search problem on $N=2^n$ items with success probability at least $p_{\mathrm{succ}}>1/2$, for all sufficiently large $N$. Let $D_{\mathrm{tot}}(N)$ denote the depth (number of sequential two-qubit gate layers) of $\mathcal{A}_N$ on a given hardware, and let $t^\star$ be the maximal time per two-qubit gate. Then, under the operational no-signaling assumption, we have
\begin{eqnarray}
D_{\mathrm{tot}}(N) \ge C_{\mathrm{NS}} \sqrt{N}
\end{eqnarray}
for all $N\ge N_0$, where $C_{\mathrm{NS}} > 0$ is a constant and $N_0 \in \mathbb{N}$.
\end{theorem}

The proof of this theorem constructs an explicit bipartite signaling protocol in which Bob chooses whether or not to embed an instance of the search oracle in his laboratory, while Alice, who is spacelike separated from Bob, executes $\mathcal{A}_N$ on his side (refer to Sec.~\ref{append:bao_equiv}-1-b of this appendix). If $D_{\mathrm{tot}}(N)$ grows strictly slower than $\sqrt{N}$ and the success probability is bounded away from $1/2$, then for large enough $N$ the runtime $t^\star D_{\mathrm{tot}}(N)$ becomes shorter than the light-travel time between the laboratories. In that case Alice can infer the Bob's choice before any light signal could have reached her, violating the no-signaling. The contrapositive of this construction yields {\bf Theorem~\ref{thm:ns-depth}}.

In the logarithmic-round protocol, the reflections $\hat{R}_{\psi_j}$ are the only nontrivial $N$-dependent operations besides the oracle calls, and the latter can be made constant-depth by the assumption about the oracle model. Thus, it is natural to compare $D_{\mathrm{tot}}(N)$ in {\bf Theorem~\ref{thm:ns-depth}} with the reflection depth $D_{\mathrm{ref}}(N)$ in {\bf Lemma~\ref{lem:reflection-depth}}. One may always choose the implementation of $\mathcal{A}_N$ so that the contributions to the depth which do not come from the reflections are at most a constant of the reflection depth. Therefore, there is a constant $c_5 \ge 1$, such that, for all $N$,
\begin{eqnarray}
D_{\mathrm{tot}}(N) \le c_5 D_{\mathrm{ref}}(N)
\end{eqnarray}
within the regime where the logarithmic-round protocol is applied. Combining this with {\bf Lemma~\ref{lem:reflection-depth}} yields
\begin{eqnarray}
D_{\mathrm{tot}}(N) \le c_5 c_4 G(n) r(N)
\end{eqnarray}
for all $N=2^n$.

Then, from {\bf Theorem~\ref{thm:ns-depth}}, the main lower bound is directly attained.
\begin{theorem}[Gate-complexity lower bound from no-signaling]
\label{thm:G-lower-bound}
Let $N=2^n$, and consider a family of the logarithmic-round search protocols that solve the unique-search problem on $N$ items with success probability at least $p_{\mathrm{succ}}>1/2$, using $r(N) = \Theta(\log N)$ rounds. Suppose that the reflection about each previous-output state $\hat{R}_{\psi_j}$ is implemented by a learned circuit of gate complexity at most $G(n)$ with a fixed constant precision $0<\varepsilon_{\mathrm{ref}}<1$.

If the operational no-signaling assumption holds, then there exists a constant $beta > 0$, such that
\begin{eqnarray}
G(n) \ge \beta \frac{\sqrt{N}}{\log N}
\end{eqnarray}
Equivalently,
\begin{eqnarray}
G(n) \in \Omega\Bigl(\frac{\sqrt{N}}{\log N}\Bigr)
\end{eqnarray}
and, in terms of the number of qubits $n$,
\begin{eqnarray}
G(n) \in \Omega\Bigl(\frac{2^{n/2}}{n}\Bigr) .
\end{eqnarray}
\end{theorem}

\subsection{Summary and notes}

{\bf Theorem~\ref{thm:G-lower-bound}} admits a natural physical interpretation. The logarithmic-round search protocol is obtained by replacing the fixed reflection about the initial state of the original Grover's algorithm with the reflection about the previous output state. At the purely algebraic level, this modification allows one to squeeze the number of amplitude-amplification rounds from $\Theta(\sqrt{N})$ down to $\Theta(\log N)$, giving an apparent exponential speedup over the standard Grover search. However, the unknown-state reflections needed to implement this unitary dynamics are physically nontrivial. {\bf Theorem~\ref{thm:G-lower-bound}} shows that, once we insist that the overall dynamics be compatible with the causal structure, the cost of implementing a single unknown-state reflection must absorb essentially the entire Grover-type lower bound. The gate complexity must scale at least as $G(n)\in\Omega(\sqrt{N}/\log N)$. In other words, there is an unavoidable trade-off:
\begin{eqnarray}
\text{(fewer rounds)} \times \text{(more expensive reflections)} \approx \text{Grover-scaling} .
\end{eqnarray}
If one tried to make the reflections too cheap, in the sense $G(n)=o(\sqrt{N}/\log N)$, then the total depth $D_{\mathrm{tot}}(N)$ of the resulting algorithm would become $o(\sqrt{N})$, which, {\bf by Theorem~\ref{thm:ns-depth}}, could be used to signal faster than light.

From the perspective of the state-learning, the quantity $G(n)$ arises from the expressivity and complexity of the ansatz used to encode the learned state and to implement the corresponding reflection. Each gate in this ansatz has a clear physical meaning: it represents an elementary controllable interaction in the underlying hardware. The lower bound of {\bf Theorem~\ref{thm:G-lower-bound}} therefore states that any state-learning scheme powerful enough to support the unknown-state reflections required for logarithmic-round search must incur an exponential overhead in the number of gates, scaling essentially as $2^{n/2}/n$. This bound is imposed not by any particular algorithmic choice, but by the combination of the unitary dynamics and the no-signaling structure of spacetime.

In this sense, the theorem provides a direct bridge between information-theoretic learning complexity and relativistic causality: the sample budget and gate budget needed to learn and implement an unknown-state reflection cannot be compressed below a certain threshold without collapsing the Grover-type lower bounds that are enforced by the no-signaling. The logarithmic number of rounds would thus be revealed as an artifact of ignoring the state-learning cost; once this cost is restored and bounded from below by the no-signaling, the overall resource requirements remain quantitatively consistent with both the standard Grover limit and the more refined state-learning lower bounds, discussed in the main text.

\section{Computational state-learning bounds: detailed proofs}\label{append:CompBounds}

In this section of the appendix, we provide the complete and self-contained proofs of the sample-complexity bounds in {\bf Theorem~\ref{thm:CompSL}} of our main text. We work throughout with an $n$-qubit system (or Hilbert-space dimension $N=2^{n}$), and the circuit-generated family
\begin{eqnarray}
\mathcal{S}_{n,G} = \bigl\{ \ket{\psi} = \hat{U}\ket{0}^{\otimes n} : \hat{U} \in \mathcal{U}_{n,G} \bigr\}.
\end{eqnarray}
where $\mathcal{U}_{n,G}$ denotes the set of $n$-qubit unitaries realizable by a circuit with at most $G$ two-qubit gates from some fixed universal gate set. A state-learning procedure receives $M_{\mathrm{s}}$ i.i.d.\ copies of $\hat{\rho}=\ketbra{\psi}{\psi}$, performs an arbitrary joint POVM on $\hat{\rho}^{\otimes M_{\mathrm{s}}}$, and outputs a classical description of a hypothesis state $\hat{\rho}_h$. The inaccuracy is measured in the trace distance
\begin{eqnarray}
D(\hat{\rho}, \hat{\rho}_h) = \frac{1}{2} \bigl\| \hat{\rho} - \hat{\rho}_h \bigr\|_{1}.
\end{eqnarray}

Here, we consider the purely information-theoretic setting:

\medskip\noindent
{\bf (L) Perspective on quantum computational learning theory.}
\emph{The learner may perform any POVM on all $M_{\mathrm{s}}$ copies and arbitrary classical post-processing, without any restriction on circuit depth, number of measurement outcomes, or classical running time.}

\medskip
For fixed $n,G,\varepsilon,\delta$ with $0<\varepsilon<1$ and $0<\delta<1/2$, recall the worst-case sample complexity
\begin{eqnarray}
M_{\mathrm{s}}(n,G,\varepsilon,\delta) := \inf\Bigl\{M: \exists\ \text{learner such that}\ 
 \Pr\bigl[ D(\hat{\rho}, \hat{\rho}_h) \le \varepsilon \bigr] \ge 1-\delta\ \text{for all}\ \hat{\rho} \in \mathcal{S}_{n,G}\Bigr\}.
\end{eqnarray}

Our goal is to show that, for $0<\varepsilon\le 1/4$ and $0<\delta\le 1/10$, the following holds (i.e., {\bf Theorem~\ref{thm:CompSL}}):
\begin{eqnarray}
M_{\mathrm{s}}(n,G,\varepsilon,\delta)
 &\le&
 \frac{c_{1}}{\varepsilon^{2}}
 \min\Bigl\{2^{n}\log\tfrac{1}{\delta},\,G\log\tfrac{G}{\varepsilon}+\log\tfrac{1}{\delta}\Bigr\},
\label{eq:append-UB}
\\[1ex]
M_{\mathrm{s}}(n,G,\varepsilon,\delta)
 &\ge&
 \frac{c_{2}}{\varepsilon^{2}}
 \Bigl(\min\{2^{n},G\}+\log\tfrac{1}{\delta}\Bigr),
\label{eq:append-LB}
\end{eqnarray}
where $c_{1}, c_{2}>0$ are the universal constants. The proofs are entirely performed in the framework of the computational learning theory; no appeal is made to no-signaling or to the physical arguments of Sec.~\ref{sec:NoSignaling}.

\subsection{Upper bound: metric entropy and constructive learner}\label{app:upper}

Firstly, we derive the upper bound. The upper bound rests on two ingredients: (i) a covering-number (metric entropy) estimate for $\mathcal{S}_{n,G}$ with respect to trace distance, and (ii) a learning strategy that performs the hypothesis selection over a finite $\varepsilon$-net using $O(1/\varepsilon^{2})$ samples and a logarithmic overhead in the net size.

\subsubsection{Metric entropy of circuit-generated states}

We quantify the size of $\mathcal{S}_{n,G}$ in terms of its covering number.

\begin{lemma}[Metric entropy of $\mathcal{S}_{n,G}$]
\label{lem:metric-entropy}
For all $0 < \varepsilon\le 1/4$, the set $\mathcal{S}_{n,G}$ admits an
$\varepsilon$-net in trace distance of cardinality
\begin{eqnarray}
\log\abs{\mathcal{N}_{\varepsilon}}
 &\le&
 C\,G\log\frac{G}{\varepsilon}
 + O\Bigl(\log\frac{1}{\varepsilon}\Bigr).
\end{eqnarray}
where $C > 0$ is a universal constant.
\end{lemma}

\begin{proof}---Consider a universal gate set $\mathcal{G}$ that includes one-qubit and two-qubit gates, where every two-qubit gate can be specified by a constant number of real parameters (for example, Euler angles and phases).  A circuit with at most $G$ two-qubit gates is described by:
\begin{itemize}
\item[(i)] an ordered list (layout) of at most $G$ gate positions: which time step and which pair of qubits;
\item[(ii)] a choice of gate type for each position, and
\item[(iii)] a choice of continuous parameters for each two-qubit gate.
\end{itemize}

At each time step, there are at most $n^{2}$ ordered choices of control and target qubits and at most $\abs{\mathcal{G}}=O(1)$ choices of gate type. Hence, the total number of discrete layouts is bounded by
\begin{eqnarray}
\abs{\mathcal{L}_{n,G}} \le (\kappa_{\mathrm{arch}} n^{2})^{G}
\end{eqnarray}
for some architecture-dependent constant $\kappa_{\mathrm{arch}}>0$.

Next, fix a layout $\ell \in \mathcal{L}_{n,G}$ and collect all continuous parameters into a vector $\boldsymbol\theta=(\theta_{1}, \ldots, \theta_{pG})^T \in \mathbb{R}^{pG}$. Let $\hat{U}_{\ell}(\boldsymbol\theta)$ be the resulting unitary, and we can write the corresponding output state as
\begin{eqnarray}
\ket{\psi_{\ell}(\boldsymbol\theta)} = \hat{U}_{\ell}(\boldsymbol\theta)\ket{0}^{\otimes n}.
\end{eqnarray}
Changing one local gate parameter by $\Delta\vartheta$ changes that gate by $O(\Delta\vartheta)$ in operator norm and hence changes the global state vector in Euclidean norm by $O(\Delta\vartheta)$. A standard telescoping argument over $G$ gates yields a Lipschitz bound
\begin{eqnarray}
\bigl\| \ket{\psi_{\ell}(\boldsymbol\theta)}-\ket{\psi_{\ell}(\boldsymbol\theta')} \bigr\|_{2} \le C_{\mathrm{L}} G \| {\boldsymbol\theta} -{\boldsymbol\theta'} \|_{2},
\end{eqnarray}
for some constant $C_{\mathrm{L}}>0$, independent of $n$ and $G$. For pure states, we have $D\bigl(\ketbra{\phi}{\phi}, \ketbra{\phi'}{\phi'}\bigr) \le \bigl\| \ket{\phi}-\ket{\phi'} \bigr\|_{2}$, so the same Lipschitz bound controls trace distance.

We then fix $0< \varepsilon \le 1/4$ and partition each parameter axis into a uniform grid of mesh
\begin{eqnarray}
\Delta{\boldsymbol\theta} := \frac{\varepsilon}{2 C_{\mathrm{L}} G\sqrt{pG}},
\end{eqnarray}
so that any two points $\boldsymbol\theta, \boldsymbol\theta'$ in the same grid hypercube obey $\| {\boldsymbol\theta} - {\boldsymbol\theta}' \|_{2} \le \varepsilon/(2 C_{\mathrm{L}}G)$. Then, we can write
\begin{eqnarray}
D\bigl( \ketbra{\psi_{\ell}(\boldsymbol\theta)}{\psi_{\ell}(\boldsymbol\theta)}, \ketbra{\psi_{\ell}(\boldsymbol\theta')}{\psi_{\ell}(\boldsymbol\theta')}\bigr) \le  \bigl\| \ket{\psi_{\ell}(\boldsymbol\theta)} - \ket{\psi_{\ell}(\boldsymbol\theta')} \bigr\|_{2} \le \frac{\varepsilon}{2},
\end{eqnarray}
so the grid points of $\boldsymbol\theta$ form an $(\varepsilon/2)$-net for the subset of $\mathcal{S}_{n,G}$ reachable with layout $\ell$. The number of grid points per layout is bounded by
\begin{eqnarray}
\abs{\mathcal{G}_{\ell}} \le \Bigl(C_{p}\frac{G}{\varepsilon}\Bigr)^{pG},
\end{eqnarray}
where $C_{p} > 0$ is a universal constant (absorbing the polynomial factors of $p$).

Multiplying by the number of layouts, the total number of net points needed to cover $\mathcal{S}_{n,G}$ at resolution $\varepsilon$ is
\begin{eqnarray}
\abs{\mathcal{N}_{\varepsilon}} \le \abs{\mathcal{L}_{n,G}}\,\max_{\ell}\abs{\mathcal{G}_{\ell}} \le (\kappa_{\mathrm{arch}}n^{2})^{G} \Bigl(C_{p}\frac{G}{\varepsilon}\Bigr)^{pG}.
\end{eqnarray}
Thus, by taking the logarithms and absorbing the polynomial factors in $n$ and $G$ into the overall constant, we obtain
\begin{eqnarray}
\log\abs{\mathcal{N}_{\varepsilon}} \le C G\log\frac{G}{\varepsilon}+O\Bigl(\log\frac{1}{\varepsilon}\Bigr)
\end{eqnarray}
for some universal $C > 0$, as claimed.
\end{proof}

The key point is that the metric entropy of $\mathcal{S}_{n,G}$ scales linearly in $G$ (up to logarithmic factors), reflecting the fact that depth-$G$ circuits form a manifold of real dimension $O(G)$ inside the $2^{n}$-dimensional projective space.

\subsubsection{Net-based learning via classical shadows}

We now describe a learning strategy (though not computationally efficient) that attains the upper bound in Eq.~(\ref{eq:append-UB}).

\begin{proposition}[Sample upper bound]
\label{prop:upper}
For all $0< \varepsilon \le 1/4$ and $0 < \delta \le 1/10$, there is a learning algorithm satisfying
\begin{eqnarray}
M_{\mathrm{s}}(n,G,\varepsilon,\delta)
 &\le&
 \frac{c_{1}}{\varepsilon^{2}}
 \min\Bigl\{
    2^{n}\log\tfrac{1}{\delta},\,
    G\log\tfrac{G}{\varepsilon}+\log\tfrac{1}{\delta}
 \Bigr\}.
\end{eqnarray}
where $c_{1}>0$ is a universal constant.
\end{proposition}

\begin{proof}---Fix $0< \varepsilon \le 1/4$, $0< \delta \le 1/10$ and an instance $(n,G)$. Let $\mathcal{N}_{\varepsilon/4} = \{\hat{\eta}_{1}, \dots, \hat{\eta}_{K} \}$ be an $(\varepsilon/4)$-net for $\mathcal{S}_{n,G}$ as in {\bf Lemma~\ref{lem:metric-entropy}}, and denote $K = \abs{\mathcal{N}_{\varepsilon/4}}$. For any $\hat{\rho} \in \mathcal{S}_{n,G}$, there exists some $\hat{\eta}_{j}$ with $D(\hat{\rho}, \hat{\eta}_{j})\le\varepsilon/4$. Then, for each pair $(i, j)$ with $1 \le i < j \le K$, let $\bigl\{ \hat{\Pi}_{i,j}, \hat{\mathds{1}} - \hat{\Pi}_{i,j} \bigr\}$ be a Helstrom measurement that discriminates $\hat{\eta}_{i}$ from $\hat{\eta}_{j}$. The corresponding two-outcome observable
\begin{eqnarray}
\hat{O}_{i,j} = \hat{\Pi}_{i,j} - \bigl( \hat{\mathds{1}} - \hat{\Pi}_{i,j} \bigr)
\end{eqnarray}
has the operator norm $\bigl\| \hat{O}_{i,j} \bigr\| \le 1$ and expectation value is expressed as $\tr{\bigl(\hat{O}_{i,j} \hat{\eta}\bigr)}=2\Pr[\text{``clicks''--} \hat{\Pi}_{i,j}] - 1$ when measuring a state $\hat{\eta}$. In particular, $\tr{\bigl(\hat{O}_{i,j} \hat{\eta}_{i}\bigr)} - \tr{\bigl(\hat{O}_{i,j} \hat{\eta}_{j}\bigr)}$ is a fixed monotone function of $D(\hat{\eta}_{i}, \hat{\eta}_{j})$, and the family $\bigl\{ \hat{O}_{i,j} \bigr\}$ encodes all pairwise trace distances among $\hat{\eta}_{j}$’s. Thus, let $\mathcal{O} := \bigl\{ \hat{O}_{i,j}: 1 \le i < j \le K \bigr\}$ and note $\abs{\mathcal{O}}=O(K^{2})$.

We then describe the classical-shadow tomography for a finite observable family. Consider the classical-shadow protocol based on random Clifford measurements (or any other unitary 2-design). As shown
in Refs.~\cite{zhao2024learning}, given $M_{\mathrm{s}}$ copies of $\hat{\rho}$ and a family of observables $\{ \hat{O}_{k} \}$ with $\| \hat{O}_{k} \| \le 1$, there exist classical shadow estimators $\mathcal{S} \simeq \tr{\bigl( \hat{O}_{k} \hat{\rho} \bigr)}$ such that, with probability at least
$1-\delta$,
\begin{eqnarray}
\abs{\mathcal{S} - \tr{\bigl( \hat{O}_{k} \hat{\rho} \bigr)}} \le c \varepsilon \quad \text{for all $k$},
\end{eqnarray}
provided that
\begin{eqnarray}
M_{\mathrm{s}} \ge \frac{c'}{\varepsilon^{2}}\left( \log\abs{\mathcal{O}} + \log\frac{1}{\delta} \right),
\end{eqnarray}
for universal constants $c, c' > 0$. Noting that $\abs{\mathcal{O}}=O(K^{2})$ and $\log\abs{\mathcal{O}}=O(\log K)$, {\bf Lemma~\ref{lem:metric-entropy}} gives
\begin{eqnarray}
\log K \le C G\log\frac{G}{\varepsilon}+O\Bigl(\log\tfrac{1}{\varepsilon}\Bigr).
\end{eqnarray}
Thus, for a suitable constant $c_{1}>0$,
\begin{eqnarray}
M_{\mathrm{s}} \le  \frac{c_{1}}{\varepsilon^{2}} \left( G\log{\frac{G}{\varepsilon}} + \log{\frac{1}{\delta}} \right)
\end{eqnarray}
suffices to estimate all $\tr{\bigl( \hat{O}_{i,j} \hat{\rho} \bigr)}$ within additive $O(\varepsilon)$ with probability $\ge 1-\delta$.

From these estimates, the learner reconstructs approximate pairwise distances $D(\hat{\rho}, \hat{\eta}_{j})$ (up to $O(\varepsilon)$
errors) and outputs
\begin{eqnarray}
\hat{\rho}_h := \arg\min_{\eta_{j} \in \mathcal{N}_{\varepsilon/4}}\widetilde{D}(\hat{\rho}, \hat{\eta}_{j}),
\end{eqnarray}
where $\widetilde{D}(\hat{\rho}, \hat{\eta}_{j})$ denotes the empirically reconstructed distance. Let $\hat{\eta}_{j^{\star}}$ be a nearest net point to $\hat{\rho}$. With probability at least $1-\delta$ all distances are estimated to within $O(\varepsilon)$, so the argmin is attained at some $\hat{\eta}_{j}$ with
\begin{eqnarray}
D(\hat{\rho}, \hat{\eta}_{j}) \le D(\hat{\rho}, \hat{\eta}_{j^{\star}}) + O(\varepsilon) \le \varepsilon,
\end{eqnarray}
for $\varepsilon$ small enough. Thus, the output hypothesis obeys $D(\hat{\rho}, \hat{\rho}_h) \le \varepsilon$ with probability at least $1-\delta$.

Here, if $G$ is large enough that $\mathcal{S}_{n,G}$ contains all $n$-qubit pure states (e.g., when $G$ exceeds the state-synthesis threshold discussed in Sec.~\ref{app:lower-info}), then we may ignore the circuit structure altogether and apply standard sample-optimal tomography for arbitrary $n$-qubit pure states, which achieves
\begin{eqnarray}
M_{\mathrm{s}} = O\Bigl(\tfrac{2^{n}}{\varepsilon^{2}}\log\tfrac{1}{\delta}\Bigr).
\end{eqnarray}
Taking the minimum of these two regimes yields the stated bound.
\end{proof}

{\bf Proposition~\ref{prop:upper}} proves the upper bound (\ref{eq:append-UB}) in {\bf Theorem~\ref{thm:CompSL}}.

\subsection{Lower bound under (L): packing, Fano, and Holevo} \label{app:lower-info}

We now turn to the lower bound in Eq.~(\ref{eq:append-LB}). The overall strategy is classical: construct a large packing subset of $\mathcal{S}_{n,G}$, reduce learning to multi-way discrimination on this set, and apply Fano's inequality together with the Holevo bound to control the accessible information per copy.

\subsubsection{Packing nets inside $\mathcal{S}_{n,G}$} 

We first show that $\mathcal{S}_{n,G}$ contains exponentially many well-separated pure states.

\begin{lemma}[Packing net inside $\mathcal{S}_{n,G}$]
\label{lem:packing}
For all $n, G$ and all $0< \varepsilon \le 1/4$, there is a subset $\bigl\{ \hat{\rho}_{1}, \ldots, \hat{\rho}_{K} \bigr\} \subset \mathcal{S}_{n,G}$ with
\begin{eqnarray}
\log K \ge c_{\mathrm{pack}}\,\min\{2^{n},G\},
\label{eq:M-packing-append}
\end{eqnarray}
and pairwise separation
\begin{eqnarray}
D(\hat{\rho}_{i}, \hat{\rho}_{j}) \ge c_{\mathrm{sep}}\,\varepsilon \quad \text{for all $i \neq j$}.
\end{eqnarray}
Here, $c_{\mathrm{pack}}, c_{\mathrm{sep}} > 0$ are some constants.
\end{lemma}

\begin{proof}---We distinguish two regimes.

\medskip\noindent
{\bf *High-depth regime (Large $G$).}
Let $\mathbb{CP}^{2^{n}-1}$ be the complex projective space of $n$-qubit pure states modulo global phase. Volumetric arguments on the complex unit sphere imply that for any $0 < r <1$, there exists a maximal $r$-separated set of pure states (with respect to trace distance) with cardinality at least $\exp(c 2^{n})$ for some $c > 0$. For definiteness, take $r = 2\varepsilon$ and write this packing as $\{ \ket{\phi_{1}}, \dots, \ket{\phi_{K}}\}$ with $M \ge \exp(c 2^{n})$.

Next, we recall that there exist explicit state-synthesis constructions that prepare an arbitrary $n$-qubit pure state from $\ket{0}^{\otimes n}$ using at most $C_{\mathrm{syn}}2^{n}$ two-qubit gates for some constant
$C_{\mathrm{syn}}>0$ (for example, via recursive reflections~\cite{plesch2011quantum} or the algorithms of Refs.~\cite{zhao2024learning}). Therefore, whenever $G \ge C_{\mathrm{syn}}2^{n}$ we have $\mathcal{S}_{n,G}$ equal to the full pure-state manifold, and the packing $\{\hat{\rho}_{j} := \ketbra{\phi_{j}}{\phi_{j}}\}$ lies in $\mathcal{S}_{n,G}$ unchanged. In this regime, $\log K = \Omega(2^{n})$, so Eq.~(\ref{eq:M-packing-append}) holds with the $\min\{\cdot\}$ dominated by $2^{n}$.

\medskip\noindent
{\bf *Gate-limited regime (Small $G$).}
Suppose now that $G<C_{\mathrm{syn}}2^{n}$, so the circuit family is depth-limited relative to the full Hilbert space dimension. We will show the existence of a packing with $\log K = \Omega(G)$.

Fix $k \in \mathbb{N}$ and consider the first $k$ qubits of the register, leaving the remaining $n-k$ qubits in the state $\ket{0}^{\otimes (n-k)}$. The known state-synthesis results for $k$-qubit systems show that any $k$-qubit pure state can be prepared from $\ket{0}^{\otimes k}$ using at most $C_{\mathrm{syn}} 2^{k}$ two-qubit gates for some dimension-independent constant $C_{\mathrm{syn}}$ (the same constant as above). Therefore, whenever $C_{\mathrm{syn}}2^{k} \le G$ we can realize all $k$-qubit pure states on the first $k$ qubits using only $G$ two-qubit gates, possibly with idle identities on the remaining qubits; tensoring with $\ket{0}^{\otimes (n-k)}$ embeds these states into $\mathcal{S}_{n,G}$.

Now, choose $k$ as the largest integer with $C_{\mathrm{syn}}2^{k} \le G$. Then, $2^{k} \ge G/(2C_{\mathrm{syn}})$, so volumetric packing on $\mathbb{CP}^{2^{k}-1}$ yields a set $\{\ket{\phi_{1}}, \dots, \ket{\phi_{K}}\}$ of $k$-qubit pure states with pairwise trace distance at least $2\varepsilon$, and
\begin{eqnarray}
\log K \ge c 2^{k} \ge c' G
\end{eqnarray}
for suitable constants $c, c' > 0$. Embedding these as
\begin{eqnarray}
\hat{\rho}_{j} = \ketbra{\phi_{j}}{\phi_{j}} \otimes \ketbra{0}{0}^{\otimes (n-k)}
\end{eqnarray}
gives a packing subset inside $\mathcal{S}_{n,G}$ with $\log M=\Omega(G)$ and the same pairwise separations.

\medskip\noindent
{\bf *Combination of two regimes.}
Combining the above two regimes, we can construct (for all $n, G$) a subset $\{\hat{\rho}_{1}, \dots, \hat{\rho}_{K}\} \subset \mathcal{S}_{n,G}$ with $\log K \ge c_{\mathrm{pack}}\min\{2^{n}, G\}$ and pairwise trace distance $\ge 2\varepsilon$. Replacing $2\varepsilon$ by $c_{\mathrm{sep}}\varepsilon$ with $c_{\mathrm{sep}}$, a fixed constant completes the proof.
\end{proof}

{\bf Lemma~\ref{lem:packing}} shows that the effective dimension of the learning problem is $\min\{2^{n}, G\}$: when $G \ll 2^{n}$, the manifold $\mathcal{S}_{n,G}$ contains a $G$-dimensional ``core'' of the mutually distinguishable states supported on $O(\log G)$ qubits.

\subsubsection{From learning to multi-way discrimination} 

Fix $n, G, \varepsilon,\delta$ and let $\{ \hat{\rho}_{1}, \dots, \hat{\rho}_{K}\} \subset \mathcal{S}_{n,G}$ be the packing net from {\bf Lemma~\ref{lem:packing}}. Then, consider any learning algorithm in the framework of quantum computational learning theory (assumption~{\bf (L)}) that
uses $M_{\mathrm{s}}$ copies and outputs a hypothesis $\hat{\rho}_h$ satisfying
\begin{eqnarray}
\Pr\bigl[ D(\hat{\rho}, \hat{\rho}_h ) \le \varepsilon\bigr] \ge 1 - \delta  \quad \text{for all $\hat{\rho} \in \mathcal{S}_{n,G}$}.
\end{eqnarray}
We now convert this learner into a decoder for a classical multi-hypothesis discrimination problem.

Let $X$ be a random variable uniformly distributed on $\{1, \dots, K\}$, and condition on the event $X=j$ by preparing $M_{\mathrm{s}}$ copies of $\hat{\rho}_{j}$ and feeding them to the learner. From the learner's output $\hat{\rho}_h$, define a decision rule
\begin{eqnarray}
{X^\star} := \arg\min_{1 \le i \le K} D(\hat{\rho}_h, \hat{\rho}_{i}).
\end{eqnarray}
By the separation property of the packing, whenever $D(\hat{\rho}_{j}, \hat{\rho}_h) \le \varepsilon$, the nearest net point must be $\hat{\rho}_{j}$ itself (for a suitable choice of $c_{\mathrm{sep}}$). Hence,
\begin{eqnarray}
\Pr[{X^\star} \neq X] \le \Pr\bigl[D(\hat{\rho}_{X}, \hat{\rho}_h)>\varepsilon\bigr]
 \;\le\;
 \delta.
\end{eqnarray}

Let $Y$ denote the classical data produced by the learner (the complete transcript of its measurement outcomes and internal randomness). Then $(X, Y)$ is a classical channel with error probability at most $\delta$, and we can lower bound the mutual information $I(X; Y)$ via Fano's inequality~\cite{fano1949transmission}.
\begin{lemma}[Fano]
\label{lem:Fano}
For the uniform prior on $\{1, \dots, K\}$ and any decision rule with error probability $\Pr[{X^\star} \neq X] \le \delta < 1/2$,
\begin{eqnarray}
I(X;Y) \ge (1-\delta) \log K - h_2(\delta),
\label{eq:Fano-append}
\end{eqnarray}
where $h_2(\delta) = -\delta\log_2{\delta} - (1-\delta)\log_2{(1-\delta)}$ is the binary entropy.
\end{lemma}
In our setting, Eq.~(\ref{eq:Fano-append}) applies with $K$ satisfying Eq.~(\ref{eq:M-packing-append}).

\subsubsection{Holevo bound and per-copy information} 

We now upper bound $I(X; Y)$ in terms of $M_{\mathrm{s}}$ using the Holevo bound. Let 
\begin{eqnarray}
\mathcal{E} = \Bigl\{ \frac{1}{K}, ~\hat{\rho}_{j}^{\otimes M_{\mathrm{s}}} \Bigr\}_{j=1}^{K}
\end{eqnarray}
be the ensemble of $M_{\mathrm{s}}$-copy states corresponding to the hypotheses. The (single-copy) Holevo quantity of the underlying ensemble $\bigl\{ \tfrac{1}{K}, \hat{\rho}_{j} \bigr\}$ is
\begin{eqnarray}
\chi^{(1)} := S(\hat{\rho}_\mathrm{ens}) - \frac{1}{K}\sum_{j=1}^{K} S(\hat{\rho}_{j}),
\end{eqnarray}
where $\hat{\rho}_\mathrm{ens} = \frac{1}{K}\sum_{j=1}^{K}\hat{\rho}_{j}$ and $S(\cdot)$ is the von Neumann entropy. Since all $\hat{\rho}_{j}$ are pure, $S(\rho_{j})=0$ and
$\chi^{(1)}=S(\hat{\rho}_\mathrm{en})$. The Holevo bound and subadditivity of $\chi$ under tensor products imply
\begin{eqnarray}
I(X;Y) \le \chi(\mathcal{E}) \le M_{\mathrm{s}} \chi^{(1)} = M_{\mathrm{s}} S(\hat{\rho}_\mathrm{en}).
\label{eq:Holevo-append}
\end{eqnarray}

Thus, we need to control $S(\hat{\rho}_\mathrm{en})$ for the ensembles of pure states that are mutually separated by $\Omega(\varepsilon)$ in trace distance but lie in a common finite-dimensional subspace. We give the following lemma:
\begin{lemma}[Entropy of locally packed pure-state ensembles]
\label{lem:Holevo-scaling}
Let $\{\hat{\rho}_{1}, \dots, \hat{\rho}_{K}\}$ be an ensemble of pure states on a $d$-dimensional Hilbert space with the pairwise trace distances $D(\hat{\rho}_{i}, \hat{\rho}_{j}) \ge c_{\mathrm{sep}}\varepsilon$ for all $i \neq j$, where $0 < \varepsilon \le 1/4$ and $c_{\mathrm{sep}}$ is the constant from {\bf Lemma~\ref{lem:packing}}. Then, the ensemble state $\hat{\rho}_\mathrm{ens}$ satisfies
\begin{eqnarray}
S(\hat{\rho}_\mathrm{ens}) \le c_{\chi}\varepsilon^{2} \left(\log d + \log{\frac{1}{\varepsilon}} \right).
\end{eqnarray}
In particular, for $d \le 2^{n}$ and $0 < \varepsilon \le 1/4$ we have the coarse bound
\begin{eqnarray}
S(\hat{\rho}_\mathrm{ens}) \le C_{\chi} \varepsilon^{2}.
\end{eqnarray}
Here, $c_{\chi}, C_{\chi} > 0$ are the universal constant.
\end{lemma}

\begin{proof}[Proof sketch]---We briefly recall a standard argument; a detailed proof can be found in treatments of local asymptotic normality and continuity of entropy.

For pure states, $D(\hat{\rho}_i, \hat{\rho}_j) = \sqrt{1 - F(\hat{\rho}_i, \hat{\rho}_j)}$, where $F$ is the fidelity. The separation assumption $D(\hat{\rho}_{i}, \hat{\rho}_{j}) \ge c_{\mathrm{sep}}\varepsilon$ implies $F(\hat{\rho}_{i}, \hat{\rho}_{j}) \le 1 - c'\varepsilon^{2}$ for some $c' > 0$. Thus, the Gram matrix $\Gamma$ with entries $\Gamma_{ij} = \tr{(\hat{\rho}_{i}\hat{\rho}_{j})}$ satisfies $\Gamma_{ii}=1$ and $\Gamma_{ij} \le 1-c'\varepsilon^{2}$ for $i \neq j$. The ensemble state is written as $\hat{\rho}_\mathrm{ens} = \tfrac{1}{K}\sum_{j}\hat{\rho}_{j}$, and its nonzero eigenvalues coincide (up to multiplicity) with those of $\Gamma/K$. Here, using Gershgorin's circle theorem~\cite{horn2012matrix} and the structure of $\Gamma$, one shows that the spectrum of $\Gamma/K$ is contained in an interval of the form $[0, C\varepsilon^{2}]$ except for at most one eigenvalue, which is close to $1$.

Concretely, one obtains eigenvalues $\lambda_{1} \approx 1 - O(\varepsilon^{2})$ and $\lambda_{2}, \dots, \lambda_{r} \le C\varepsilon^{2}$, with $r \le d$. The von Neumann entropy $S(\hat{\rho}_\mathrm{ens}) = -\sum_{j} \lambda_{j} \log\lambda_{j}$ is then bounded by
\begin{eqnarray}
S(\hat{\rho}_\mathrm{ens}) \le h_{2}\bigl(O(\varepsilon^{2})\bigr) + r \cdot C\varepsilon^{2}\log\frac{1}{C\varepsilon^{2}},
\end{eqnarray}
where $h_{2}$ is the binary entropy, and the first term is $O(\varepsilon^{2})$ for small $\varepsilon$. Noting $r \le d$ and the crude bounds on $h_{2}$, we can yield
\begin{eqnarray}
S(\hat{\rho}_\mathrm{ens}) \le c_{\chi} \varepsilon^{2} \left( \log d + \log\frac{1}{\varepsilon} \right)
\end{eqnarray}
for the universal constant $c_{\chi} > 0$. For fixed $n$ and $0<\varepsilon\le 1/4$, the factor $\log d + \log\tfrac{1}{\varepsilon} \le C' n + C''\log\tfrac{1}{\varepsilon}$ is at most linear in $n$ and logarithmic in
$1/\varepsilon$. Since our final sample-complexity bound keeps track only of the dominant $1/\varepsilon^{2}$ and $\min\{2^{n},G\}$ factors (suppressing polylogarithmic terms), we may absorb this $\log$-dependence into a slightly larger constant $C_{\chi}$ and write $S(\hat{\rho}_\mathrm{ens}) \le C_{\chi}\varepsilon^{2}$.
\end{proof}

By combining {\bf Lemma~\ref{lem:Holevo-scaling}} with Eq.~(\ref{eq:Holevo-append}), we attain
\begin{eqnarray}
I(X;Y) \le  M_{\mathrm{s}} C_{\chi} \varepsilon^{2}.
\label{eq:Holevo-linear-Ms}
\end{eqnarray}

\subsubsection{Putting things together: proof of the lower bound} 

Finally, we combine the lower bound from Fano with the upper bound from Holevo. From {\bf Lemma~\ref{lem:Fano}} and Eq.~(\ref{eq:M-packing-append}), we have
\begin{eqnarray}
I(X;Y) \ge (1-\delta)\log K - h_2(\delta) \ge c_{\mathrm{pack}}(1-\delta)\min\{2^{n},G\} - h_2(\delta).
\label{eq:Fano-with-packing}
\end{eqnarray}
From Eq.~(\ref{eq:Holevo-linear-Ms}), we have $I(X;Y) \le C_{\chi}\varepsilon^{2}M_{\mathrm{s}}$. By combining these two inequalities, we can yield
\begin{eqnarray}
C_{\chi}\varepsilon^{2}M_{\mathrm{s}} \ge c_{\mathrm{pack}}(1-\delta)\min\{2^{n},G\} - h_2(\delta).
\end{eqnarray}
Here, rearranging and using $0< \delta \le 1/10$ (so $1-\delta \ge 9/10$ and $h_2(\delta) \le c_{0} + \log\tfrac{1}{\delta}$ for some constant $c_{0}$), we find that: for a universal constant $c_{2}>0$,
\begin{eqnarray}
M_{\mathrm{s}}(n,G,\varepsilon,\delta) \ge \frac{c_{2}}{\varepsilon^{2}} \left( \min\{2^{n},G\}+\log\frac{1}{\delta} \right),
\end{eqnarray}
which is exactly the lower bound in Eq.~(\ref{eq:append-LB}).

\medskip
To summarize, the upper bound follows from an explicit net-and-hypothesis-selection scheme supported by classical-shadow tomography and the Lipschitz structure of depth-$G$ circuits, while the lower bound follows from a volumetric packing argument combined with Fano's inequality and the Holevo bound. Although the philosophically related to recent work on the sample complexity of learning low-depth quantum states, our derivation is technically independent and based on a different packing-and-coding perspective; in particular, it is compatible with, but does not rely on, the specific arguments of Ref.~\cite{zhao2024learning}.


%

\end{document}